\documentclass[preprint2]{emulateapj}
\usepackage{graphicx}
\usepackage{longtable}
\usepackage{mdwlist}
\usepackage{color}
\usepackage{enumitem}
\newenvironment{QandA}{\begin{enumerate}}
                      {\end{enumerate}}
\newenvironment{answered}{\par\normalfont}{}
\usepackage{lipsum}
\slugcomment{version \today: fm}
\shorttitle{Deciphering the large-scale environment of radio galaxies in the local Universe}
\shortauthors{F. Massaro et al.}

\begin{document}
\title{Deciphering the large-scale environment of radio galaxies in the local Universe: where do they born, grow and die?}
\author{F. Massaro\altaffilmark{1,2,3}, N. \'Alvarez-Crespo\altaffilmark{1,2,3}, A. Capetti\altaffilmark{2}, \\ R. D. Baldi\altaffilmark{4}, I. Pillitteri\altaffilmark{5}, R. Campana\altaffilmark{6} \& A. Paggi\altaffilmark{1,2,3}}
\altaffiltext{1}{Dipartimento di Fisica, Universit\`a degli Studi di Torino, via Pietro Giuria 1, I-10125 Torino, Italy.}
\altaffiltext{2}{INAF-Osservatorio Astrofisico di Torino, via Osservatorio 20, 10025 Pino Torinese, Italy.}
\altaffiltext{3}{Istituto Nazionale di Fisica Nucleare, Sezione di Torino, I- 10125 Torino, Italy.}
\altaffiltext{4}{Department of Physics and Astronomy, University of Southampton, Highfield, SO17 1BJ, UK.}
\altaffiltext{5}{INAF-Osservatorio Astronomico di Palermo G.S. Vaiana, Piazza del Parlamento 1, 90134, Italy.}
\altaffiltext{6}{INAF/OAS, via Piero Gobetti 101, I-40129, Bologna, Italy.}

\begin{abstract} 
The role played by the large-scale environment on the nuclear activity of radio galaxies (RGs), is still not completely understood. Accretion mode, jet power and galaxy evolution are connected with their large-scale environment from tens to hundreds of kpc. Here we present a detailed, statistical, analysis of the large-scale environment for two samples of RGs up to redshifts $z_\mathrm{src}$=0.15. The main advantages of our study, with respect to those already present in the literature, are due to the extremely homogeneous selection criteria of catalogs adopted to perform our investigation. This is also coupled with the use of several clustering algorithms. We performed a direct search of galaxy-rich environments around RGs using them as beacon. To perform this study we also developed a new method that does not appear to suffer by a strong $z_\mathrm{src}$ dependence as other algorithms. We conclude that, despite their radio morphological (FR\,I $vs$ FR\,II) and/or their optical (HERG $vs$ LERG) classification, RGs in the local Universe tend to live in galaxy-rich large-scale environments having similar characteristics and richness. We highlight that the fraction of FR\,Is-LERG, inhabiting galaxy rich environments, appears larger than that of FR\,IIs-LERG. We also found that 5 out of 7 FR\,II-HERGs, with $z_\mathrm{src}\leq$0.11, lie in groups/clusters of galaxies. However, we recognize that, despite the high level of completeness of our  catalogs, when restricting to the local Universe, the low number of HERGs ($\sim$10\% of the total FR\,IIs investigated) prevent us to make a strong statistical conclusion about this source class.
\end{abstract}


\section{Introduction}
\label{sec:intro} 
In the early 70's Fanaroff \& Riley proposed to classify extragalactic radio sources, having an extended structure resolved in two or more components at 1.4 GHz. Their scheme is based on the ratio $R_\mathrm{FR}$ of the angular separation between regions of highest surface brightness on the opposite sides of the central radio galaxy or quasar, to the total extent of the source measured up to the lowest contour level. Any compact component located on the central galaxy, like the radio core, was not taken into account. Radio sources having $R_\mathrm{FR}\leq$0.5 (i.e, edge-darkened) where placed in class I, namely FR\,Is, while those for which $R_\mathrm{FR}>$0.5 (i.e, edge-brightened) in class II, known as FR\,IIs \citep{fanaroff74}. 

This radio morphological distinction corresponds to a sharp division in luminosities. Radio sources having $L_\mathrm{178MHz}$ lower than 2$\times$ 10$^{25}$ W\,Hz$^{-1}$ s$^{-1}$ appeared to be almost all FR\,Is while those above this threshold being FR\,IIs. This luminosity threshold was remarkably close to the dividing line between radio sources with strong and weak cosmological evolution \citep[see e.g.,][]{longair71}.

This FR classification scheme was then linked to the environment on Mpc scale of the extragalactic radio galaxies a few decades later. It was found that FR\,Is generally inhabiting galaxy-rich environments, being members of groups or galaxy clusters, while FR\,IIs tend to live more isolated \citep[see e.g.,][]{zirbel97}, with well-known exceptions were already known \citep[see e.g., ][for a recent analysis of the X-ray observations of FR\,II]{hardcastle00}, as the archetypal Cygnus A \citep[see e.g.,][for a review]{carilli96}. 

In the last decade a firm link between optical emission, accretion mode and host galaxy properties, including star formation rate, was estabilished fofr the radio galaxy population \citep[see e.g.,][]{baldi08,balmaverde08,tasse08,smolcic09,baldi10,hardcastle13,mingo14}. An additional classification was developed for radio galaxies in the 80's. This was based on the properties of their optical emission lines \citep{hine79}, distinguishing between high and low excitation radio galaxies \citep[HERGs and LERGs, respectively; see also][]{laing94}. Their differences are not simply related to the orientation of the active galaxy with respect to the line of sight but are also related to their accretion modes (i.e., radiatively efficient $vs$ inefficient) \citep[see e.g.,][ and reference therein]{chiaberge02,hardcastle06,hardcastle09,best12}. In addition, HERGs appear to have, almost exclusively, an FR\,II radio morphology, while LERGs can be FR\,I or FR\,II \citep[see e.g.,][]{hine79,laing94}. Hence accretion mode does not directly determine radio morphological class \citep{heckman14}.  

As occurs for FR\,Is and FR\,IIs, LERGs are preferentially low luminosity radio sources, mostly lying at low redshifts $z_\mathrm{src}$\footnote{Here we adopt the symbol $z_\mathrm{src}$ to indicate the source redshift rather than the usual $z$ to distinguish it from the redshift of a possible nearby galaxy group or cluster, labelled as $z_\mathrm{cl}$.}, while HERGs dominate the high radio luminosity sky, being at higher $z_\mathrm{src}$. This appears clear even considering extragalactic sources selected out of radio surveys with high flux limits and large beams that, as recently shown, are also not representative of the whole radio galaxy population \citep[see e.g.,][]{capetti17a}. Therefore it is crucial to consider both radio and optical classifications for the radio galaxy population while investigating their large-scale environments.

Using the tenth-nearest-neighbour estimator in the $z_\mathrm{src}$ range between 0.02 and. 0.10, Best et al. (2004) found that radio-loud active galaxies are preferentially located in galaxy groups and poor-to-moderate richness galaxy clusters, consistent with previous results \citep[see e.g.,][]{prestage88,hill91}. In particular, the flux ratio of absorption-line to emission-line changes dramatically with the environment, having essentially all radio-loud active galaxies in rich environments showing no emission lines \citep[see e.g.,][]{best04}. Thus a considerable care must be put in selecting samples of radio-loud active galaxies from their optical emission-line properties (LERGs $vs$ HERGs), since, investigating how environment properties are related to their optical spectra, selection criteria should not be related to their optical properties.

Recently, Gendre et al. (2013) showed that at a given radio luminosity $L_R$ at 1.4 GHz, the FR morphological dichotomy is consistent with both accretion modes even when restricting to only rich or only poor environments. This could imply that radio morphology is independent of the accretion mode and depends on the jet power and its interactions with the larger-scale environment. Thus, FR\,Is lie in higher density environments than FR\,IIs. This picture is therefore consistent with FR\,Is having jets disrupted by a denser surrounding medium \citep{bicknell94,laing08}. Gendre et al. (2013) also claimed that accretion modes could be linked to the large-scale environment, with HERGs living almost exclusively in low-density environments and LERGs inhabiting a wider range of galaxy densities, independently of their radio morphology. 

Using X-ray observations, Ineson et al. (2013) performed a systematic study of cluster environments of radio galaxies at $z_\mathrm{src}\sim$0.5. They found tentative evidence for a correlation between radio luminosity and cluster X-ray luminosity, possibly driven by the LERG sub-population. Then, at  $z_\mathrm{src}\sim$0.1, Ineson et al. (2015) claimed a stronger link between radio luminosity and richness, as in Best et al. (2004), and between radio luminosity and central density for LERGs, but not for HERGs, although there are less HERGs at low $z_\mathrm{src}$. No differences in LERGs were found between the two analyses.

In contrast with both Best et al. (2004) and Ineson et al. (2015) results, Belsole et al. (2007) found no link between radio luminosity and galaxy density at higher $z_\mathrm{src}$; however their sample could be biased towards a selection of HERGs.

More recently Miraghei \& Best (2017) compared FR\,I LERGs with FR\,II LERGs at fixed stellar mass and radio luminosity showing that the former ones typically reside in richer environments and are hosted by smaller galaxies with higher mass surface density. This picture is again consistent with jet disruption effects, a possible driver of the FR dichotomy.

Finally, adopting the fifth nearest neighbor density $\Sigma_5$, as in Best et al. (2004) analysis, Ching et al. (2017) confirmed previous results with a larger sample. LERGs and HERGs exist in different large-scale environments depending on their radio luminosity, with high radio luminosity LERGs more likely to be in galaxy groups. In contrast, the environments of HERGs and low luminosity LERGs are indistinguishable from that of a radio-quiet control sample.

Comparing claims and results from different analyses, carried out with different techniques and on different samples, requires extreme caution. Methods to estimate the cluster richness or procedures to associate a source with a galaxy group or cluster or differences in the region sizes selected for galaxy counts, could introduce biases. In addition, the possible evolution of the environments with $z_\mathrm{src}$, changes in HERG and LERG populations with $z_\mathrm{src}$, lack of powerful sources in our local Universe and the Malmquist bias in flux limited catalogs could also affect analyses and comparisons.  Nevertheless, analyses based on ill-defined small groups of sources and, as recently shown, conclusions based on samples selected from  radio surveys with high flux limit and large beam, as the Third Cambridge catalog \citep[3C; see e.g.,][]{edge59,bennett62,spinrad85}, could be also strongly affected by selection biases \citep{capetti17a,capetti17b}.

To shed a light on the role played by the large-scale environment on the nuclear activity of radio galaxies, here we present a detailed study of the large-scale environment of radio galaxies using well-defined and statistically homogenous catalogs of FR\,I (LERGs) and FR\,II (LERGs and HERGs) radio galaxies at $z_\mathrm{src}\leq$0.15. We highlight differences and advantages of the analysis carried out here in comparison with literature studies.

The paper is organized as follows. In \S~\ref{sec:samples} we present all samples and catalogs used to carry out our analysis while in \S~\ref{sec:definition} we outline definitions adopted to perform our investigation. Then in \S~\ref{sec:analysis} we describe $step-by-step$ the clustering procedure used. In \S~\ref{sec:results} we discuss results obtained having then \S~\ref{sec:summary} devoted to our summary and conclusions. A comparison with literature claims is presented in \S~\ref{sec:compare}. Finally, in \S~\ref{sec:future} we discuss on future perspectives and possible developments of our analysis achievable with dedicated X-ray observations. Technical details on clustering algorithms used here are fully described in the Appendix.

We adopt cgs units for numerical results and we also assume a flat cosmology with $H_0=69.6$ km s$^{-1}$ Mpc$^{-1}$, $\Omega_\mathrm{M}=0.286$ and $\Omega_\mathrm{\Lambda}=0.714$ \citep{bennett14}, unless otherwise stated. Thus, 1\arcsec\ corresponds to 0.408 kpc at $z_\mathrm{src}=$0.02 and 2.634 kpc at $z_\mathrm{src}=$0.15, given the values of the cosmological constants previously reported.

\section{Sample Selection}
\label{sec:samples} 
Several source samples and catalogs have been used to carry out our analysis: (i) two catalogs of radio galaxies, extremely homogenous and carefully selected on the basis of multifrequency observations; (ii) a catalog of random positions in the Sloan Digitail Sky Survey (SDSS) footprint; (iii) a sample of quiescent elliptical galaxies and (iv) two catalog of groups and clusters of galaxies, again based on the SDSS observations. Here we described them briefly with particular attention to their selection criteria.

\subsection{Radio Galaxies}
We recently created two catalog of FR\,I and FR\,II radio galaxies \citep[i.e., FRICAT and FRIICAT respectively;][]{capetti17a,capetti17b} combining observations available in the SDSS Data Release 9 \citep{ahn12}, the National Radio Astronomy Observatory (NRAO) Very Large Array (VLA) Sky Survey \citep{condon98} and the Faint Images of the Radio Sky at Twenty-centimeters (FIRST) survey \citep{white97}. All sources in these catalogs have optical spectra that allowed us to obtain their $z_\mathrm{src}$ and determine their LERG $vs$ HERG classification, precisely and unambiguously.

Radio galaxy catalogs were selected starting from the original sample of Best \& Heckman (2012). We first consider only those sources classified as AGNs, and we then performed a cut at $z_\mathrm{src}$ lower than 0.15. This led to select 3356 sources out of the original 18286. Subsequently we visually inspected all FIRST images for each individual source selecting only those having radio emission beyond 30 kpc, measured from the position of the optical host galaxy. Radio contours of surface brightness were built at the level of 0.45 mJy/beam, thus matching the FIRST sensitivity and taking into account of the cosmological dimming of the surface brightness. The total number of radio sources selected decreases to 743. Then we performed a final classification distinguishing between FR\,Is and FR\,IIs.

For the present analysis we restricted our radio galaxy catalogs to those sources lying in the central part of the SDSS footprint \citep[see e.g.,][]{ahn12}, same area covered by the main catalog of groups and clusters of galaxies adopted in our analysis (see following sections). In this way, the FRICAT, that includes 219 radio galaxies, all optically classified as LERGs, spanning a redshift range between 0.02 and 0.15, was reduced to 195 sources, while for the FRIICAT the number of sources decreased from 129 to 115 with $z_\mathrm{src}$ between 0.045 and 0.15. In the FRIICAT there are 14 radio galaxies classified as HERGs while all the others are LERGs. 

At a given [OIII] luminosity, sources listed in the FRICAT show radio luminosities spanning about two orders of magnitude and extending to much lower ratios between radio and line power than the FR\,Is listed in the 3C catalog \citep[see][for additional details]{capetti17a}. On the other hand, the majority of the FR\,IIs listed in the FRIICAT have a radio luminosities lower, up to two orders of magnitude, than the threshold one between FR\,Is and FR\,IIs of the 3C catalog \citep[see e.g.][for more details]{capetti17b}. For both catalogs the relation between the morphological classification and radio luminosity disappears considering low power radio sources.

\subsection{Mock sources}
In our analysis we used a catalog of mock sources (labeled as MOCK hereinafter) to estimate the efficiency of our procedures and to estimate their uncertainties. This has been created shifting the position of all FRICAT and FRIICAT radio galaxies by a random radius between 2 and 3 degrees in a random direction of the sky, up to obtain 5000 fake sources/positions. The range of values for the random shift were chosen larger the maximum angular separation corresponding to 2\,Mpc in the radio galaxy catalogs (i.e. 1.1 degrees) and smaller than 3 degrees to preserve the sky distribution of sources in the SDSS footprint. Similar procedures has been already successfully adopted in previous analyses with optical and infrared catalogs \citep[see e.g.,][]{massaro11,dabrusco14,massaro14}.

We then removed from the MOCK sample all sources having a radio counterpart within 5\arcsec. To preserve a redshift distribution of the MOCK catalog, similar to that of that of radio galaxies, we verified that the source ratios between the two catalogs, per bin of $z_\mathrm{src}$ equal to 0.01, is at least 15. Finally we highlight that to create the MOCK catalog we kept all optical magnitudes associated with $z_\mathrm{src}$ to preserve also the luminosity distribution similar to that o radio galaxies. 

The final MOCK sample lists 4056 sources, more than an order of magnitude larger than the total number of radio galaxies considered (i.e., 310).

\subsection{Quiescent Elliptical Galaxies}
We also built a catalog of {\it quiescent elliptical galaxies} (hereinafter ELL). This allow us to investigate optical colors of sources in the large-scale environment of radio galaxies. This catalog will be used only to search for elliptical galaxies surrounding our radio galaxies and to estimate their local source density but not for a comparison with the radio galaxy catalogs.

\begin{enumerate} 
\item We first considered all sources listed in the Galaxy Zoo\footnote{https://www.galaxyzoo.org} data release 1, including 667944 sources \citep{lintott08}. 
\item We then those having a single counterpart in the SDSS data release 9 within 5\arcsec. We considered only galaxies with SDSS flags: {\it spType}, {\it spClass} equal to GALAXY and {\it subclass} NULL. We chose only those objects having an elliptical classification based on at least 45 votes, according to the Galaxy Zoo analysis, and with spectroscopic $z_\mathrm{src}$ smaller than 0.15, as for radio galaxies.
\item We included only elliptical galaxies with a clean photometry (i.e., SDSS flags {\it q\_mode}=1 and {\it Q}$>$2) and classified as {\it galaxies} (i.e., SDSS flag {\it cl} equal to 3).
\item We excluded sources for which the Galaxy Zoo classification is uncertain.
\item We did not include galaxies having an uncertain estimate of $z_\mathrm{src}$.
\item Sources with a radio counterpart within 5\arcsec\ were also excluded to avoid a possible radio galaxy contamination.
\end{enumerate}

\subsection{Catalogs of groups and clusters of galaxies}
Several catalogs of galaxy clusters and groups are available for the SDSS footprint. We selected the one created by Tempel et al. (2012, hereinafter T12) To carry out our analysis since it has the largest number of cluster/group detections with spectroscopic redshifts. This catalog of groups and clusters was created using a modified version of the Friends-of-Friends (FoF) algorithm \citep{hucra82,tago10}. Its redshift $z_\mathrm{cl}$ distribution spans a range between 0.009 and 0.20 peaking around 0.08 and thus becoming less efficient at larger $z_\mathrm{cl}$ values. 

We considered only groups and clusters with a spectroscopic redshift estimate listed in the T12 catalog, for a total of 77858 sources for which the galaxy density, indicated by the {\it N$_\mathrm{gal}$} parameter, was also computed.

Then, we also considered a second catalog of galaxy groups and clusters built using a new Gaussian Mixture Brightest Cluster Galaxy (GMBCG) algorithm \citep{hao10}. This was created using the {\it red sequence} \citep{visvanathan77,gladders98} combined with the search for a Brightest Cluster Galaxy (BCG). The GMBCG catalog was chosen since it is more efficient that the T12 at $z_\mathrm{cl}$ larger than 0.08. The GMBCG catalog, including only  1296 with spectroscopic $z_\mathrm{cl}$ below 0.15 out of 55424 clusters/groups, allowed us to verify the number of BCG candidates in our radio galaxy and quiescent elliptical catalogs.

\section{Cosmological neighbors and \\ candidate elliptical galaxies}
\label{sec:definition} 
For each radio galaxy we downloaded a table listing all optical sources, detected in the SDSS DR9, with a clean photometry (i.e., SDSS flags: {\it q\_mode}=1 and {\it Q}=3 and {\it mode}=1), lying within a 2\,Mpc radius, computed at $z_\mathrm{src}$ of the central source. A radius of 2\,Mpc was chosen to be slightly larger than the  typical size of massive galaxy clusters \citep[i.e., $R_\mathrm{200} \sim$ 1.4\,Mpc;][]{rines13}. We then defined two types of sources in their environment.
\begin{center} 
{\it A. Cosmological neighbors:}
\end{center}
all optical sources lying within the 2\,Mpc radius computed at $z_\mathrm{src}$ of the central object with all the SDSS magnitude flags indicating a galaxy-type object (i.e., {\it uc=rc=gc=ic=zc=}3), and having a spectroscopic redshift $z$ with $\Delta\,z=|z_\mathrm{src}-z|\leq$0.005 (i.e., $\sim$1500 km/s). This $\Delta\,z$ choice corresponds to the maximum velocity dispersion in groups and clusters of galaxies \citep[see e.g.,][]{moore93,eke04,berlind06}.

\begin{center} 
{\it B. Candidate elliptical galaxies:}
\end{center}
all optical sources, lying within the 2\,Mpc distance from the central radio galaxy, estimated at $z_\mathrm{src}$, and having $u-r$ and the $g-z$ colors consistent with those of the ELL sample within $\Delta\,z$$<$0.005. This color-color selection is based on the iso-density contours, computed adopting the Kernel density Estimation \citep[KDE: see e g.,][]{richards04,dabrusco09,massaro13a}, at 90\% a level of confidence. Source selected as {\it candidate elliptical galaxies} do not necessarily have spectroscopic redshifts, they have only the same colors of elliptical galaxies at the radio galaxy $z_\mathrm{src}$. We only perform this selection for elliptical-type galaxies, in the large-scale environment of our radio galaxies, because their fraction in galaxy groups or cluster is much larger than that of spirals \citep[see e.g.,][]{biviano00}.

In Figure~1 we show the color-color plot ($u-r$ and $g-z$) for optical sources surrounding SDSSJ080113.28+344030.8. Sources in the ELL sample within a $\Delta\,z$$<$0.005 centered at $z_\mathrm{src}$ of SDSSJ080113.28+344030.8 are reported as cyan circles together with their KDE iso-density contours (black) while blue crosses are the {\it candidate elliptical galaxies} in its 2\,Mpc field. 
\begin{figure}[h!]
\begin{center}
\includegraphics[height=8.cm,width=8.cm,angle=0]{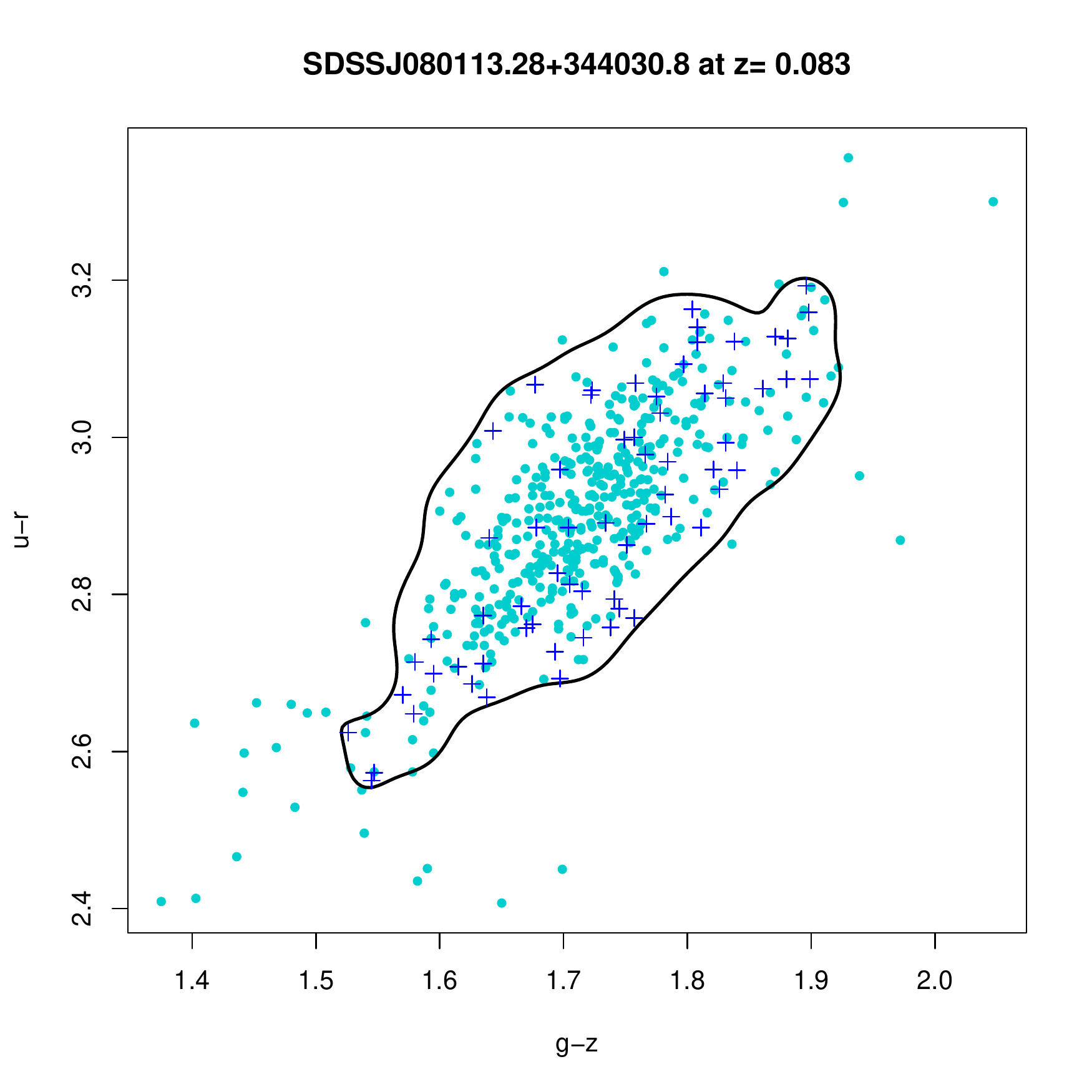}
\caption{The panel shows the $u-r$ $vs$ $g-z$ color-color plot with cyan circles representing quiescent elliptical galaxies, in the ELL sample, within a $\Delta\,z$$<$0.005 centered at $z_\mathrm{src}$ of SDSSJ080113.28+344030.8 and the black line being their 90\% contour level computed with the KDE. Then blue crosses are the {\it candidate elliptical galaxies}, selected among those optical sources lying within the angular separation correspondent to 2\,Mpc around the central radio galaxy.}
\end{center}
\label{fig:figure1}
\end{figure}

\begin{figure*}[t!]
\begin{center}
\includegraphics[height=8.6cm,width=8.6cm,angle=0]{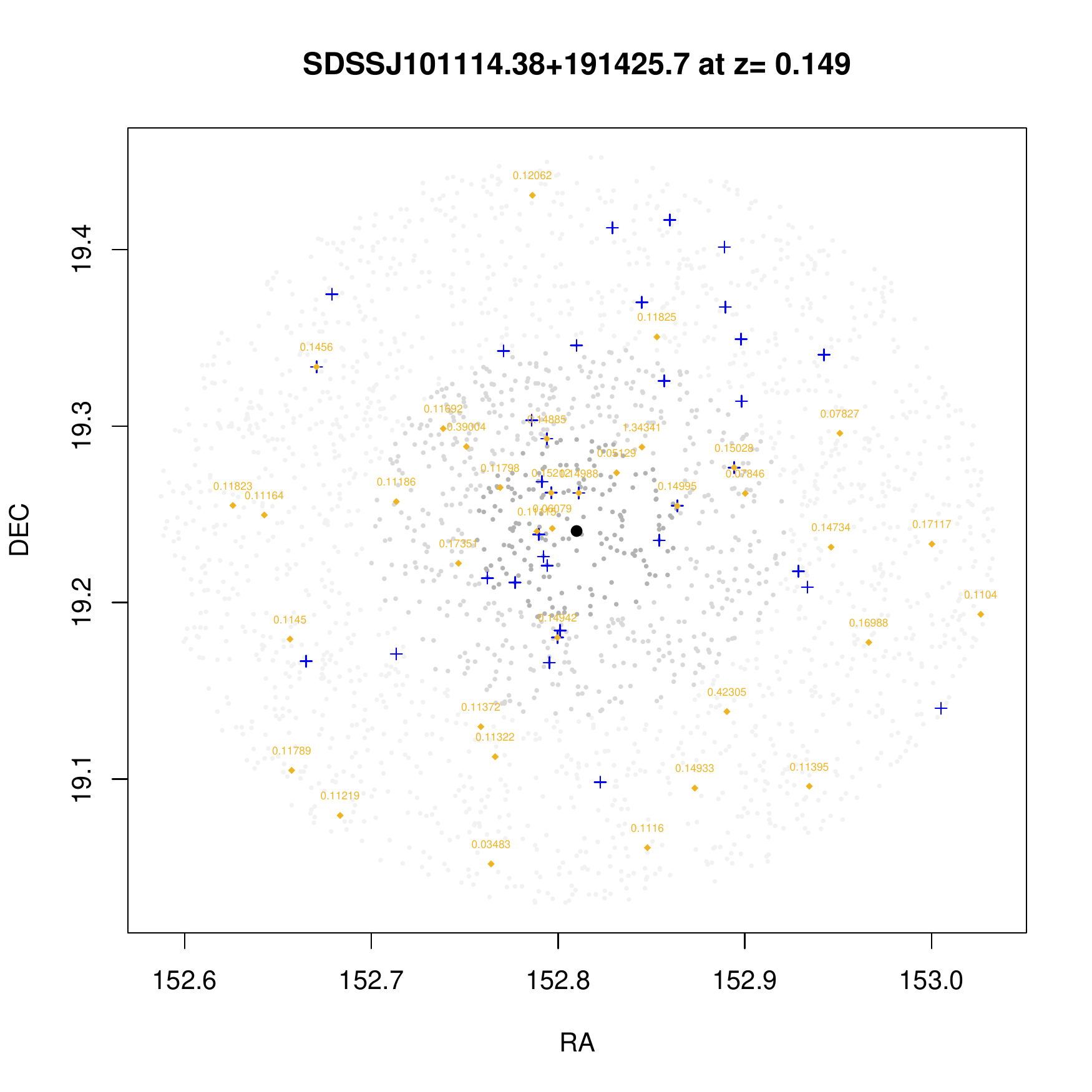}
\includegraphics[height=8.6cm,width=8.6cm,angle=0]{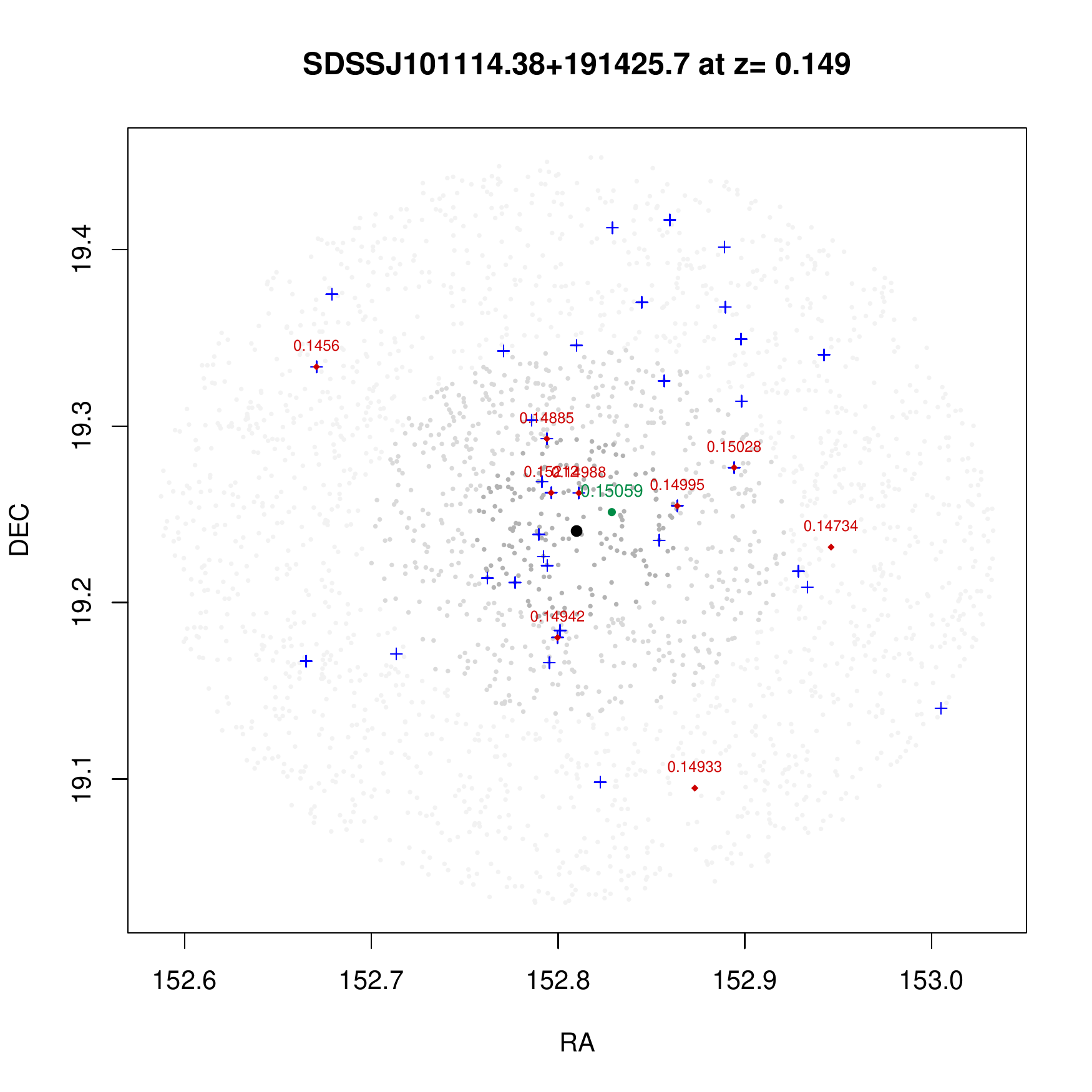}
\caption{a) The left panel shows the position of all SDSS sources within 2\,Mpc distance computed at $z_\mathrm{src}$=0.149 for the radio galaxy  SDSSJ101114.38+191425.7. Different intensities of grey indicated those lying within 500kpc, 1\,Mpc and 2\,Mpc, respectively. All SDSS objects having a spectroscopic $z$ are shown as orange circles and their $z$ value is also reported close to their location. Blue crosses, visible in both left and right panel mark {\it candidate elliptical galaxies}, i.e. SDSS sources in the field with optical colors similar to {\it quiescent elliptical galaxies} at $z_\mathrm{src}$=0.149 and within a $\Delta\,z$ of 0.005. b) In the right panel {\it cosmological neighbors} are shown as red circles, while the green point marks the location of the closest group or cluster of galaxies, again within a $\Delta\,z$ of 0.005, listed in the T12 galaxy cluster/group catalog.}
\label{fig:figure2}
\end{center}
\end{figure*}

In Figure~\ref{fig:figure2} we show the FR\,I radio galaxy: SDSSJ101114.38+191425.7 (central black circle in both panels) where all the SDSS sources lying within 2\,Mpc (grey background circles in both panels), computed at the $z_\mathrm{src}$ of the central object, are shown together with (i) SDSS sources with spectroscopic $z$ (orange circles in the left panel), (ii) {\it cosmological neighbors} (red circles in the right panel) and (iii) {\it candidate elliptical galaxies} (blue crosses in both panels). In the same figure we also show the location of the closest galaxy cluster/group in the T12 catalog, labelled with its $z_\mathrm{cl}$ (green circle in the right panel).

\begin{figure}[b!]
\begin{center}
\includegraphics[height=7.2cm,width=8.2cm,angle=0]{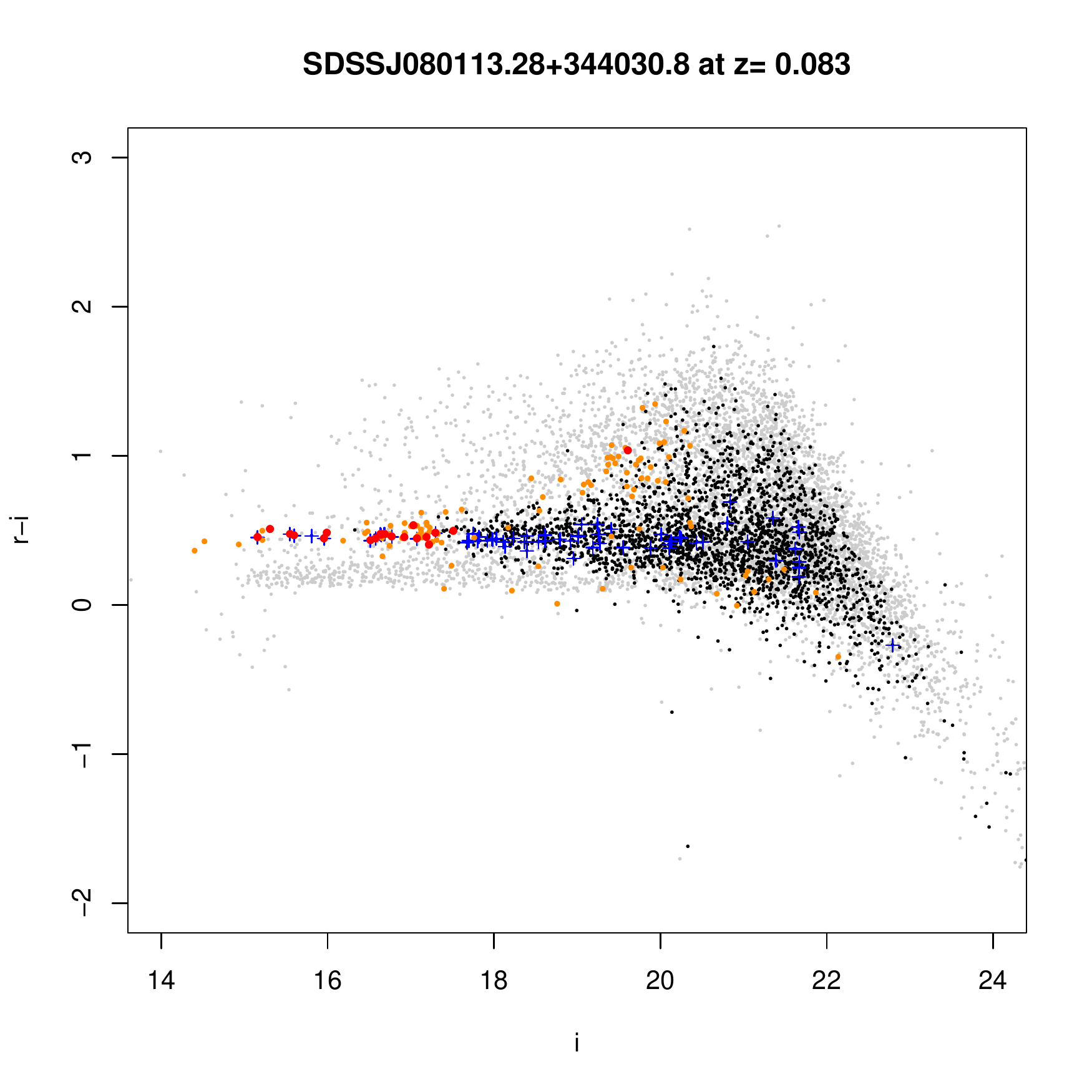}
\caption{We show the color-magnitude plot using the SDSS $r,i$ magnitudes for a radio galaxy in our sample. Background/foreground SDSS sources, within 2\,Mpc distance from the central source, are marked with black circles while {\it cosmological neighbors} are shown in red and {\it candidate elliptical galaxies} as blue crosses. Generic SDSS sources with spectroscopic $z$ are shown as orange circles. It is quite evident how both {\it cosmological neighbors} and {\it candidate elliptical galaxies} follow the ``red sequence''.}
\label{fig:figure3}
\end{center}
\end{figure}

Given our color-color selection of {\it candidate elliptical galaxies}, based on four SDSS magnitudes, we built a color-magnitude plots to verify that selected {\it cosmological neighbors} and {\it candidate elliptical galaxies} also belong to a well known feature of galaxy clusters: the red sequence. This is just an additional check to verify the presence of galaxy-rich large-scale environment around radio galaxies investigated, since galaxies that are members of groups and clusters of galaxies tend to be redder than background and foreground galaxies in the same field. In Figure~\ref{fig:figure3} we show the plot built with the $r$ and $i$ magnitudes (i.e., the same used in the GMBCG) for the FR\,I radio galaxy: SDSSJ080113.28+344030.8. It is clear that both {\it cosmological neighbors} (red circles) and {\it candidate elliptical galaxies} (blue crosses) belong to the red sequence. This color-code for both {\it cosmological neighbors} and {\it candidate elliptical galaxies} will be maintained for the rest of the figures reported in the paper.

Finally, we note that the whole analysis reported above was not only performed for both the radio galaxy catalogs but it was also carried out for the MOCK catalog, adopting exactly the same criteria and thresholds, to quantify the ``noise'' of our procedures, as described in the following.

\section{The Step-by-Step Clustering Analysis}
\label{sec:analysis} 

\subsection{Step 1: Positional cross-matches with catalogs of groups and clusters of galaxies}
The first step to test if a radio galaxy live in galaxy-rich large-scale environments was performed searching for groups and/or clusters of galaxies listed in the T12 catalog and within a 2\,Mpc radius and having $\Delta\,z=|z_\mathrm{src}-z_\mathrm{cl}|$$\leq$0.005, computed using only spectroscopic redshifts. the same analysis was then carried out for the MOCK catalog where the $z_\mathrm{src}$ value corresponds to that of the fake source listed therein.

\begin{figure}[t!]
\begin{center}
\includegraphics[height=6.2cm,width=8.2cm,angle=0]{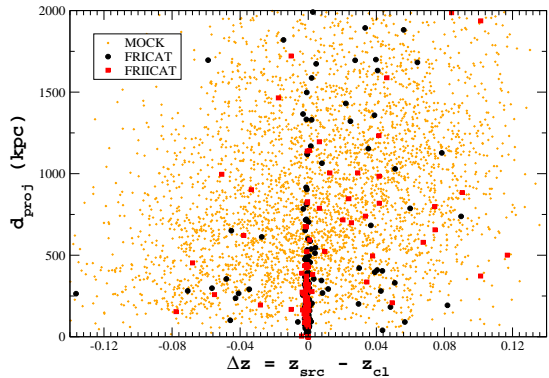}
\caption{The projected distance $d_\mathrm{proj}$ as function of the $\Delta\,z$ (i.e., the redshift difference between $z_\mathrm{src}$ of the radio galaxy or the MOCK source, and the $z_\mathrm{cl}$ of the positionally closest galaxy group/cluster in the T12 catalog. FR\,Is are marked with black circles while FRIIs are shown as red squares. MOCK sources are indeed orange diamonds.}
\label{fig:figure4}
\end{center}
\end{figure}

Figure~\ref{fig:figure4} shows one of the results of the cross-matching analysis, plotting the projected distance $d_\mathrm{proj}$ between each radio galaxy and the closest galaxy group or cluster as function of $\Delta\,z$. The same is shown for the MOCK catalog. More than 70\% of the total number of FR\,Is and more than 55\% of all FR\,IIs lie in galaxy-rich large-scale environments, being within 2\,Mpc and within $\Delta\,z\leq$0.005 from a galaxy group/cluster. We also noticed that a large fraction of radio galaxies lie in a $\Delta\,z$ range even smaller than the adopted threshold. 

In this cross-matching analysis we initially considered as member of a group/cluster of galaxies only those radio galaxies for which the galaxy density $N_\mathrm{gal}$, in the T12 catalog, is larger than 3 (i.e., $N_\mathrm{gal}>3$). However, we immediately noticed that clustering algorithms, such as the FoF and those described in the Appendix, are not able to find large-scale structures on hundreds of kpc unless they are extremely rich. This is mainly due to the high number of optical sources in the background and/or in the foreground. Unfortunately clustering algorithms could find separate groups/clusters, that lying at the same $z_\mathrm{cl}$ are indeed linked/related/connected to each other. 

\begin{figure}[t!]
\begin{center}
\includegraphics[height=8.2cm,width=8.2cm,angle=0]{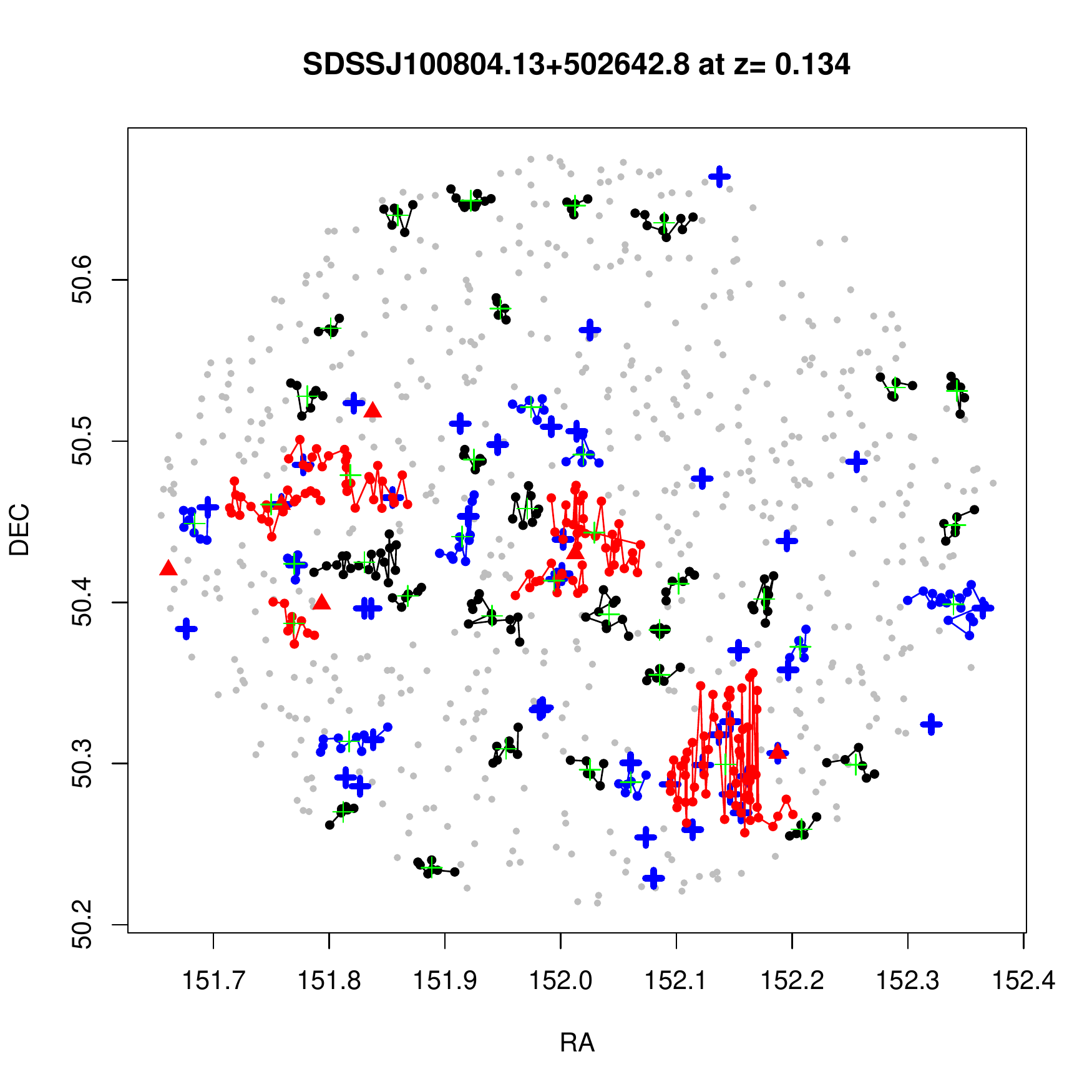}
\caption{All source clusters found and identified adopting the Density-Based Spatial Clustering of Applications with Noise (DBSCAN) algorithm (see Appendix for more details) for a radio galaxy in our sample. Those clusters including at least a {\it candidate elliptical galaxy} are highlighted in blue and those marked in red have at least one {\it cosmological neighbor} as a member while clusters in black lack both. Begin within a $\Delta\,z\leq$0.005 all the red clusters could belong to the same structure but they are indeed found separately by the algorithm.}
\label{fig:figure5}
\end{center}
\end{figure}

An example of this problem is shown in Figure~\ref{fig:figure5}. Here results of one of the clustering algorithms adopted in our analysis: the Density-Based Spatial Clustering of Applications with Noise (DBSCAN, see Appendix for additional details), are reported for the radio galaxy SDSS J100804.13+502642.8. All source clusters identified by the DBSCAN that include at least a {\it candidate elliptical galaxy} are highlighted in blue while those marked in red have at least one {\it cosmological neighbor} as a member. Source clusters found marked in black lack both {\it candidate elliptical galaxy} and {\it cosmological neighbor}. All red clusters could belong to the same cosmological structure, being within a $\Delta\,z\leq$0.005 from the central radio galaxy, a but are found and identified, by the DBSCAN algorithm, as separate source clusters.

Since clustering algorithms could potentially split a source cluster in smaller groups, the simple cross-matching analysis with the T12 catalog could be biased and the following criterion has been finally adopted to carry it out. We considered sources lying in galaxy rich large-scale environments those having {\it more than one} galaxy group/cluster within 2\,Mpc and with $\Delta\,z\leq$0.005 listed in the T12 catalog. Since the minimum value of {\it N$_\mathrm{gal}$} reported therein is 2, having at least 2 galaxy groups/clusters within $\Delta\,z\leq$0.005 corresponds to threshold previously adopted.

Finally, we performed cross-matches between radio galaxy catalogs with the GMBCG catalog. The main advantages here are that this galaxy cluster catalog  is based on a procedure more efficient than the T12 at larger $z_\mathrm{cl}$, and permit us also to search for FR\,Is and FR\,IIs that could be BCG candidates. 

Cross-matches with the GMBCG were computed adopting the same procedure used for the T12. We assumed that a radio galaxy is associated with a group/cluster of galaxies, that includes a BCG candidate, when the redshift difference $\Delta\,z$ computed between that of the central source $z_\mathrm{src}$ and the $z_\mathrm{cl}$ reported in the GMBCG is less than 0.005. However, for these cross-matches $\Delta\,z$ was computed using spectroscopic redshifts of both radio galaxy and GMBCG catalogs, when available, and with photometric estimates for GMBCG only in all other cases. The average uncertainty, at redshifts lower than 0.15, in the estimates of the photometric redshifts reported in the GMBCG is of the order of 0.023, evaluated as the mean difference between the spectroscopic and the photometric values available for more than 1200 sources listed in the catalog. We noticed a posteriori that cross-matches with the GMBCG catalog provide only a negligible improvement with respect to other methods, but, as previously stated, it was useful to identify radio galaxies potentially being BCG candidates.

\subsection{Step 2: Cosmological Over-densities}
Cross-matching analysis has an additional problem with respect to the previously mentioned one of having source clusters split in the T12 catalog. The  efficiency of the algorithm used to create the galaxy group/cluster catalogs typically decreases at higher $z_\mathrm{src}$ thus potentially biasing our analysis. 

\begin{figure}[]
\begin{center}
\includegraphics[height=7.2cm,width=8.2cm,angle=0]{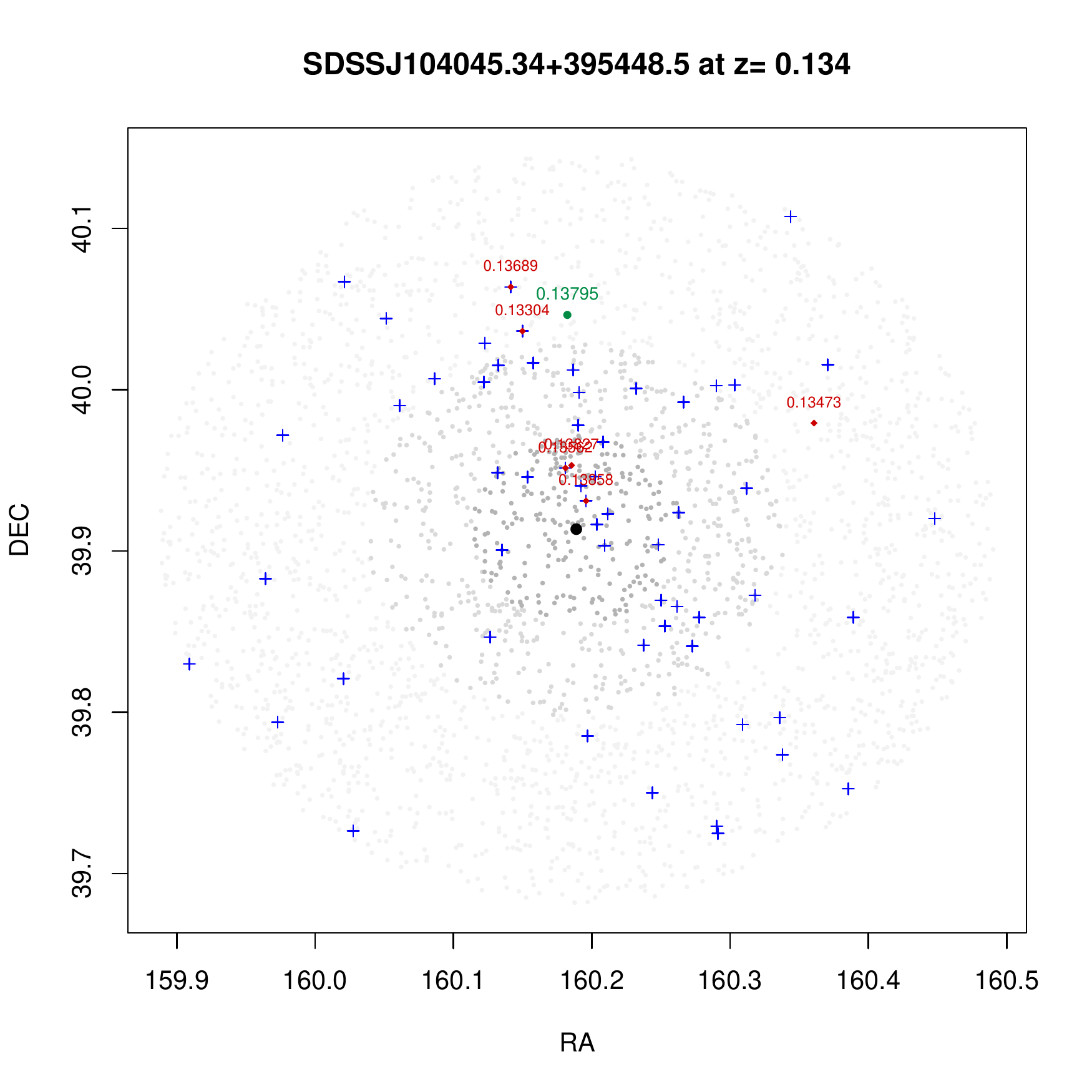}
\caption{The FR\,I radio galaxy SDSSJ104045.34+395448.5 at z=0.134 for which the T12 galaxy groups and clusters indicates a single cluster within 2\,Mpc with an environmental density, {\it $N_\mathrm{gal}$}=2 and located more than 1\,Mpc from the central source. In this case the number of {\it cosmological neighbors} are at least 5, three of which lie within 500\,kpc. {\it Cosmological neighbors} are shown as red circles, with their spectroscopic redshifts reported, while the green point mark the location of the closest group or cluster of galaxies within $\Delta\,z$$<$0.005 and blue crosses mark the location of {\it candidate elliptical galaxies}. SDSSJ104045.34+395448.5 lies in the center of the field marked with a black circle.
}
\label{fig:figure6}
\end{center}
\end{figure}

There are radio galaxies, as for example SDSSJ104045.34+395448.5, classified as FR\,I and shown in Figure~\ref{fig:figure6}, having a large number of {\it cosmological neighbors} within 2\,Mpc and even within 500\,kpc but associated with a T12 group/cluster with $N_\mathrm{gal}=$2. This galaxy density is too small to be considered a galaxy-rich large-scale environment according to our thresholds. However, sources as SDSSJ104045.34+395448.5 certainly lie in galaxy-rich large-scale environments and thus an additional criterion and/or method must be used to recover similar cases.

We then carried out the following Monte Carlo procedure estimating the {\it cosmological over-density}. 

We also indicated as sources lying in galaxy-rich, large-scale environments, those having the number of {\it cosmological neighbors} within 500\,kpc within the 5\% of those measured for fake sources belonging to the MOCK catalog, in a $z_\mathrm{src}$ bin of 0.01. For redshifts larger than 0.1 we additionally imposed to have more than 2 {\it cosmological neighbors} within 1\,Mpc. For example, in the $z_\mathrm{src}$ range between 0.13 and 0.14, as for SDSSJ104045.34+395448.5 shown in Figure~\ref{fig:figure6}, there are 493 sources in the MOCK catalog, but only 18 (i.e. less than 4\% in this redshift bin) show a number of {\it cosmological neighbors} larger than this radio galaxy. Thus we claimed that SDSSJ104045.34+395448.5 also lies in a galaxy-rich large-scale environment even if it cannot be found adopting the cross-matching analysis. 

The threshold of 5\% on the fraction of {\it cosmological neighbors} around each radio galaxy was set arbitrarily. However results of our analysis do not change if we consider a more conservative value. Changing this threshold can only decrease the fraction of sources claimed to be in galaxy-rich large-scale environments but preserve our main results. Estimating cosmological over-density was indeed necessary at redshifts larger than $\sim$0.1 where the efficiency of the T12 cluster catalog strongly decreases \citep{tempel12}. In addition, the cosmological over-density method appears to be redshift-independent with respect to cluster crossmatches, as detailed discussed in the following sections.

\subsection{Threshold summary}
In summary, we claim that a source (i.e., radio galaxy or MOCK) lies in a galaxy-rich large-scale environment when at least one of the following statements is verified.
\begin{enumerate} 
\item There is a group/cluster of the T12 catalog within 2\,Mpc, with $N_\mathrm{gal}$ larger than 3 and with $\Delta\,z\leq$0.005 or more than one group/cluster of galaxies with the same constraints but also having $N_\mathrm{gal}=$2.
\item When the redshift difference $\Delta\,z$ computed between that of the central source and the $z_\mathrm{cl}$ reported in the GMBCG is less than 0.005. 
\item The number of {\it cosmological neighbors} is more than expected in random positions of the sky within a 5\% threshold for the same redshift bin which the source belongs to.
\end{enumerate}

The first two criteria are related to the cluster cross-matching analysis performed with the T12 and GMBCG catalogs while the third one helps us to recover cases for split sources clusters working where the efficiency of the FoF algorithm in the T12 catalog significantly decreases. It is worth highlighting that all the above criteria are equivalent in identifying galaxy-rich large-scale environments. Thus, as shown in the following, below $z_\mathrm{cl}$=0.08, where the T12 catalog has the higher efficiency detecting galaxy clusters and groups, radio galaxies are found in galaxy-rich large scale environment adopting either the first or the second criterion with only a few exceptions.

The most important characteristic of using the cosmological over-densities is that selecting our thresholds on the basis of the MOCK catalog it is adaptive and it does not appear to have $z_\mathrm{src}$ dependence, being only affected by the SDSS spectroscopic completeness. Nevertheless, it is worth mentioning that the cosmological over-density allows us to mitigate the bias due to the large number of galaxy clusters found at low redshift in the T12 catalog (see next sections for more details).

Finally, to prove that our results are independent by the clustering algorithms chosen to carry out the analysis, as they are by the thresholds of $\Delta\,z$ and the 2\,Mpc radius, we tried three additional methods, namely: DBSCAN, Voronoi Tessellation and Minimum Spanning Tree (MST). These are all clustering algorithms already used in large optical and infrared surveys to search for galaxy groups and clusters, as spatial over-densities, as alternatives to the FoF method\citep{hucra82}. For all these three methods we considered a radio galaxy in a galaxy-rich large-scale environment when the number of {\it cosmological neighbors} belonging to one cluster (i.e., region of high density of optical sources), found applying the algorithm, is larger than the top 5\% of those detected in the MOCK sample adopting the same method. We also run all three algorithms considering the number of {\it candidate elliptical galaxies} instead of that of {\it cosmological neighbors} (see Appendix for more details).

\subsection{Noise and uncertainties}
It is crucial to highlight that results on the MOCK catalog provide an estimate of the false positives we could get adopting our algorithms when claiming that a source belong to a galaxy-rich large-scale environment. Testing our methods over the MOCK catalog helped us to estimate their ``noise''. 

Both FR\,I and FR\,II radio galaxy catalogs have a completeness larger than 90\% in total, being almost 100\% in the low redshift bins. However the SDSS footprint covers only $\sim$1/3 of the sky and we needed to estimate the uncertainties on the ratios/percentages of radio galaxies taking into account the underlying population. Thus, assuming a binomial distribution, where, for each bin of redshift, magnitude and/or luminosity, finding a radio galaxy in a galaxy-rich large-scale environment is a ``success'', we computed binomial confidence intervals correspondent to 1$\sigma$ adopting the procedure described in Cameron (2011). 

In each plot where ratios of radio galaxies found in galaxy-rich large-scale environment is shown, we report both the comparison between confidence intervals and results obtained with the MOCK catalog as well as ratios with noise subtracted. To take into account the noise subtraction we simply define the number of ''successes'' as the number of radio galaxies found in galaxy-rich large-scale environment per bin minus the average number of MOCK sources rescaled for the number of total radio galaxies in that bin. 

For example, if the total number of radio galaxies in the range $z_1<z\mathrm{src}<z_2$ is $n$, those lying in galaxy-rich large-scale environment is $k$, and $<k_\mathrm{m}>=5.2$ is the average number of MOCK sources found in galaxy-rich large-scale environment using the same procedure on $N_s$ simulations computed with $n$ sources, the number of successes used to compute the noise subtracted confidence intervals is $k'=k-<k_\mathrm{m}>$, assuming that the uncertainty on the simulations is negligible due to their high number.

\section{Results}
\label{sec:results} 

\subsection{FR\,Is and FR\,IIs in galaxy-rich large-scale environments}
\begin{figure*}[t!]
\begin{center}
\includegraphics[height=6.2cm,width=8.2cm,angle=0]{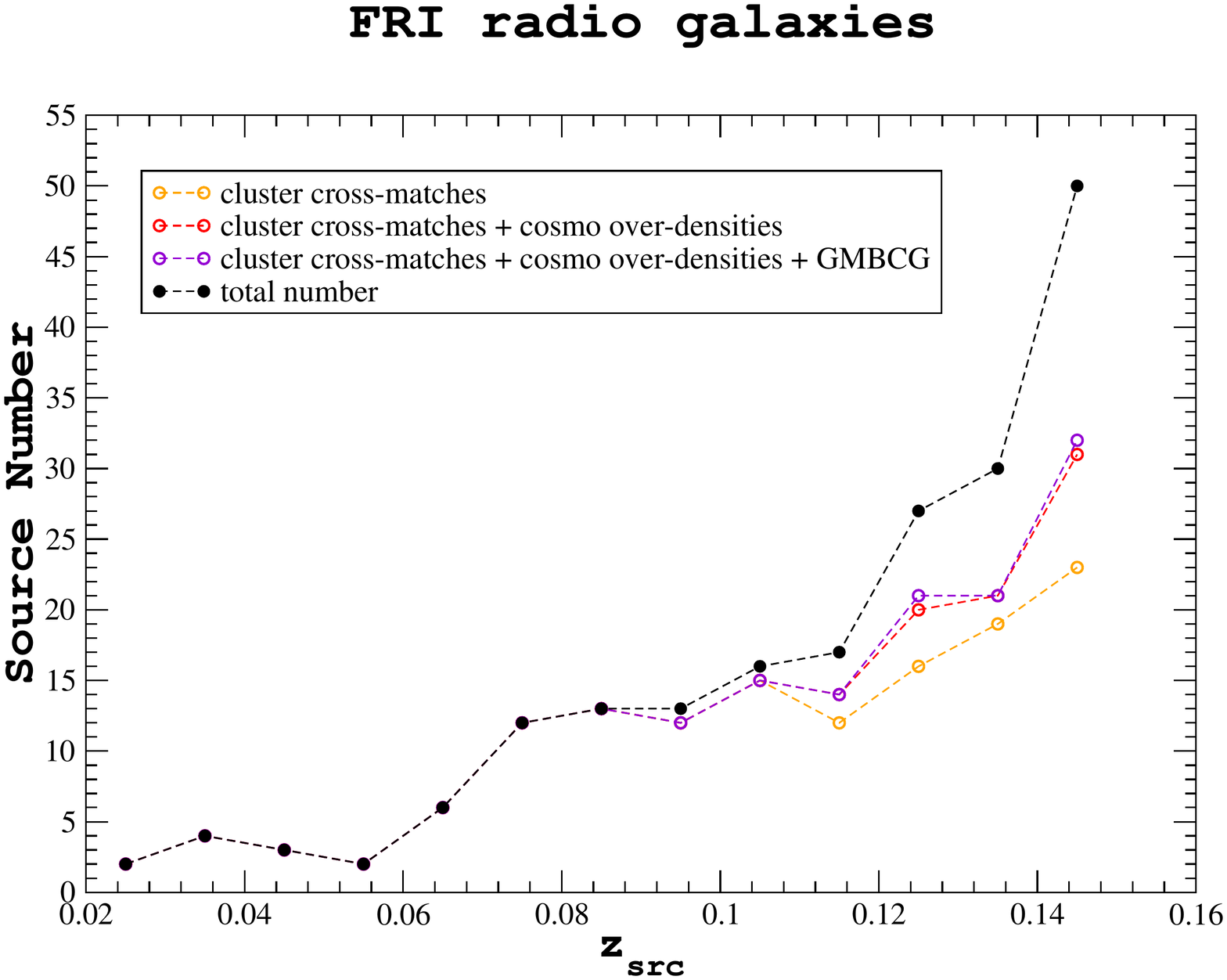}
\includegraphics[height=6.2cm,width=8.2cm,angle=0]{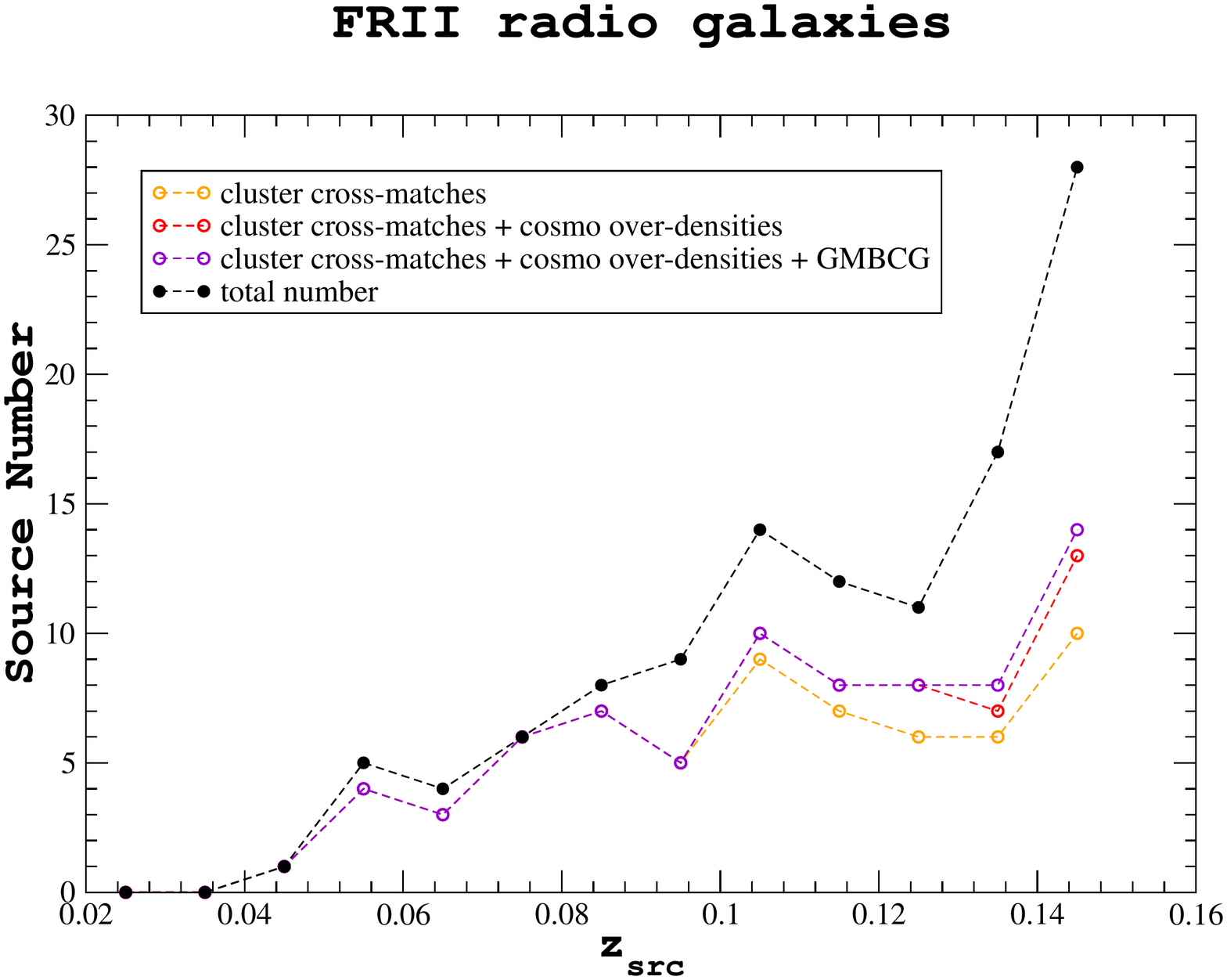}
\caption{{\it Left panel)} Number of FR\,I radio galaxies in redshift bins of 0.01. Filled black circles represent the total number of sources per bin of $z_\mathrm{src}$, while empty circles mark those sources i) having a cross-match with a cluster/group of galaxies in the T12 cluster catalog (orange circles) plus ii) a cosmological over-density (red circles) plus iii) an association with a cluster hosting a BCG candidate in the GMBCG catalog (magenta circles). {\it Right panel)} Same as left panel for the FR\,II radio galaxies.}
\label{fig:figure7}
\end{center}
\end{figure*}

\begin{figure}[]
\begin{center}
\includegraphics[height=6.2cm,width=8.2cm,angle=0]{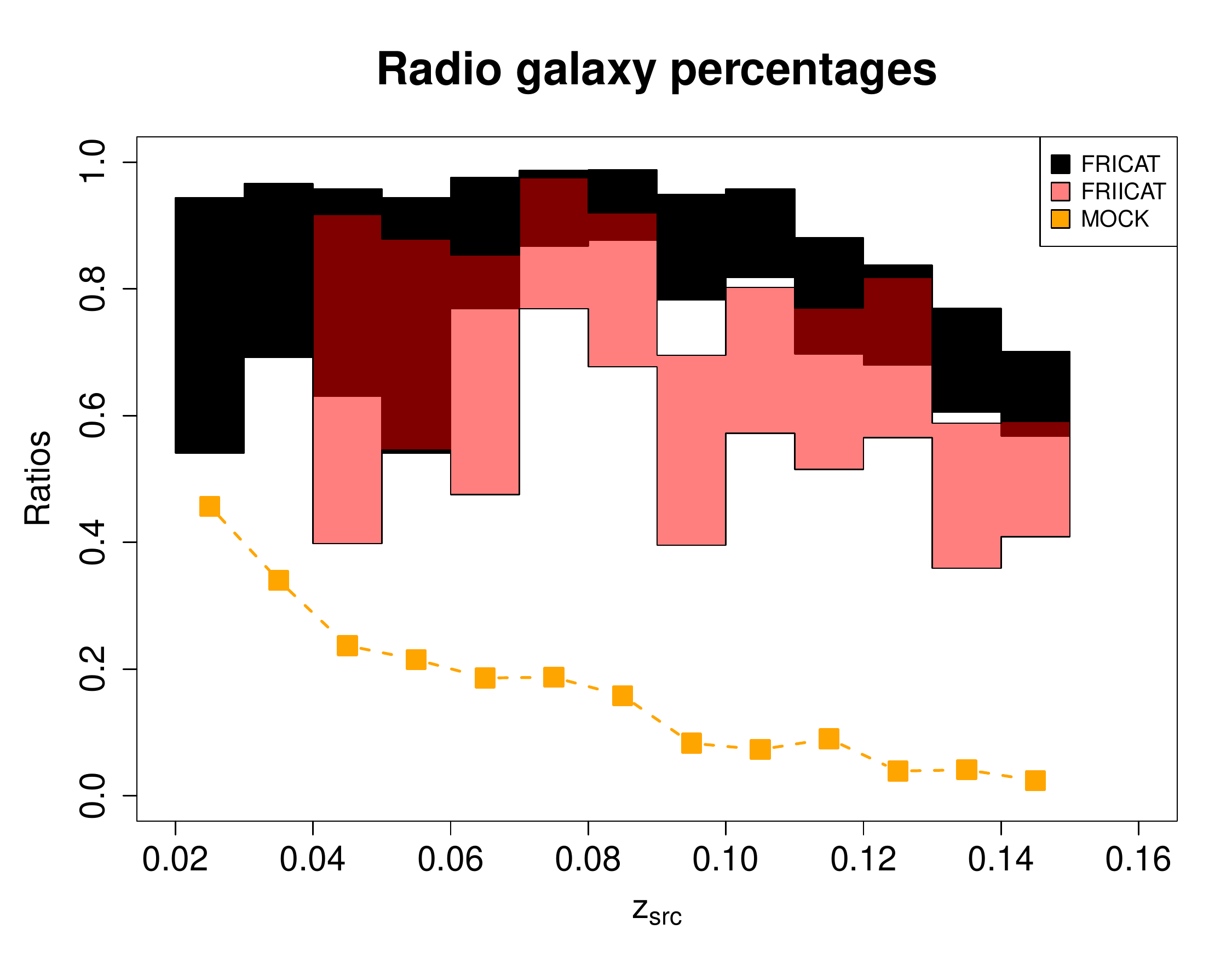}
\includegraphics[height=6.2cm,width=8.2cm,angle=0]{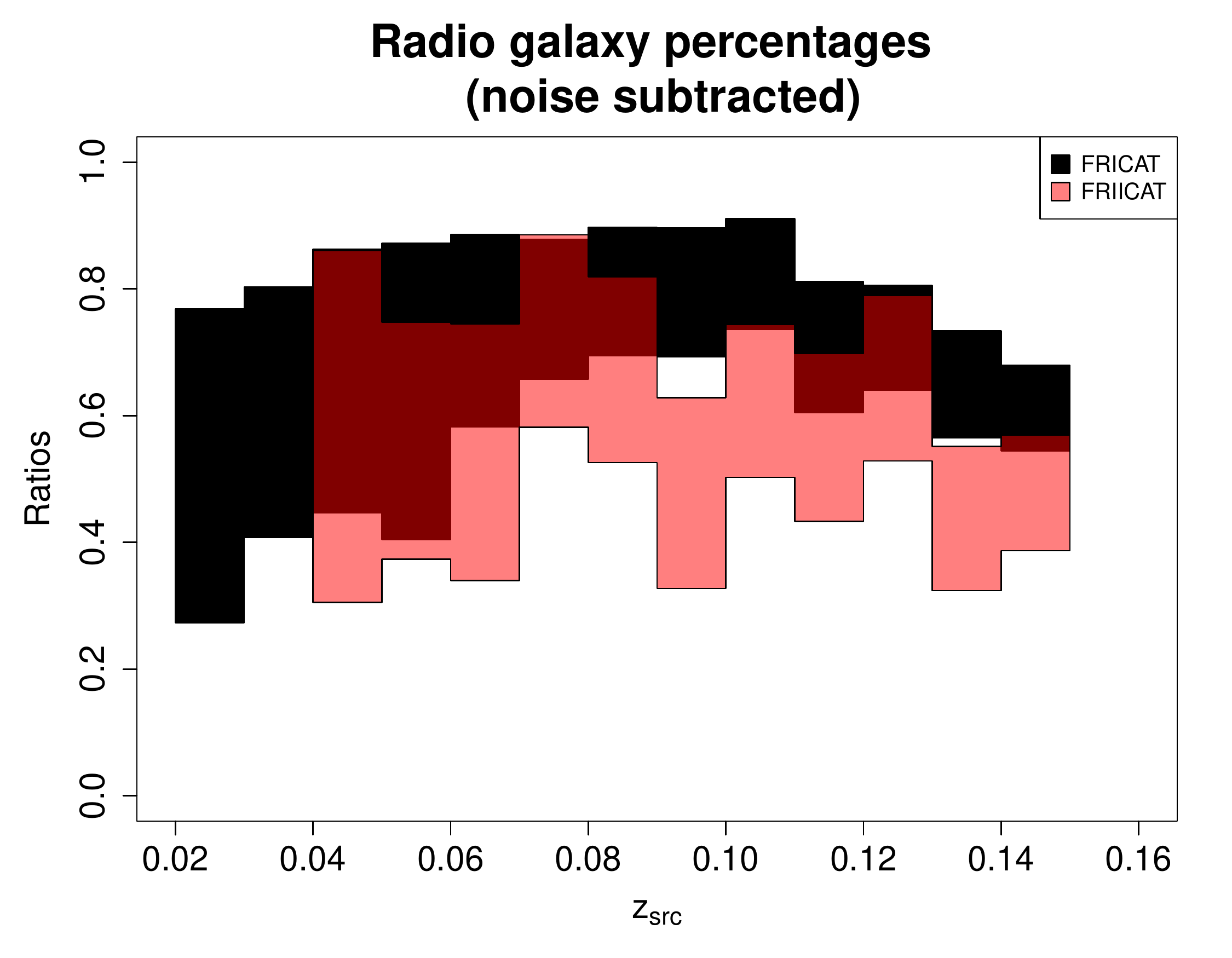}
\caption{Upper panel) The ratio between the number of radio galaxies (FR\,Is in black and FR\,IIs in red) in comparison with MOCK sources (orange) living in galaxy-rich large-scale environments over their total number as function of $z_\mathrm{src}$. These fractions where estimated adopting (i) cross-matches with the T12 cluster catalog plus (ii) cosmological over-densities and (iii) associations with the GMBCG catalog. The rise of the fraction for the MOCK sources at low redshifts is due to the T12 cross-matches. Both classes of radio galaxies appear to follow the same trend lying in galaxy-rich large-scale environment. Confidence intervals for the ratios of radio galaxies in each redshift bin are estimated as described in \S~\ref{sec:analysis}. Lower panel) the same as upper panel but having noise subtracted. We remark that there are no FR\,II radio galaxies in the first two redshift bins.}
\label{fig:figure8}
\end{center}
\end{figure}

Results of our analysis are discussed here. The number of radio galaxies lying in galaxy-rich large-scale environments with respect to their total number per bin of redshift and considering all the criteria previously described (i.e., cross-matches with cluster/group catalog, cosmological over-densities, cross-matches with the GMBCG) are shown in Figure~\ref{fig:figure7}. There are 29 FR\,Is out of 195 and 16 FR\,II out of 115 in galaxy-rich large-scale environments at $z_\mathrm{src}\leq$0.08, corresponding to the completeness limit of the T12 catalog. All these FR\,Is and 14 out of 16 FR\,IIs lie in galaxy-rich environments. The two FR\,IIs not belonging to galaxy-rich environments are optically classified as LERGs. The three FR\,II-HERGs in our sample at $z_\mathrm{src}<$0.08 all lie in galaxy groups/clusters. Ratios between the number of FR\,Is and FR\,IIs belonging to galaxy-rich environment with respect to their total number in the FRICAT and FRIICAT, respectively are then shown in Figure~\ref{fig:figure8} together with those in the MOCK sample to which the same criteria were adopted.

\begin{figure}[]
\begin{center}
\includegraphics[height=6.2cm,width=8.2cm,angle=0]{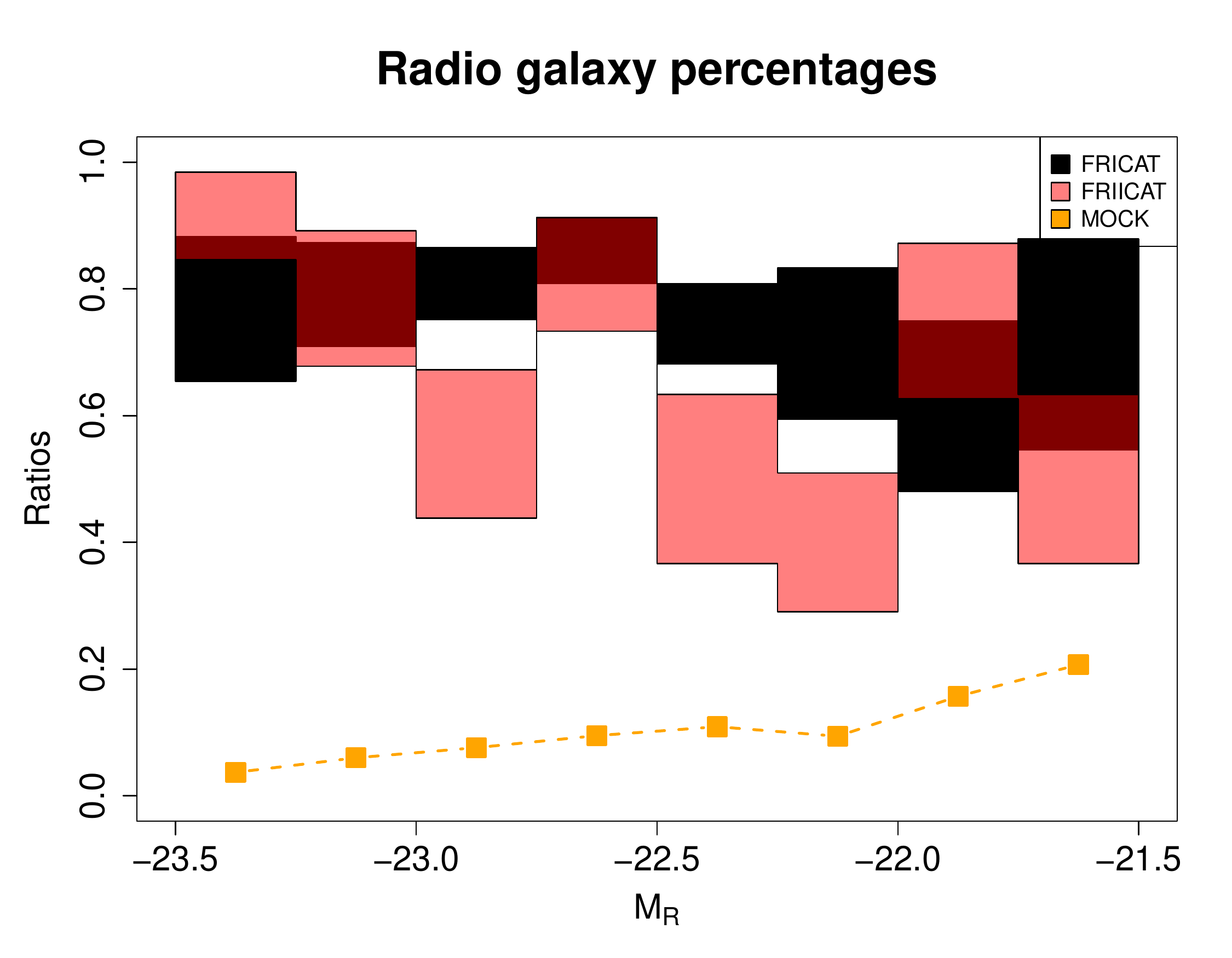}
\includegraphics[height=6.2cm,width=8.2cm,angle=0]{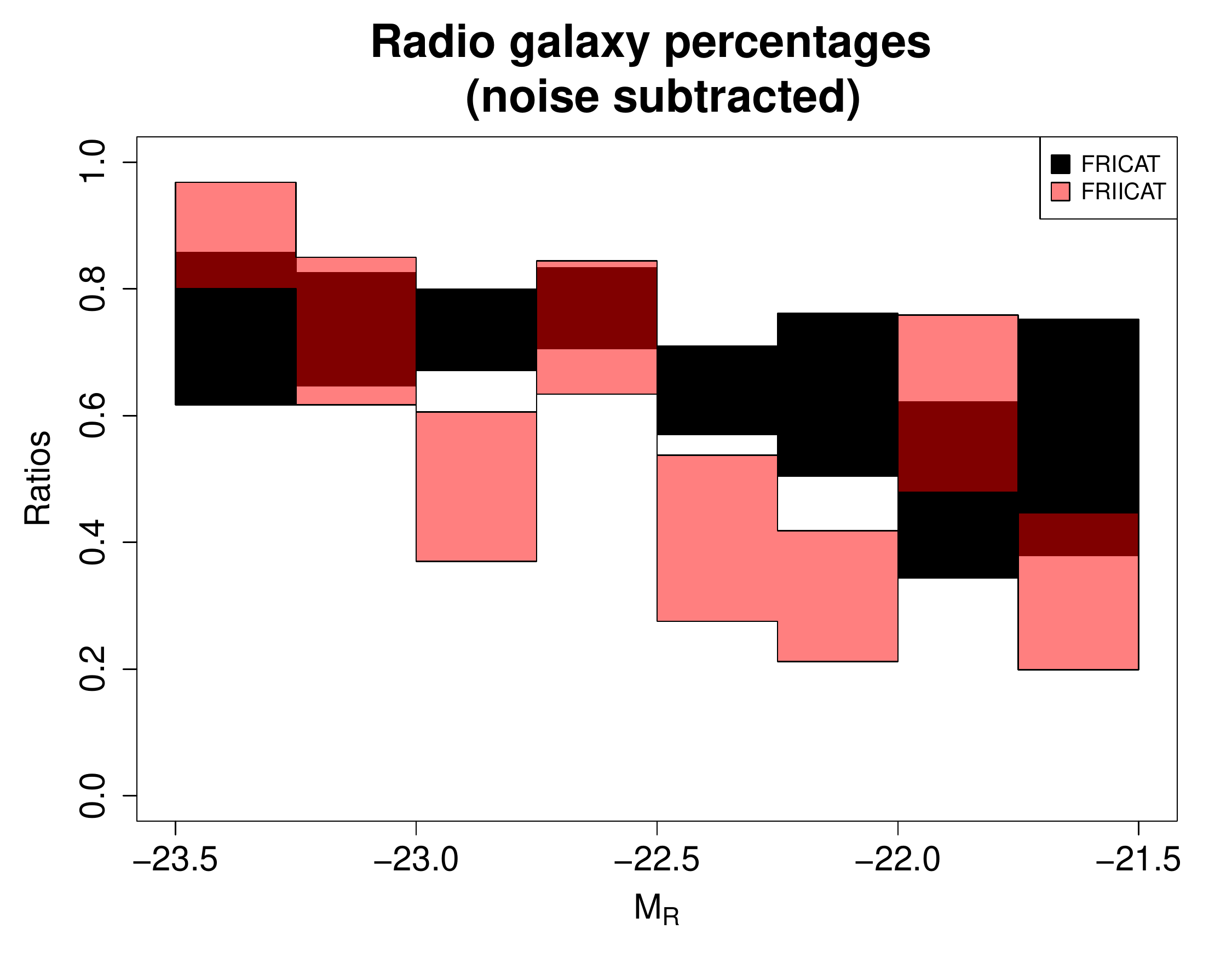}
\caption{Same as Figure~\ref{fig:figure7} where the ratios are expressed as function of the absolute magnitude in the R band: $M_\mathrm{R}$. Upper panel reports the comparison with the results obtained in the MOCK catalog while lower panle refers to the ratios with noise subtracted as described in \S~\ref{sec:analysis}.}
\label{fig:figure9}
\end{center}
\end{figure}

In Figure~\ref{fig:figure9} we also report the ratios between the number of radio galaxies (FR\,Is in black and FR\,IIs in red) and MOCK sources (orange) living in galaxy-rich large-scale environments over their total number as function of the absolute magnitude in the R band: $M_\mathrm{R}$.

\begin{figure*}[]
\begin{center}
\includegraphics[height=6.2cm,width=8.2cm,angle=0]{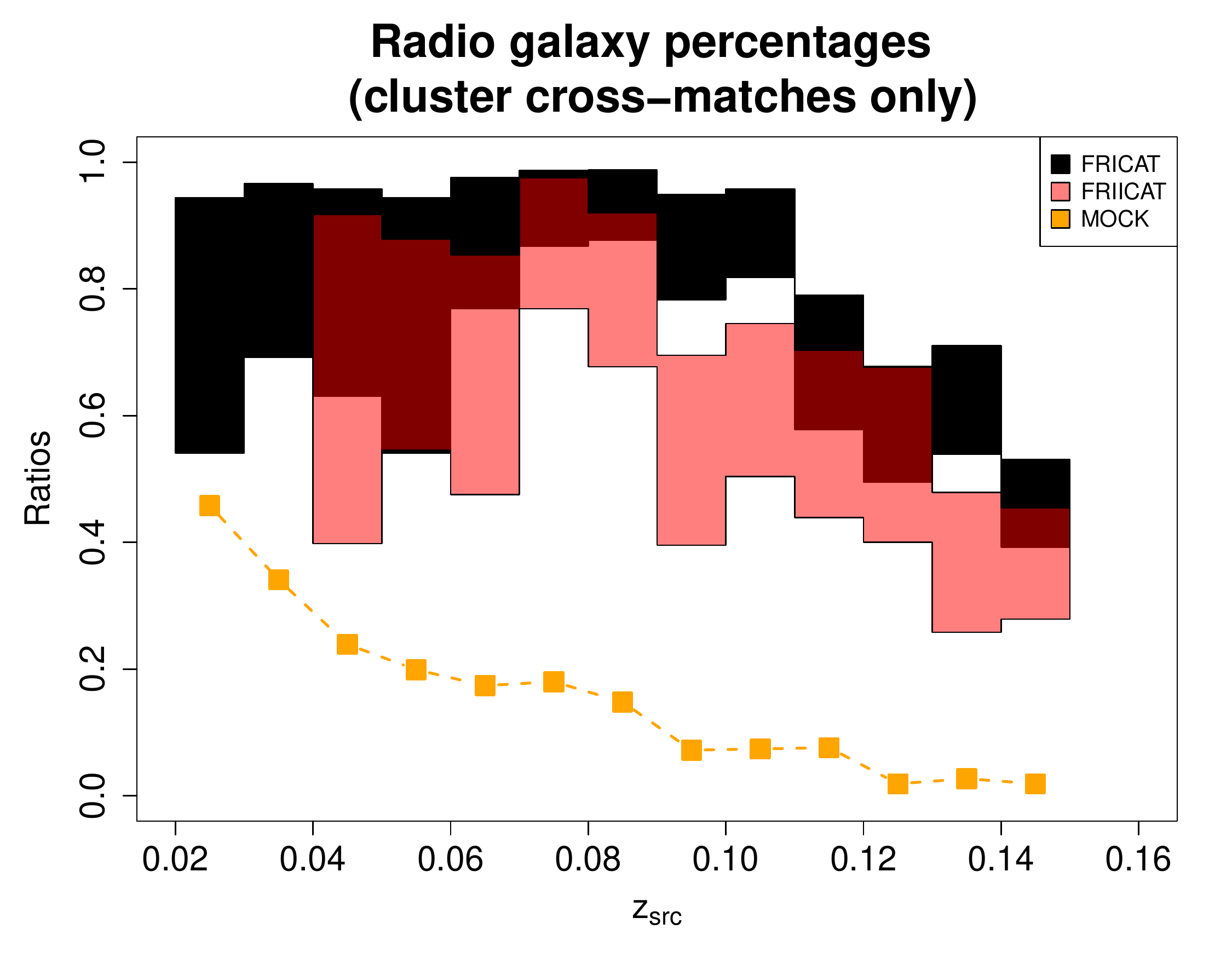}
\includegraphics[height=6.2cm,width=8.2cm,angle=0]{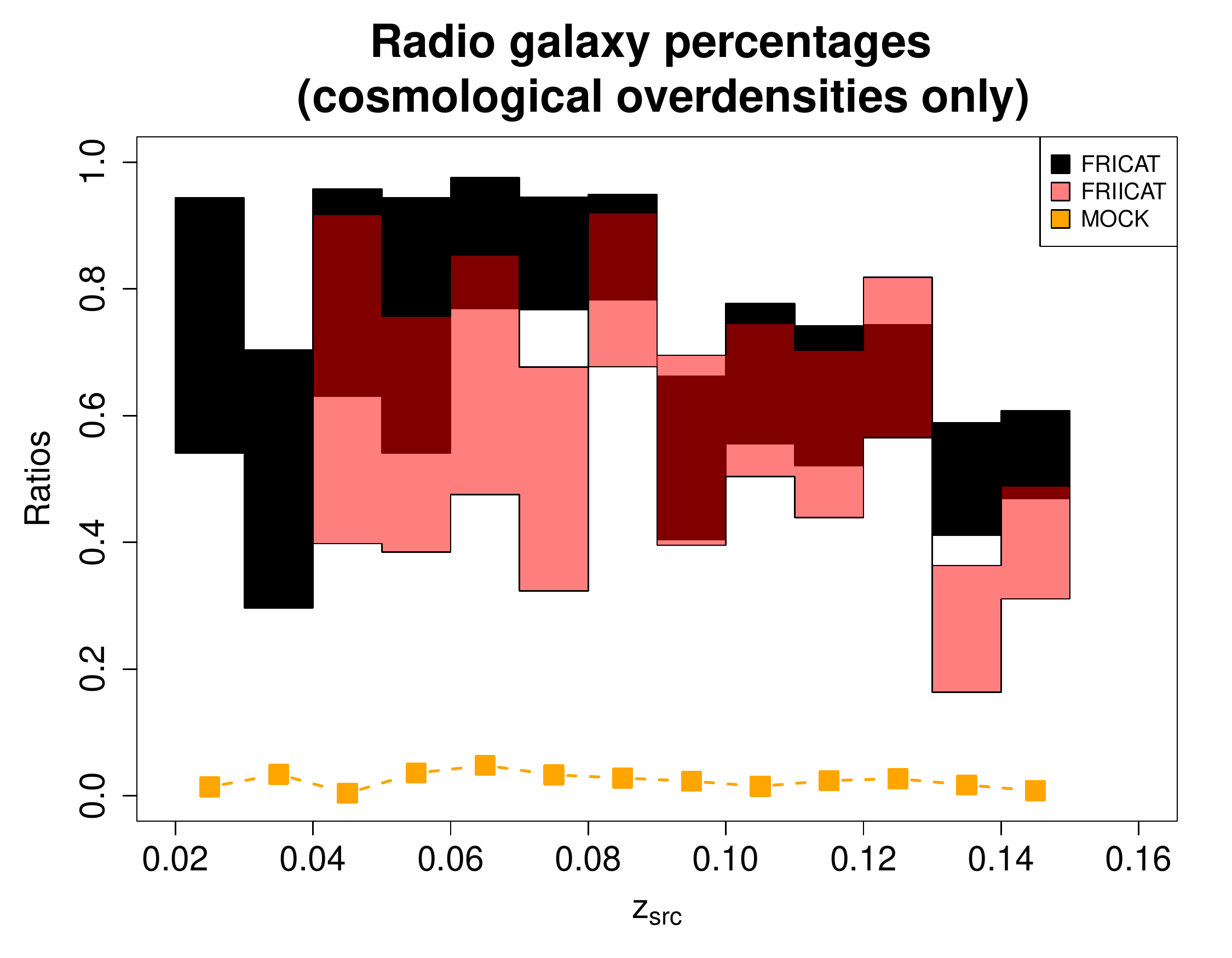}
\includegraphics[height=6.2cm,width=8.2cm,angle=0]{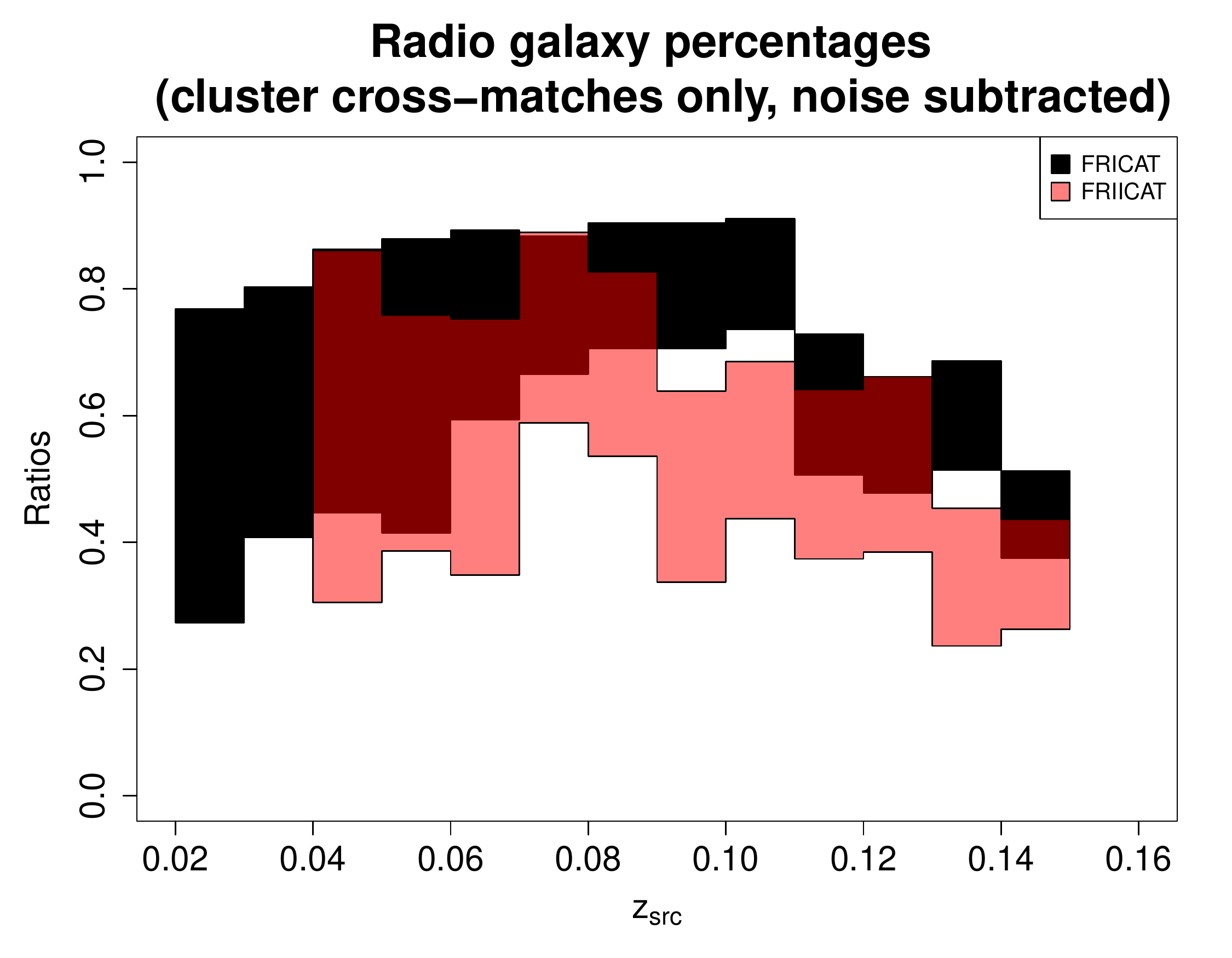}
\includegraphics[height=6.2cm,width=8.2cm,angle=0]{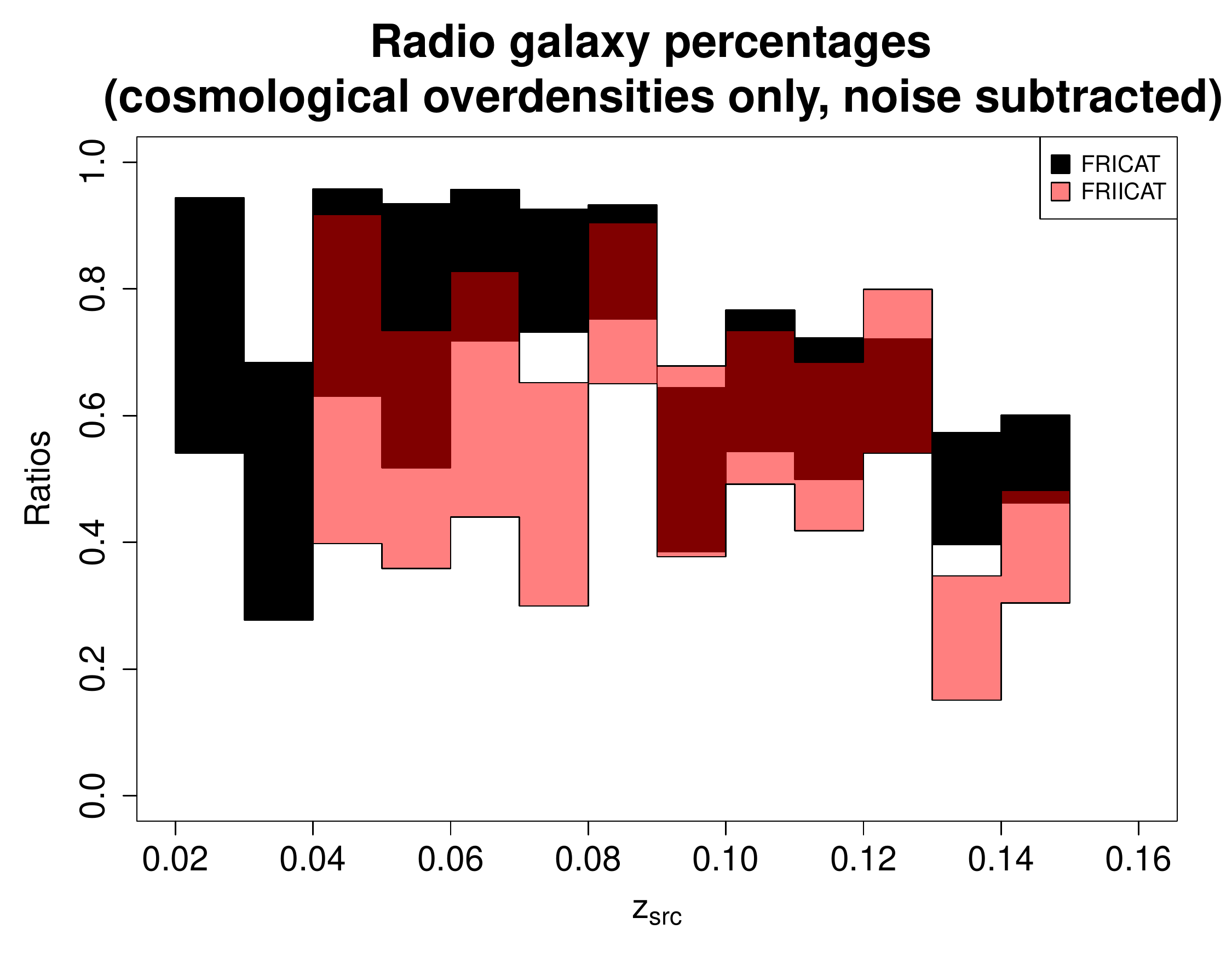}
\caption{The fraction of radio galaxies (FR\,Is marked in black while FRIIs shown in red) and MOCK sources (orange) in galaxy-rich large-scale environments as function of redshift $z_\mathrm{src}$. In the top left panel we show the ratios computed only adopting the cross-matching analysis with the T12 catalog of groups and clusters while in the top right panel those calculated using only the cosmological over-density. The efficiency of the former procedure significantly decreases with the redshift while the latter is less affected by this effect. It is also evident how both methods show the gap between the fraction for real sources in galaxy-rich large-scale environment and those in the MOCK catalog: the main result of our analysis. Lower panels show the same ratios but taking into account the noise subtraction as described in \S~\ref{sec:analysis}.}
\label{fig:figure10}
\end{center}
\end{figure*}

It is worth highlighting that in both Figure~\ref{fig:figure8} and \ref{fig:figure9} the fraction of radio galaxies found in galaxy-rich large-scale environment decreases significantly with $z_\mathrm{src}$. We therefore re-analyzed all radio galaxy samples comparing results obtained using the T12 cluster cross-match procedure with those found only searching for cosmological over-densities. The improvements given by the GMBCG cross-matches are negligible. Results from this comparison are shown in Figure~\ref{fig:figure10}. 

It is clear how both methods show the gap between the fraction for real sources in galaxy-rich large-scale environment and those in the MOCK catalog. However, it is also quite evident how the fraction of MOCK sources, claimed to be in galaxy-rich large-scale environments, raises when the $z_\mathrm{src}$ decreases. This shows how the efficiency of cosmological over-densities is not $z_\mathrm{src}$ dependent. On the other hand, this strong $z_\mathrm{src}$ dependence in the T12 cluster cross-matches turns into a higher probability to find a source lying in a galaxy-rich large-scale environment at low $z_\mathrm{src}$. This effect has to be taken into account when comparing our results with those available in the literature. 

Despite methods and procedures or criteria and thresholds adopted, that are the same for all samples and catalogs, our main result is that: the fraction of FR\,IIs in rich environments could appear systematically lower than that of FR\,Is but radio galaxies in both radio classes inhabit galaxy-rich large-scale environments in the local Universe independently by their radio morphology.

\subsection{Richness}
\begin{figure*}[]
\begin{center}
\includegraphics[height=6.2cm,width=8.2cm,angle=0]{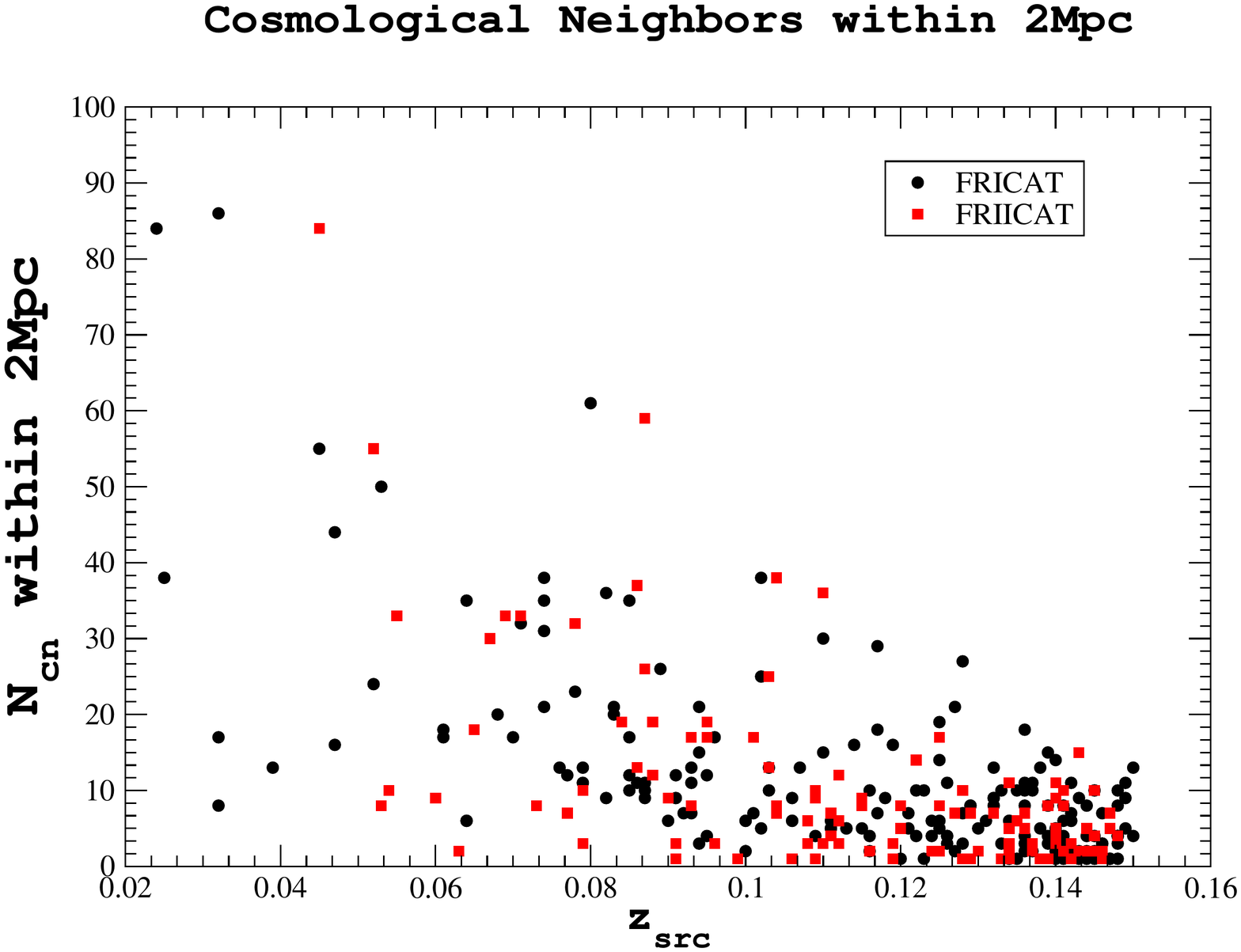}
\includegraphics[height=6.2cm,width=8.2cm,angle=0]{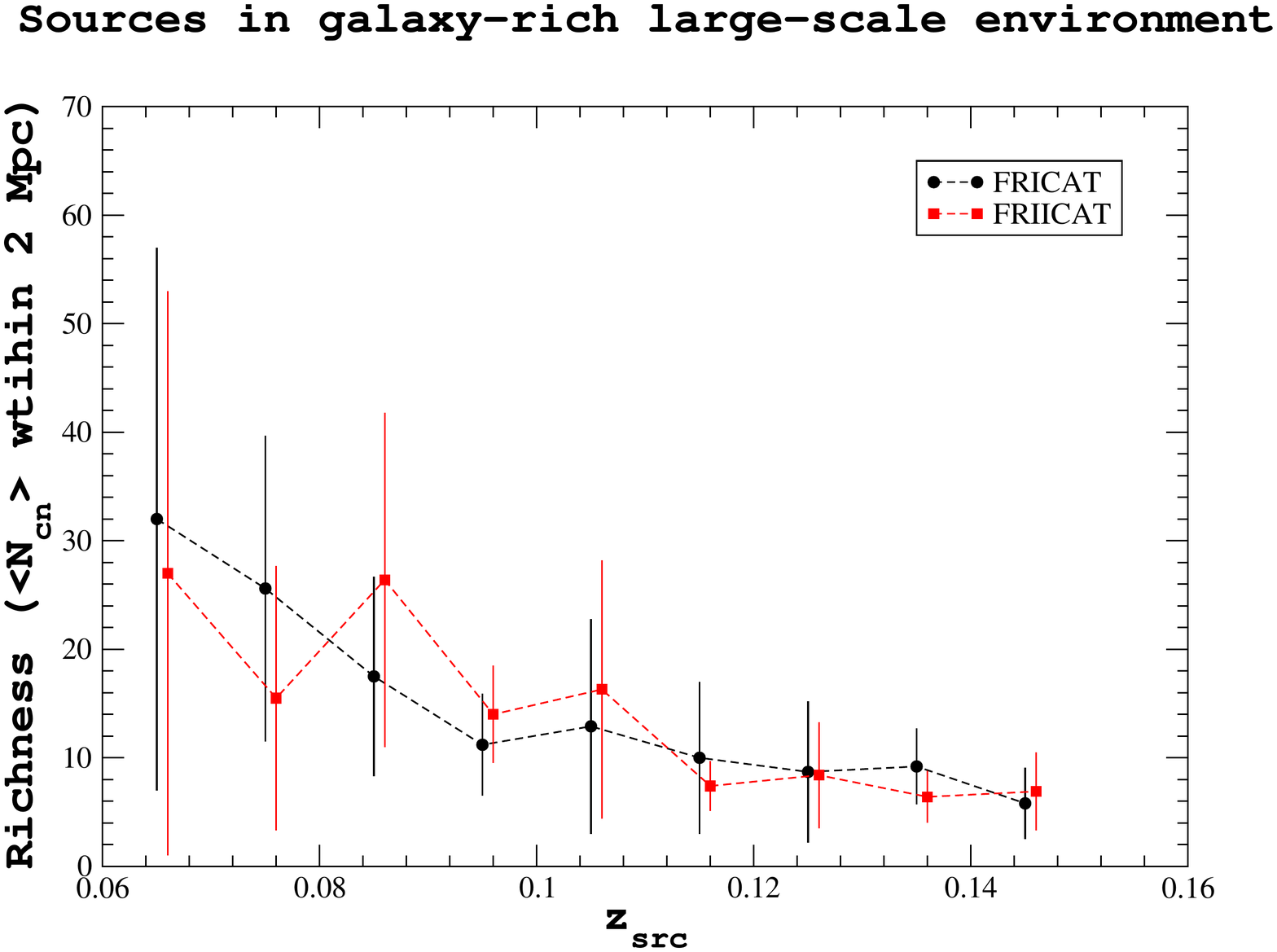}
\caption{{\it Left panel)} The total number of {\it cosmological neighbors} $N_\mathrm{cn}$ within 2\,Mpc as function of the central source redshift $z_\mathrm{src}$. There are no differences between the two FR catalogs. The {\it richness} estimated according to the cosmological over-density procedure appears the same. {\it Right panel)} The average number of {\it cosmological neighbors} $<N_\mathrm{cn}>$ counted within 2\,Mpc from the central source as function of redshift. The uncertainty on $<N_\mathrm{cn}>$ is computed from the distribution of $N_\mathrm{cn}$ within the same $z_\mathrm{src}$ bin for each source class, respectively. The first bin is larger and includes all sources up to $z_\mathrm{src}=$0.065.}
\label{fig:figure11}
\end{center}
\end{figure*}

We test if the richness of their environment is also the same. Since the galaxy density $N_\mathrm{gal}$ reported in T12 catalog of groups and clusters could be misleading and underestimated, as shown in Figure~\ref{fig:figure6}, we proceeded as described in the following. We computed the total number of {\it cosmological neighbors} $N_\mathrm{cn}$ within 2\,Mpc as function of $z_\mathrm{src}$, and we show in Figure~\ref{fig:figure11} that there are no differences between FR\,I and FR\,II radio galaxies. In the same figure we also present the average number of {\it cosmological neighbors} $<N_\mathrm{cn}>$ within 2\,Mpc as function of $z_\mathrm{src}$ to highlight the lack of differences between radio galaxies with different morphology.

\begin{figure}[]
\begin{center}
\includegraphics[height=6.2cm,width=8.2cm,angle=0]{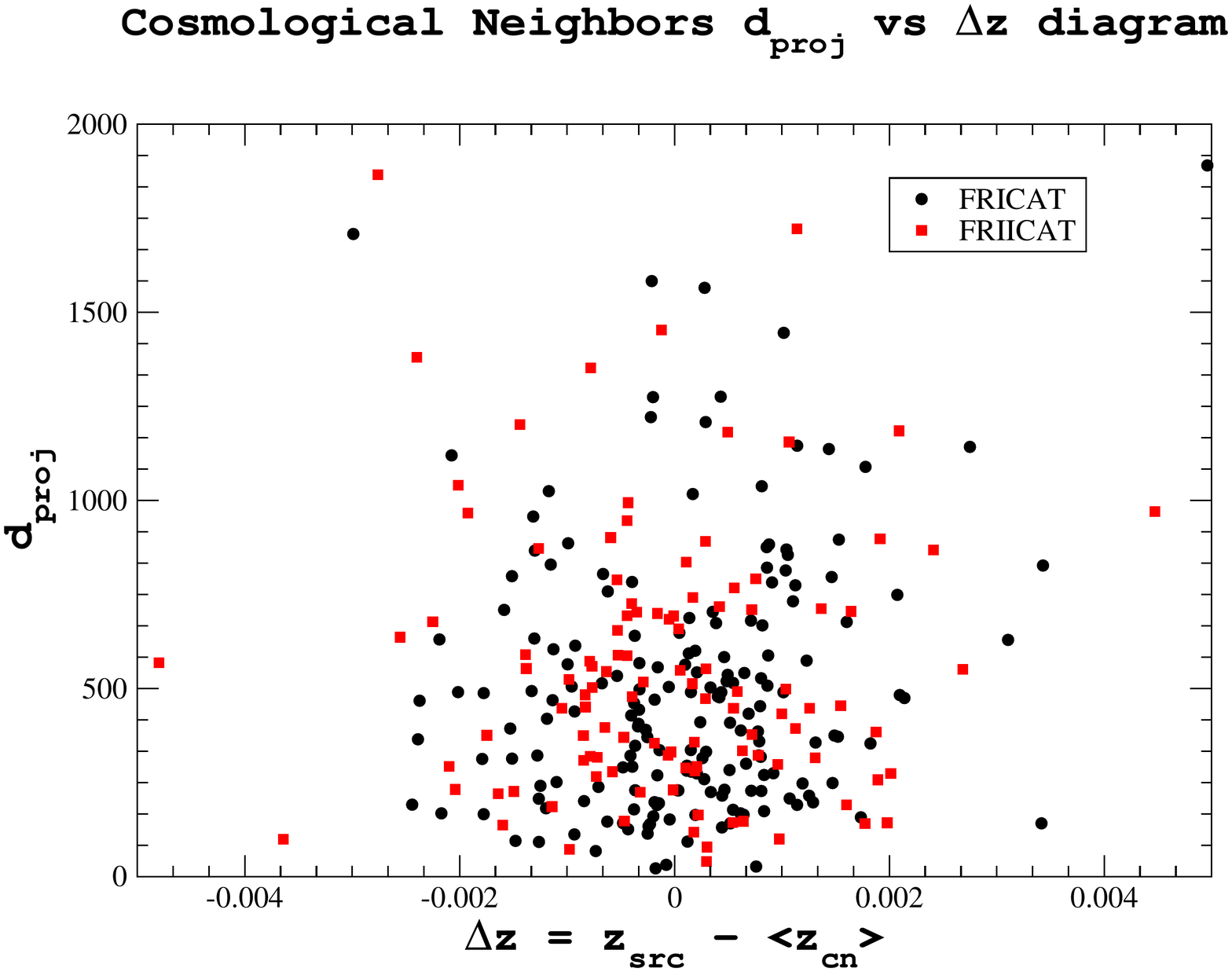}
\caption{The projected distance $d_\mathrm{proj}$ as function of the redshift difference $\Delta\,z$ between that of the central radio galaxy, $z_\mathrm{src}$ and the average values of redshifts and coordinates computed with the {\it cosmological neighbors} sample. FR\,I radio galaxies are marked with black circles while FRIIs are shown in red, as in all plots.}
\label{fig:figure12}
\end{center}
\end{figure}

Using the cosmological over-densities, we also computed the projected distance $d_\mathrm{proj}$ as function of $\Delta\,z$ between that of the central galaxy (i.e., $z_\mathrm{src}$) and the average values of redshifts $<z_\mathrm{cn}>$ and of coordinates in the {\it cosmological neighbors} sample. In Figure~\ref{fig:figure12} FR\,Is do not appear different from FR\,IIs. Furthermore, it is clear how the threshold of $\Delta\,z=$0.005 is not extremely conservative and how both classes of radio galaxies lie closer to the centers of galaxy groups and clusters. Restricting this threshold at lower values (e.g., 0.003) does not affect our main results.

\begin{figure}[]
\begin{center}
\includegraphics[height=6.2cm,width=8.2cm,angle=0]{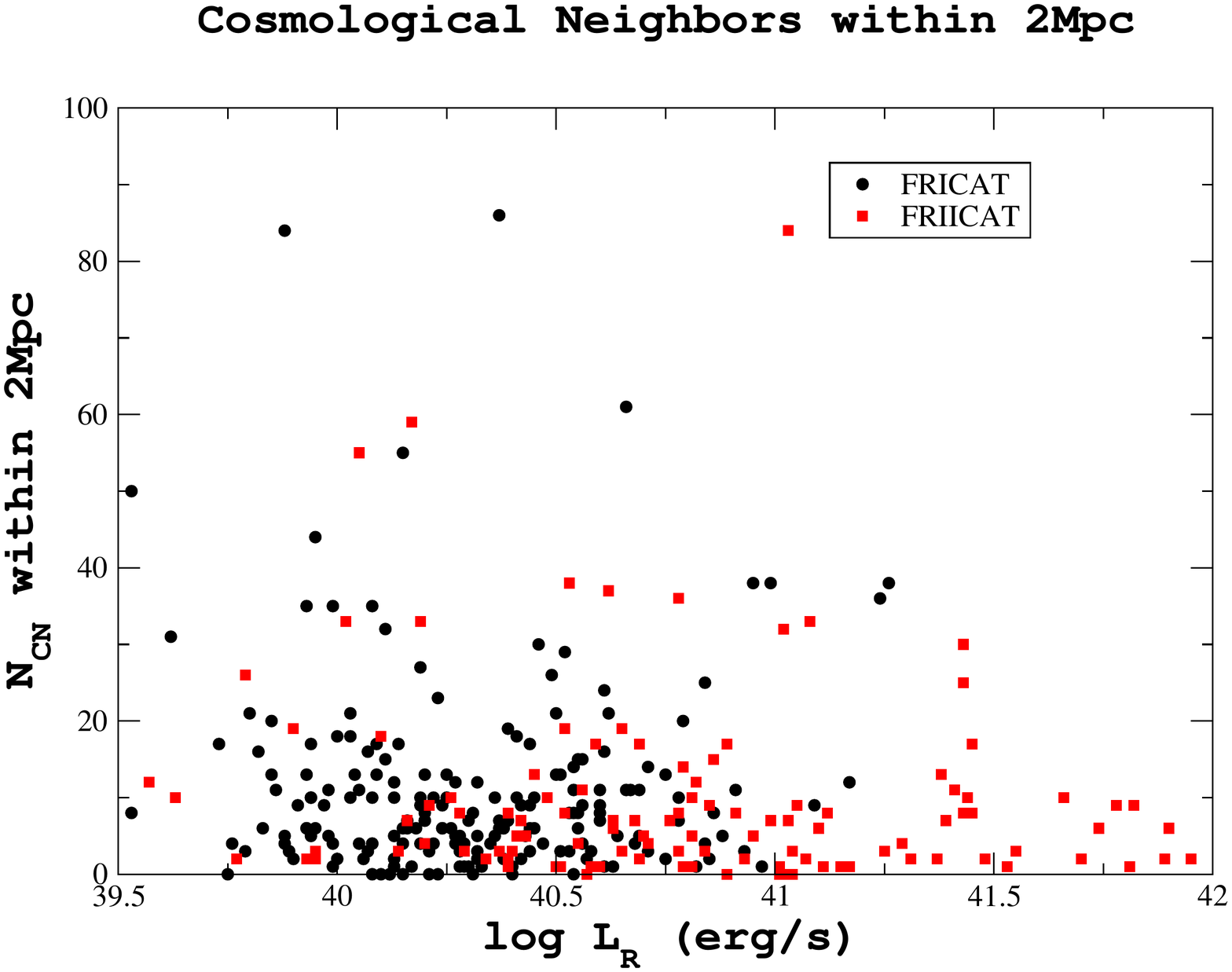}
\caption{Same of Figure~\ref{fig:figure10} where the total number of {\it cosmological neighbors} $N_\mathrm{cn}$ within 2\,Mpc is reported as function of the radio luminosity $L_R$ at 1.4 GHz.}
\label{fig:figure13}
\end{center}
\end{figure}

Finally, we also verified if $N_\mathrm{cn}$, is related to the radio luminosity $L_R$ at 1.4 GHz. As shown in Figure~\ref{fig:figure13}, even if the distribution of radio luminosity for the FRICAT and the FRIICAT is quite different \citep[see e.g.,][for more details]{capetti17a,capetti17b}, for a given value of $L_R$ we have similar value of $N_\mathrm{cn}$ for both classes.

\subsection{LERGs and HERGs in galaxy-rich large-scale environments}
We then explored the large-scale environments and its richness adopting the optical classification. 

To carry out this analysis a couple of problems related to selection effects, that could introduce biases, should be properly mentioned. We stress the fact that radio galaxies, selected in flux limited samples, at high $z_\mathrm{src}$, where it is difficult to find surrounding galaxy-rich environments, are mostly FR\,II HERGs, while at low $z_\mathrm{src}$ are almost all FR\,Is and FR\,II LERGs, generally found in environments denser of galaxies with respect to the former. The same situation occurs when considering radio and/or optical flux limited surveys, where high luminosity sources (i.e., mostly HERGs) will appear inhabiting less dense environment than low luminosity ones (i.e., LERGs). This will also appear as function of the stellar mass ($M_\mathrm{star}$) since brighter sources have also higher values of $M_\mathrm{star}$, being generally estimated from the absolute magnitude \citep{miraghei17,ching17}. Thus to test if FR\,II HERGs and FR\,II LERGs live in environments with different galaxy density it could be useful to investigate this aspect as function of redshift, as previously reported. This will also guarantee to have the analysis an independent of the efficiency of clustering algorithms with $z$ and of the cosmological evolution of the two source classes.

\begin{figure}[]
\begin{center}
\includegraphics[height=6.2cm,width=8.2cm,angle=0]{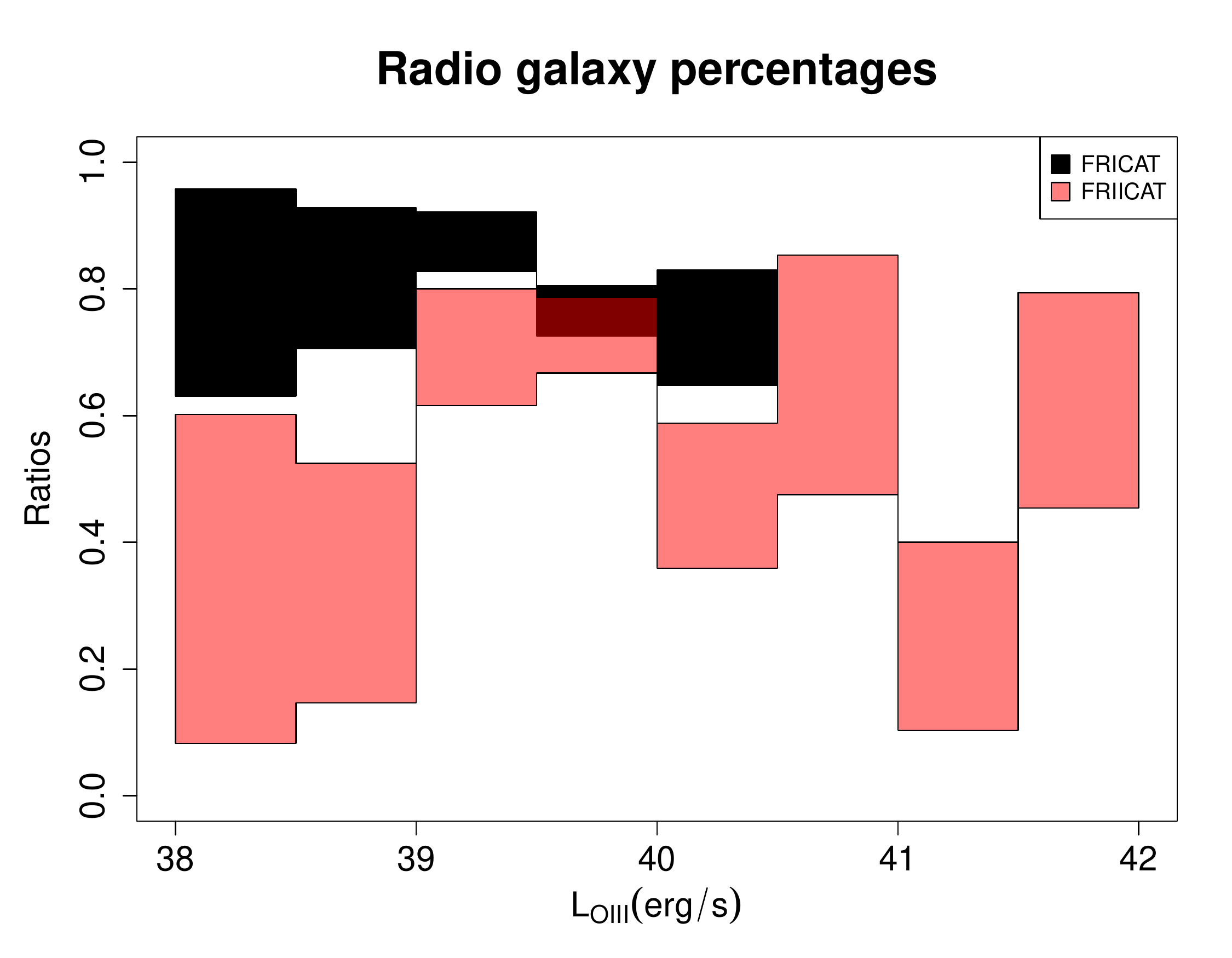}
\caption{Same of Figure~\ref{fig:figure8} and Figure~\ref{fig:figure9} where ratios are expressed as function of the [O III] luminosity $L_\mathrm{[O III]}$. This allows us to highlight the HERGs in the FRIICAT being at high values of $L_\mathrm{[O III]}$.}
\label{fig:figure14}
\end{center}
\end{figure}

\begin{figure}[]
\begin{center}
\includegraphics[height=6.2cm,width=8.2cm,angle=0]{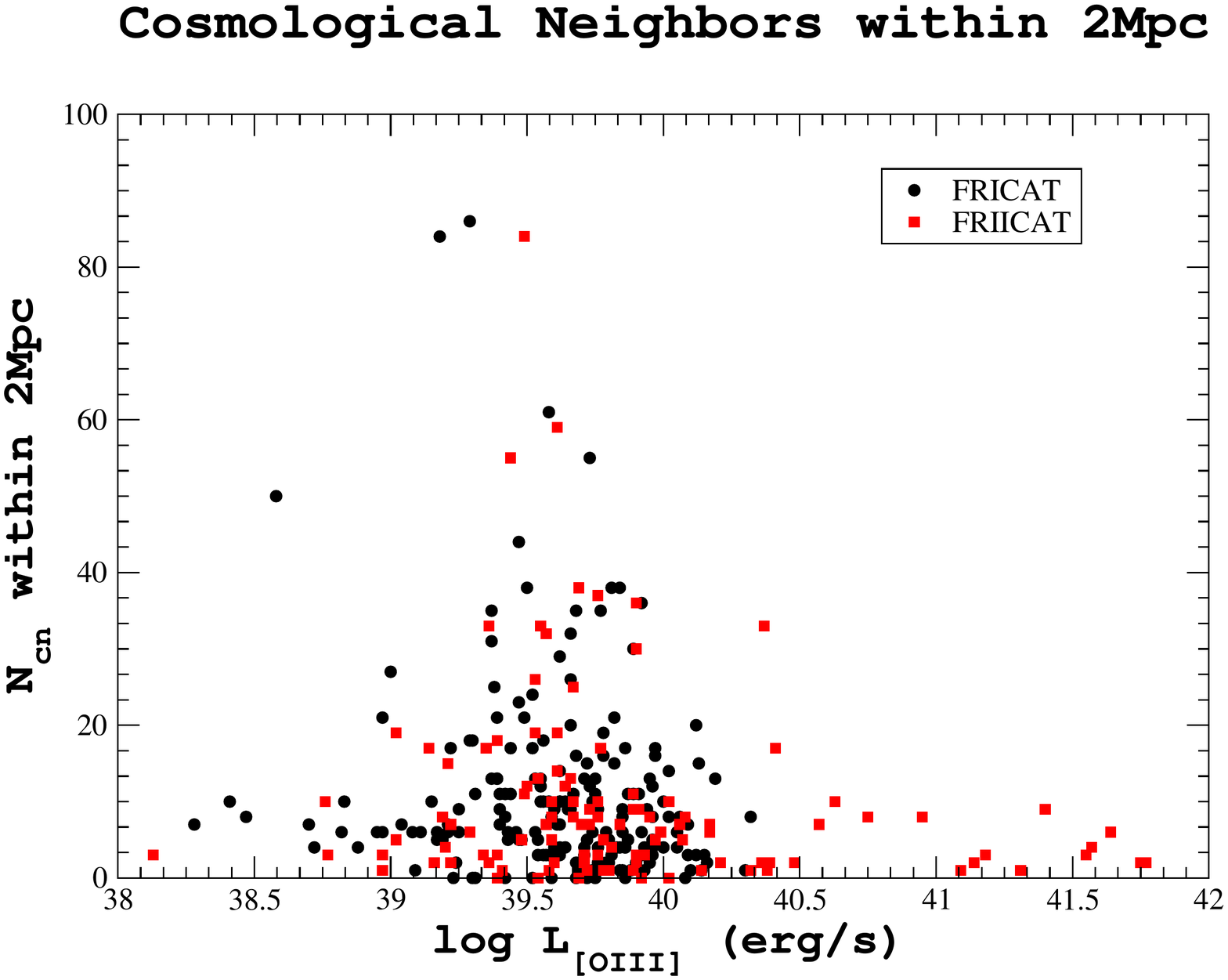}
\caption{Same of Figure~\ref{fig:figure13} where the total number of {\it cosmological neighbors} $N_\mathrm{cn}$ within 2\,Mpc is reported as function of the [OIII] emission line luminosity $L_\mathrm{[OIII]}$.}
\label{fig:figure15}
\end{center}
\end{figure}

We then plotted the ratio of FR\,Is and FR\,IIs in galaxy-rich environments over their total number as function of the [O III] luminosity $L_\mathrm{[OIII]}$ in Figure~\ref{fig:figure14}. At higher values of $L_\mathrm{[OIII]}$ (i.e., above $\sim$10$^{40}$erg\,s$^{-1}$) most of FR\,IIs are all HERGs, but again it seems that their fraction in galaxy-rich environments does not strongly depend by $L_\mathrm{[OIII]}$. The small difference at $log\,L_\mathrm{[OIII]}<$39 erg\,s$^{-1}$ is simply due to the smaller number of FR\,IIs in the FRIICAT. For the sake of completeness we also show in Figure~\ref{fig:figure15} the total number of {\it cosmological neighbors} $N_\mathrm{cn}$ within 2\,Mpc as function of the [OIII] luminosity: $L_\mathrm{[OIII]}$ and not neat differences appear in the range $40<log\,L_\mathrm{[OIII]}<$40.5 erg\,s$^{-1}$, where radio galaxy catalogs include both LERGs and HERGs.

{\it It is worth highlighting that our study is based on extremely homogeneous samples, with respect to other analyses present in the literature, but being restricted to the local Universe (i.e., at $z_\mathrm{src}\leq$0.15) has radio galaxy catalogs including only a limited number of HERGs (only 14 all in the FRIICAT). Thus results on LERGs and HERGs comparison has to be treated with caution, being less statistically strong.}  

The total number of FR\,II LERGs lying in galaxy-rich large-scale environment is 34; where 9 out of 36 (i.e., 75\%) is at $z_\mathrm{src}\leq$0.11. This is the same fraction FR\,II HERGs at similar redshifts. Below $z_\mathrm{src}=$0.11 the total number of HERGs in galaxy-rich environments is 5 out of a total of 7, while this fraction decreases as 2 out of the remaining 7 at higher redshifts.

Using the GMBCG we found that 33 FR\,Is (17\%) appear associated with a cluster hosting a BCG candidate, lying within $\Delta\,z\leq$0.005, estimated by using only spectroscopic redshifts. They also belong to galaxy clusters with more than 8 members. In particular 10\% of their total number lie at projected distance smaller than 1\,kpc, being themselves the BGC candidates. A similar situation occurs again for FR\,IIs. Eighteen (16\% of the FRIICAT) belong to a BCG cluster, as previously stated, and 11 (10\%) are BCG candidates. There are no FR\,II HERGs that are positionally associable with a BCG listed in the GMBCG, even if HERGs are always the most luminous sources with respect to all their {\it cosmological neighbors}.

\subsection{The $\Sigma_k$ comparison}
We also investigated the distribution of the $\Sigma_k$ parameter, i.e. the k-th nearest neighbor density \citep[see e.g.,][]{best04}, for both radio galaxies and MOCK sources. 

The $\Sigma_k$ parameter is defined as the ratio between the source number k and the projected area $\pi\,r_k^2$, where $r_k$ is the projected distance between the central galaxy and the k-th nearest neighbor. We  computed it for $k$=5 (i.e., $\Sigma_5$), adopting the distance (in kpc) between the central galaxy and the fifth closest {\it candidate elliptical galaxy}: \citep{ching17}. 

As extensively discussed in the literature, this parameter can be used as beacon to trace the dark matter halo density \citep{sabater13,worpel13} and it also appears to correlate with its halo mass \citep{haas12}. 
\begin{figure}[]
\begin{center}
\includegraphics[height=6.2cm,width=8.2cm,angle=0]{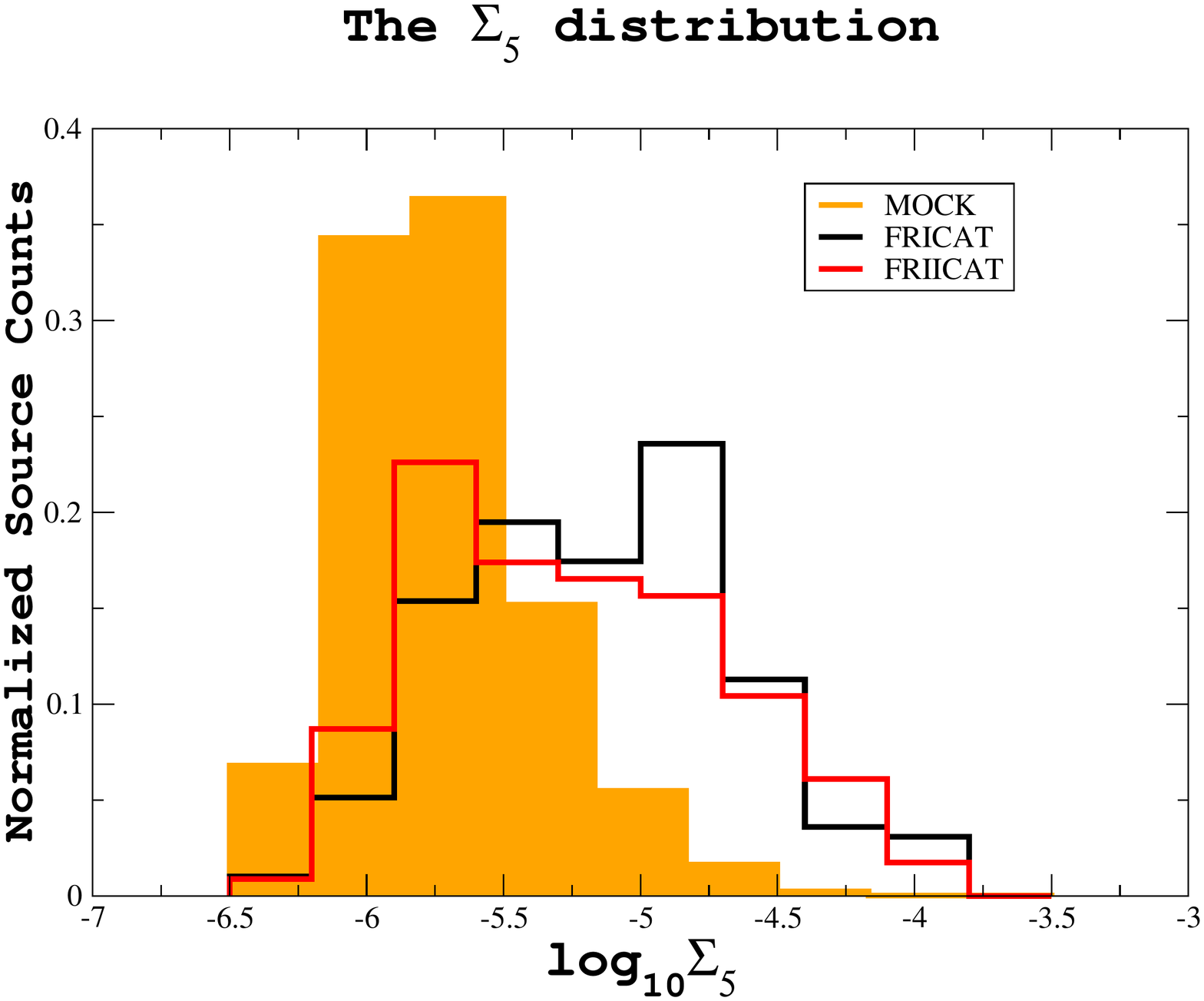}
\includegraphics[height=6.2cm,width=8.2cm,angle=0]{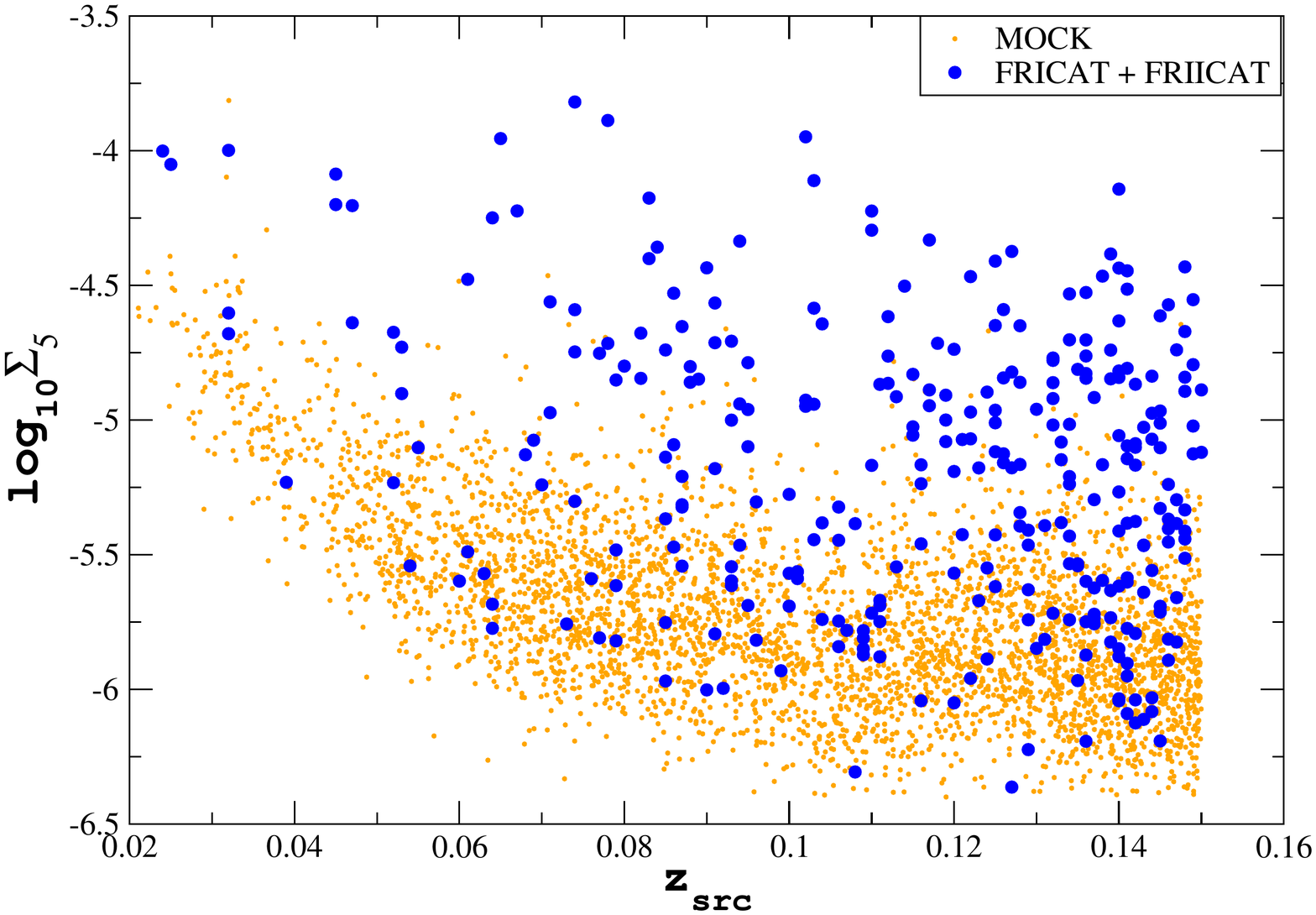}
\caption{Top panel) The normalized distribution of the $\Sigma_5$ parameter, a.k.a. the fifth nearest neighbor density, tracing the dark matter halo density on large-scale environments \citep{sabater13,worpel13}. It is evident how radio galaxies tend to have, on average, larger values of $\Sigma_5$ with respect to the values computed for fake sources in the MOCK catalog. Bottom panel) The log of the $\Sigma_5$ parameter as function of the central source redshift $z_\mathrm{src}$. In this plot radio galaxies (blue circles) are reported together in comparison with MOCK sources (orange circles). It is clear how the $\Sigma_5$ estimator is affect by a redshift dependence due to the Malmquist bias.}
\label{fig:figure16}
\end{center}
\end{figure}
According to all previous analyses the distribution of $\Sigma_5$ as reported in Figure~\ref{fig:figure16} (top panel), also shows that both FR\,Is and FR\,IIs live in galaxy-rich large-scale environments, with larger values of dark matter halo density, than random MOCK sources (see Figure~\ref{fig:figure16}). However, the lower panel of the same figure marks a $z_\mathrm{src}$ dependence of $\Sigma_5$, possibly due to the Malmquist bias that could affect the use of this estimator.

\section{Summary and Conclusions}
\label{sec:summary} 
We presented a detailed statistical analysis of the large-scale environments of radio galaxies. 

The main advantages of our study, with respect to those previously carried out in the literature, are the (i) sample selection, extremely homogenous over a wide range of frequencies and the (ii) large variety of clustering algorithms adopted for our analysis always providing consistent results.

In particular, for the radio galaxies, we used the FRICAT and FRIICAT catalogs, complete at level of confidence higher than 90\%. Thanks to their selection criteria these catalogs are not contaminated by compact radio objects, as compact steep spectrum sources and FR\,0 \citep{baldi15,baldi18}, which show a different cosmological evolution, with respect to FR\,Is and FR\,IIs and potentially lie in different environments. 

We proved that identifying a galaxy-rich environment only using a cross-matching analysis with a catalog of groups and clusters can introduce biases due to the redshift dependence of the algorithm (e.g., FoF) used to build it. Thus our analysis was carried out performing a direct search around radio galaxies using them as beacons.

We investigated the large-scale environment of radio galaxies adopting both their radio morphological \citep[FR\,I $vs$ FR\,II;][]{fanaroff74} and the optical spectroscopic classification \citep[HERG $vs$ LERG;][]{hine79}. This allowed us to to search for a link between their environment and their radio extended structure and their accretion modes. However, it is worth mentioning that due to the selection criteria of our radio galaxy catalogs, the limited number of HERGs listed in the FRIICAT does not allow us to firmly compare our results with those present in the literature.

Our main results is that, in the local Universe, FR\,Is and FR\,IIs as well as HERGs (all FR\,IIs in our catalog) and LERGs live in galaxy-rich large-scale environments having the same richness, independently on the redshift range considered or their radio luminosity or their absolute magnitude. This is also independent of thresholds and algorithms used to identify a galaxy-rich large-scale environments. 

More than 70\% of the FR\,Is and more than 55\% of all FR\,IIs in our catalogs lie in galaxy-rich large-scale environments. The probability to find an FR\,I lying in a dense neighborhood appears larger than for FR\,IIs. This claim could be also biased by the possible presence of fossil groups where FR\,IIs could reside and that could be only revealed with X-ray observations. As previously stated, the number of HERGs in our samples prevent us to make a strong statistical conclusion, when considering the optical classification, even if 5 out of 7 HERGs up to $z_\mathrm{src}=$0.11 lie in groups/clusters of galaxies and this fraction decreases up to 2 out of 7 at higher redshifts. On the other hand, finding HERGs in denser environments is consistent with optical observations carried out on high redshift radio galaxies, generally used as beacon to search for protoclusters \citep[see e.g.,][for a recent review]{miley08}, that appear to be all HERGs \citep[see e.g.,][]{rottering97,debreuk01,jarvis01,debreuk06,bornancini07}.

We also found that $\sim$17\% of the FR\,Is are associated with a cluster hosting a BCG candidate, lying within $\Delta\,z\leq$0.005, estimated by using only spectroscopic redshifts reported in the GMBCG catalog. All these FR\,Is belong to galaxy clusters with more than 8 members and 20 of them (10\% of the total number) being themselves the BGCs. A similar situation occurs again for FR\,II radio galaxies whereas $\sim$16\% belong to a BCG cluster, and 10\% are the BCGs. There are no FR\,II HERGs in our sample associated with a BCG when considering the GMBCG. However, all HERGs in our sample, that lie in galaxy-rich large-scale environments, are indeed BCGs according to the luminosity distribution of their {\it cosmological neighbors}.

Finally, it is worth highlighting that to carry out our analysis we developed a method based on number of counts of {\it cosmological neighbors} in the large-scale environments of selected sources, measuring cosmological over-densities. This method has been also successfully tested and compared with several clustering algorithms generally used to perform blind searchers of galaxy groups and clusters in large optical and infrared surveys.

\section{Literature comparison}
\label{sec:compare} 
\begin{figure}[ht]
\begin{center}
\includegraphics[height=8.2cm,width=8.2cm,angle=0]{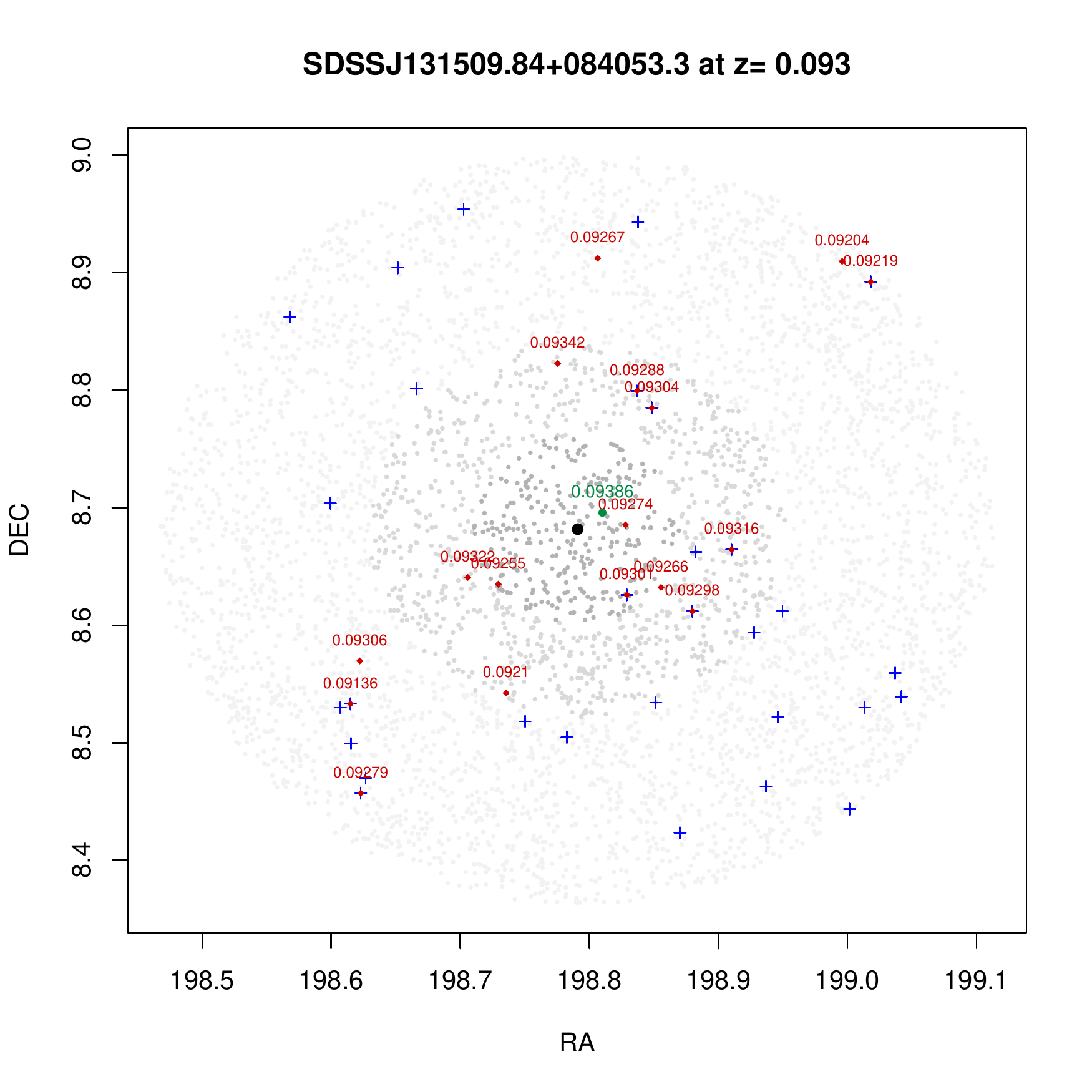}
\caption{The large-scale environment of SDSSJ131509.84+084053.3 a radio galaxy classified as FR\,II HERG. The total number of {\it cosmological neighbors} within 1\,Mpc (i.e., 11) is a clear example of a HERG associated with a galaxy group/cluster at $z_\mathrm{src}=$0.093.}
\label{fig:figure17}
\end{center}
\end{figure}

The comparison between the results achieved in our analysis, even if limited to the local Universe (i.e., $z_\mathrm{src}\leq$0.15) and the claims present in the literature (all reported below in italics) can be summarized as follows. {\it Here literature claims are reported in italic.}
\begin{QandA}
\item {\it HERGs are found almost exclusively in low-density environments while LERGs occupy a wider range of densities, independent of FR morphology \citep{gendre13}.}
         \begin{answered}
         Indeed we proved that LERGs and HERGs in the local Universe (at least up to $z_\mathrm{src}\leq$0.1) live in galaxy-rich large-scale environments with same richness. This conclusion is based on the number of {\it cosmological neighbors} within 2\,Mpc (see Figure~\ref{fig:figure11}). At higher redshifts it was not possible, for us, to establish a firm conclusion given the low number of HERGs in our catalogs. Our result is consistent with X-ray observations of FR\,IIs \citep[see e.g.,][]{hardcastle00}.
         \end{answered}

\item {\it There is a significant overlap in the environment between LERGs and HERGs, and no clear driving factor between the FR\,I and FR\,II sources is found even when combining radio luminosity, accretion mode \citep{gendre13}.}
         \begin{answered}
	We confirmed this statement.
         \end{answered}

\item {\it FR\,Is radio galaxies lie in higher density environments, on average, than FR\,IIs \citep{miraghei17}.}
         \begin{answered}
         On the contrary, we showed that fraction of FR\,Is living in galaxy-rich large-scale environments could be slightly larger than that of FR\,IIs, but the richness of their environments is certainly consistent at all redshifts sampled by our analysis (see right panel of Figure~\ref{fig:figure11}). The difference with respect to literature works could be due to radio sources that are contaminants of selected samples, a bias that does not affect our radio galaxy catalogs. The major difference with respect to literature analyses, that strengthen our result, is that the sample selection carried out for both radio galaxy catalogs is extremely homogeneous when considering their radio morphology.
         \end{answered}

\item {\it The environments of LERGs display higher density compared to the HERGs \citep{miraghei17}.}
         \begin{answered}
         No differences were found in the environments of LERGs and HERGs in our analysis. The case of the FR\,II HERG SDSSJ131509.84+084053.3 having 11 {\it cosmological neighbors} within 1\,Mpc is shown in Figure~\ref{fig:figure17} as example of a galaxy-rich environment. However, we again highlight that our claim is limited to $z_\mathrm{src}\leq$0.15, where the number of HERGs is only a tiny fraction of the whole FRIICAT (i.e., $\sim$10\% of the FRIICAT), but it does not suffer of possible selection effects due to different cosmological evolution of HERGs and LERGs.
         \end{answered}

\item {\it High-luminosity radio galaxies with weak or no emission lines (LERGs) lie in more massive haloes than non-radio galaxies of similar stellar mass and color \citep{ching17}. The HERGs are typically in lower mass haloes than LERGs.}
         \begin{answered}
         The distribution of the $\Sigma_5$ parameter for our radio galaxy catalogs is not in agreement with this statement at least in the $z_\mathrm{src}$ range considered. As in the previous case, we remark that our result, even if based on a, statistically homogeneous, sample selection is limited by the number of HERGs in the radio galaxy catalogs (i.e., about 10\% of all FR\,II sources).
         \end{answered}

\item {\it At low redshifts, there is a correlation between radio luminosity and the cluster environment for LERGs but not for HERGs \citep{ineson15}.}
         \begin{answered}
         No difference between the richness of their environments were indeed found as function of their $L_R$, not even between FR\,Is and FR\,IIs. However, we also remark that the sample selected by Inseson et al. (2015) spans a wider range of radio luminosities than our radio galaxy catalogs that could better reveal trends between radio power and richness. It is worth highlighting that in our analysis the richness is estimate by the number of {\it cosmological neighbors} instead of using X-ray observations. The advantage of our approach is that it is difficult to get X-ray observing time for a large, homogeneously selected samples of sources. This is challenging since good spatial resolution is needed to separate diffuse X-ray emission surrounding radio galaxies due to the intergalactic medium from the extended one, mainly detected along radio axis, due to inverse Compton scattering of seed photons arising from the Cosmic Microwave Background in radio lobes \citep{scharf03,celotti04,erlund07,smail12,massaro13b,massaro18,stuardi18}.
         \end{answered}         
         
\end{QandA}
Finally, it is worth mentioning that, in the future, we could also extend the radio galaxy catalogs at redshifts larger than 0.15, loosing a small fraction of its completeness, but making our analyses more suitable for comparison with others present in the literature \citep[e.g.,][]{ineson13,ineson15}.

\section{Future perspectives}
\label{sec:future} 
Our analysis is only based on radio and optical information while to obtain a more complete view of the large-scale environments of radio galaxies X-ray observations will be crucial. CArrying out these observations we could perform a unique, unbiased, survey of radio galaxies, extremely homogeneous in terms of source selection. 

An X-ray survey will allow us to: (1) know the real fraction of FR\,Is lying in groups or being in galaxy clusters taking into account fossil groups; (2) measure mass, temperature and luminosity of the IGM; (3) determine the location of the radio galaxy with respect to the group/cluster center, an information immediately obtainable from the IGM distribution traced in the X-rays; (4) discover the the fraction of radio galaxies lying in cooling core groups/clusters and test if there is a gradient of temperature close to the radio galaxy, revealing active feedback processes, to name a few examples.

\appendix
\section{Comparison with other clustering algorithms}
\label{sec:appendix} 
We carried out our analysis also considering several clustering algorithms, generally adopted to search for groups and clusters of galaxies in large optical surveys. This allowed us to (i) claim that our analysis is independent by the method adopted and to (ii) test efficiency and/or biases and/or limits of different methods.

The clustering algorithms adopted for our comparison are: (i) Density-Based Spatial Clustering of Applications with Noise (DBSCAN), (ii) Voronoi Tessellation and (iii) Minimum Spanning Tree (MST). These have been used in addition to the analyses carried out with the T12 cross-matches and/or the cosmological over-densities. As in the two previous cases, the level of significance of all clustering algorithms was determined on the basis of results achieved on the MOCK sample. 

In this comparison analysis we searched for source over-densities both hosting a {\it cosmological neighbors} and/or {\it candidate elliptical galaxy}, defining galaxy-rich large-scale environments those above a certain threshold optimized as described below. As last, additional, test we also run the KDE estimator on the {\it candidate elliptical galaxies} within 2\,Mpc of each radio galaxy. 

The gap between real and MOCK sources rises clearly when the definition of galaxy-rich large-scale environments depends on the number of {\it cosmological neighbors}, while it is only evident at redshifts larger than $\sim$0.1 when using {\it candidate elliptical galaxies}. This occurs almost for all clustering algorithms adopted due to a large scatter in the optical colors of elliptical galaxies at low redshifts. However it was crucial to test the methods also using {\it candidate elliptical galaxies} for future applications of proposed procedures at higher redshifts than $z_\mathrm{src}=$0.15, where the SDSS spectroscopic coverage decreases significantly.

Finally, we note that results of our additional tests with the clustering algorithms were run keeping FRICAT and FRIICAT separated but are shown here as a whole sample of radio galaxies given their similar galaxy-rich large-scale environment.

\subsubsection{Density-Based Spatial Clustering of Applications with Noise (DBSCAN)}
Density-Based Spatial Clustering of Applications with Noise \citep[DBSCAN;][]{ester96} is a supervised clustering algorithm able to locate regions of high source density. The main advantage of this algorithm is the possibility of discovering source clusters of an arbitrary shape handling the noise. DBSCAN depends of only two parameters: $\varepsilon$ the maximum radius of a neighborhood and $k$, the minimum number of points within a $\varepsilon$-neighborhood. A source cluster is defined as a maximal set of density-connected points. On the other hand, its mayor drawback is that results are highly dependent on the choice of $\varepsilon$ and $k$. To avoid this problem we adopted the OPTICS procedure \citep{mihael99}, proposed as an implementation of DBSCAN, to overcome the difficulty on this initial parameter selection.

OPTICS procedure works implementing the DBSCAN algorithm for an infinite number of distance parameters $\varepsilon_i$, which are smaller than a generating distance. This implementation does not produce source clusters explicitly, but generates a so-called {\it reachability plot}, i.e., an ordering of the data objects representing the density-based clustering structure. In the {\it reachability plot} points belonging to a source cluster show up as {\it valleys}, the deeper the valley the denser the source cluster is. Thus we adopted as the input $\varepsilon$ for DBSCAN the mean of 75\% all local maxima in the {\it reachability plot} and then applied the DBSCAN algorithm to locate source clusters. 

\begin{figure}
\begin{center}
\includegraphics[height=5.2cm,width=5.8cm,angle=0]{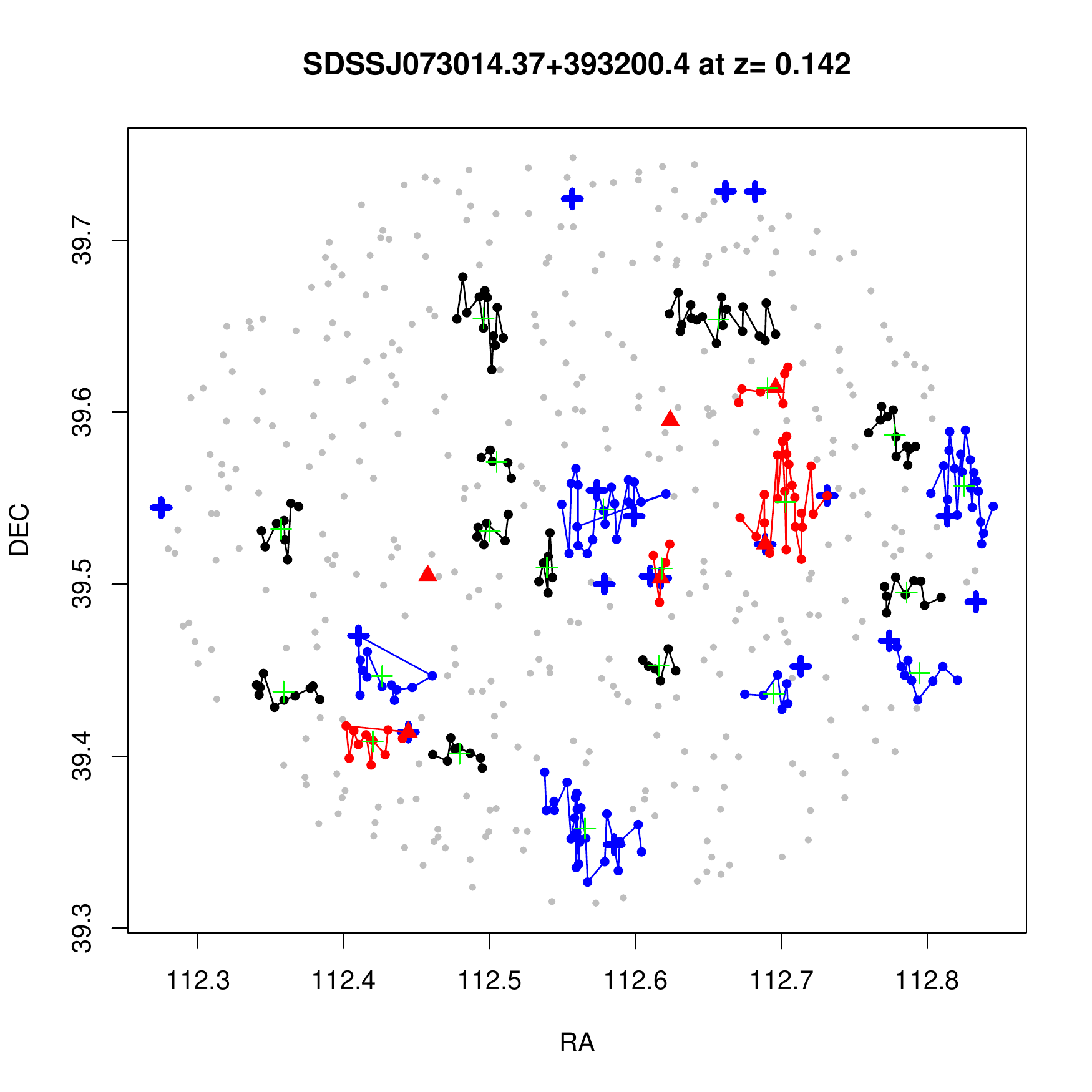}
\caption{The galaxy-rich large-scale environment detected by the DBSCAN algorithm optimized with the OPTICS procedure for a radio galaxy in our sample. We marked in red those clusters for which one of its members is a {\it cosmological neighbors} as previously defined, while in blue those hosting a {\it candidate elliptical galaxy}, while black clusters are those composed by simple SDSS galaxies in the field. Cluster centers are indicated with green crosses. The source analyzed lies in the center of the field and in this case it belongs to a cluster with, at least, a candidate elliptical galaxy.}
\label{fig:figure18}
\end{center}
\end{figure}

In Figure~\ref{fig:figure18} we show as connected points all source clusters found combining DBSCAN algorithm with the OPTICS implementation; if a {\it cosmological neighborhood} is included within a source cluster found, the cluster itself is marked in red, while blue source clusters are those where {\it candidate elliptical galaxies} belong to, all others are simply shown in black.

\begin{figure*}
\begin{center}
\includegraphics[height=6.4cm,width=7.8cm,angle=0]{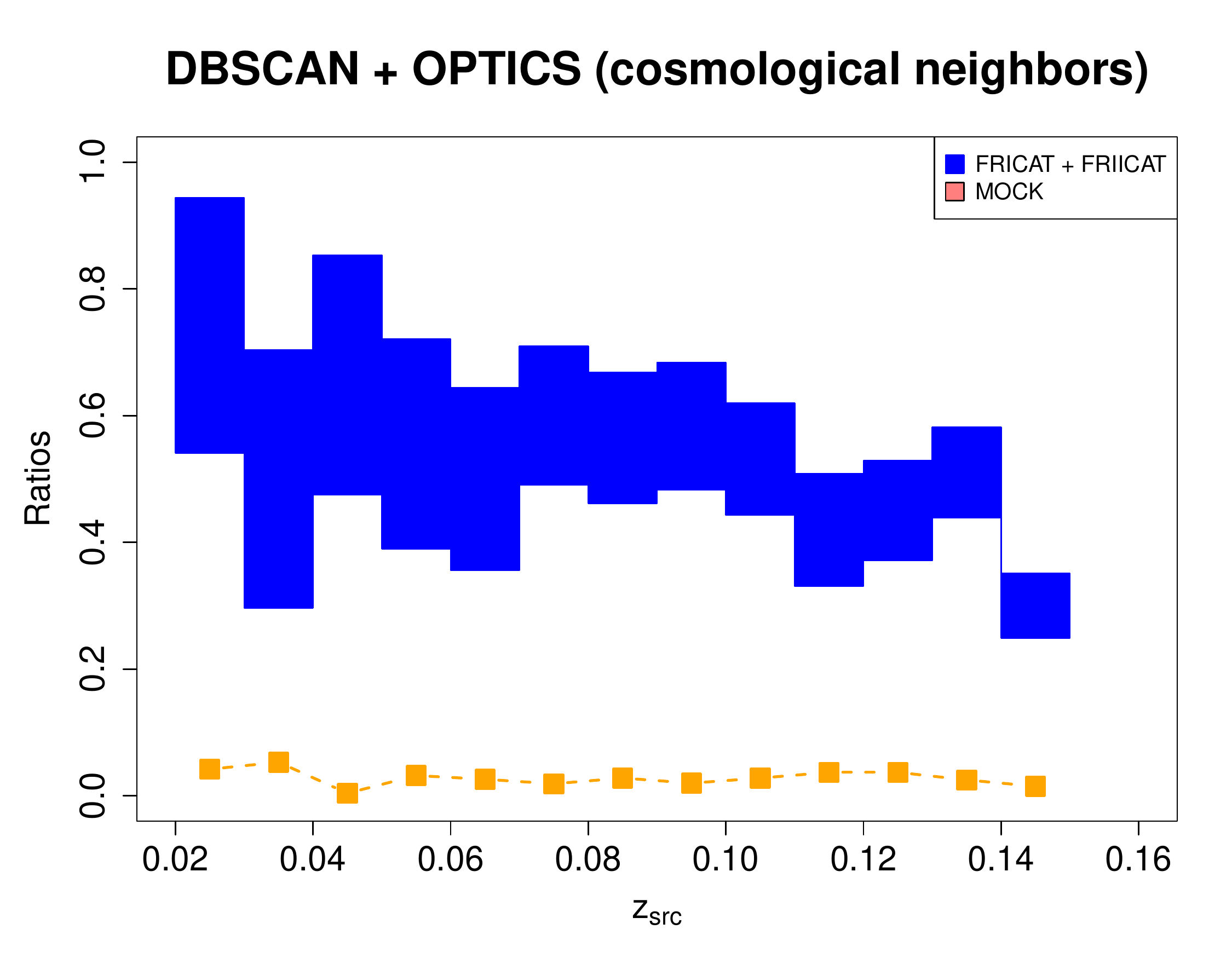}
\includegraphics[height=6.4cm,width=7.8cm,angle=0]{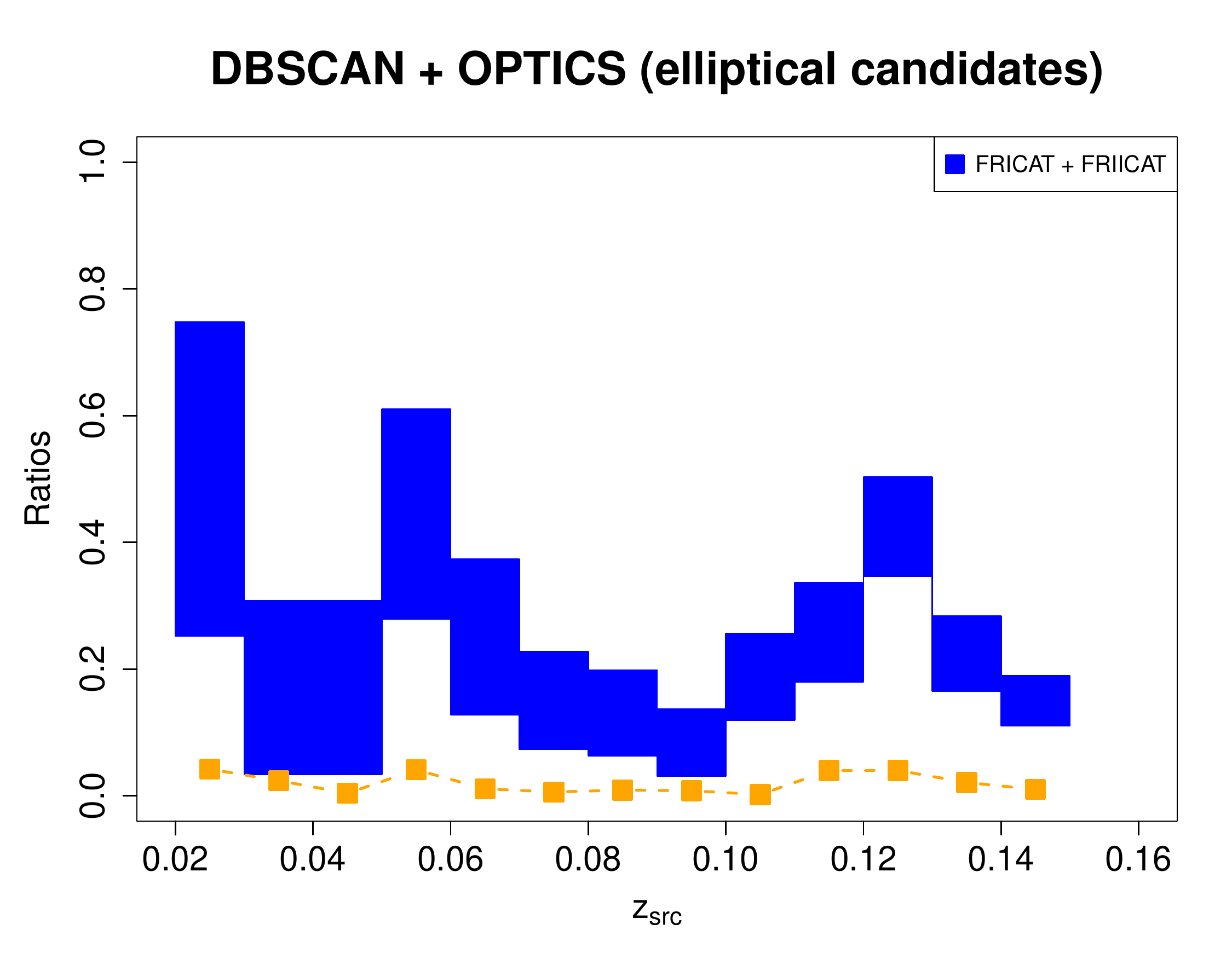}
\includegraphics[height=6.4cm,width=7.8cm,angle=0]{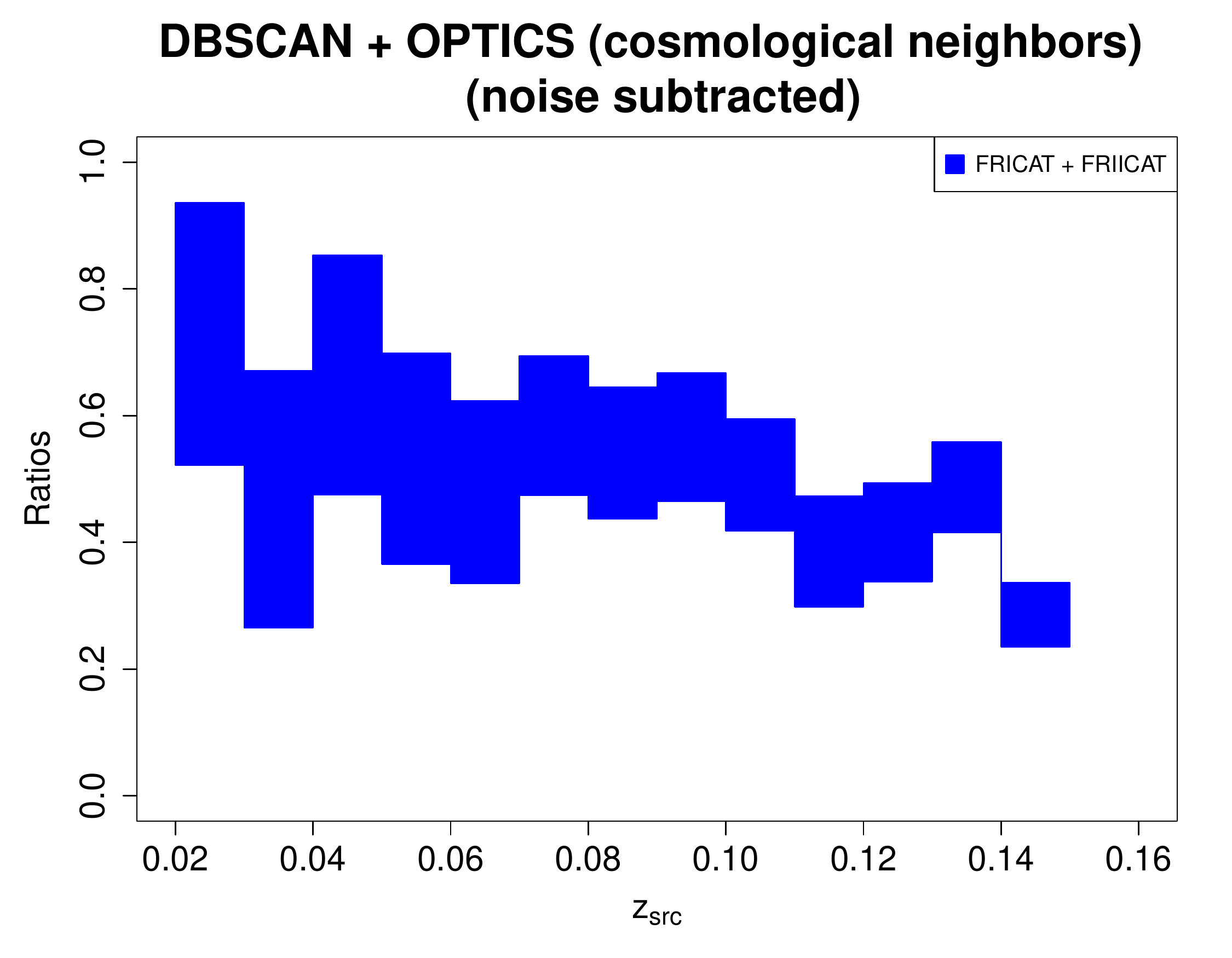}
\includegraphics[height=6.4cm,width=7.8cm,angle=0]{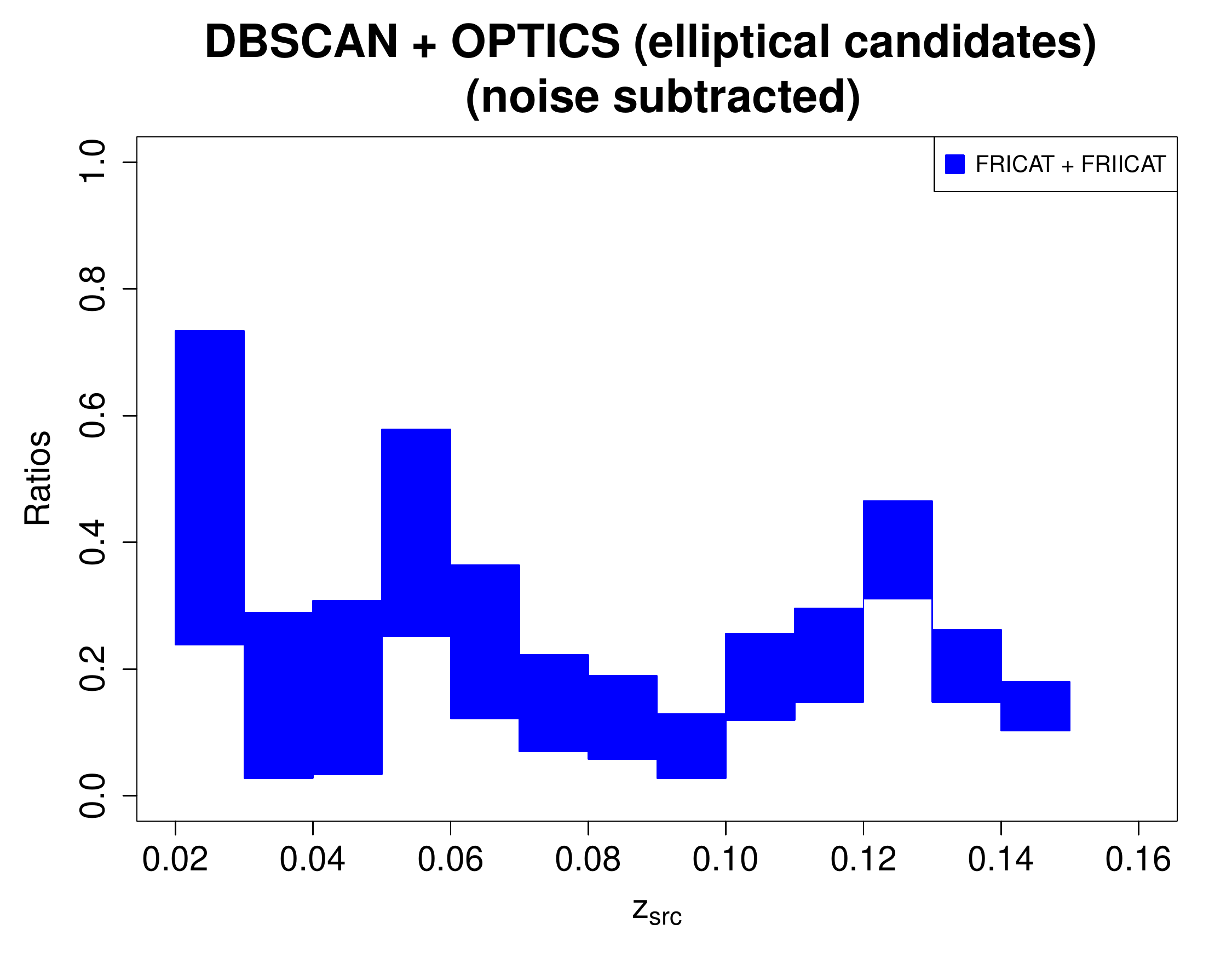}
\caption{The fractions of radio galaxies found in galaxy rich environments computed with the DBSCAN+OPTICS algorithm only. The gap between real and fake sources rises clearly when the definition of galaxy-rich large-scale environments depend on the number of {\it cosmological neighbors} (top left panel), while it is only evident at $z_\mathrm{src}$ larger than 0.1 when using {\it candidate elliptical galaxy} (top right panel). Lower panels show the same ratios but taking into account of the noise subtraction as described in \S~\ref{sec:analysis}.}
\label{fig:figure19}
\end{center}
\end{figure*}

We set thresholds to consider a source surrounded by a galaxy-rich, large-scale environment to the number of {\it cosmological neighbors} or {\it candidate elliptical galaxies}, respectively, lying within a cluster detected applying the DBSCAN+OPTICS algorithm to the MOCK sample within the top 5\% of the cases, as shown in Figure~\ref{fig:figure19}.

\subsubsection{Voronoi Tessellation}
Voronoi tessellation \citep{lee80} is a clustering algorithm that, partitioning a considered region, creates the so-called Voronoi cells on the basis of the distances between points (i.e. sources) presented in the region itself. The area of each Voronoi cell is inversely proportional to the source density in the neighborhood (i.e., smaller areas correspond to regions of higher source density). 

\begin{figure*}
\begin{center}
\includegraphics[height=5.2cm,width=5.8cm,angle=0]{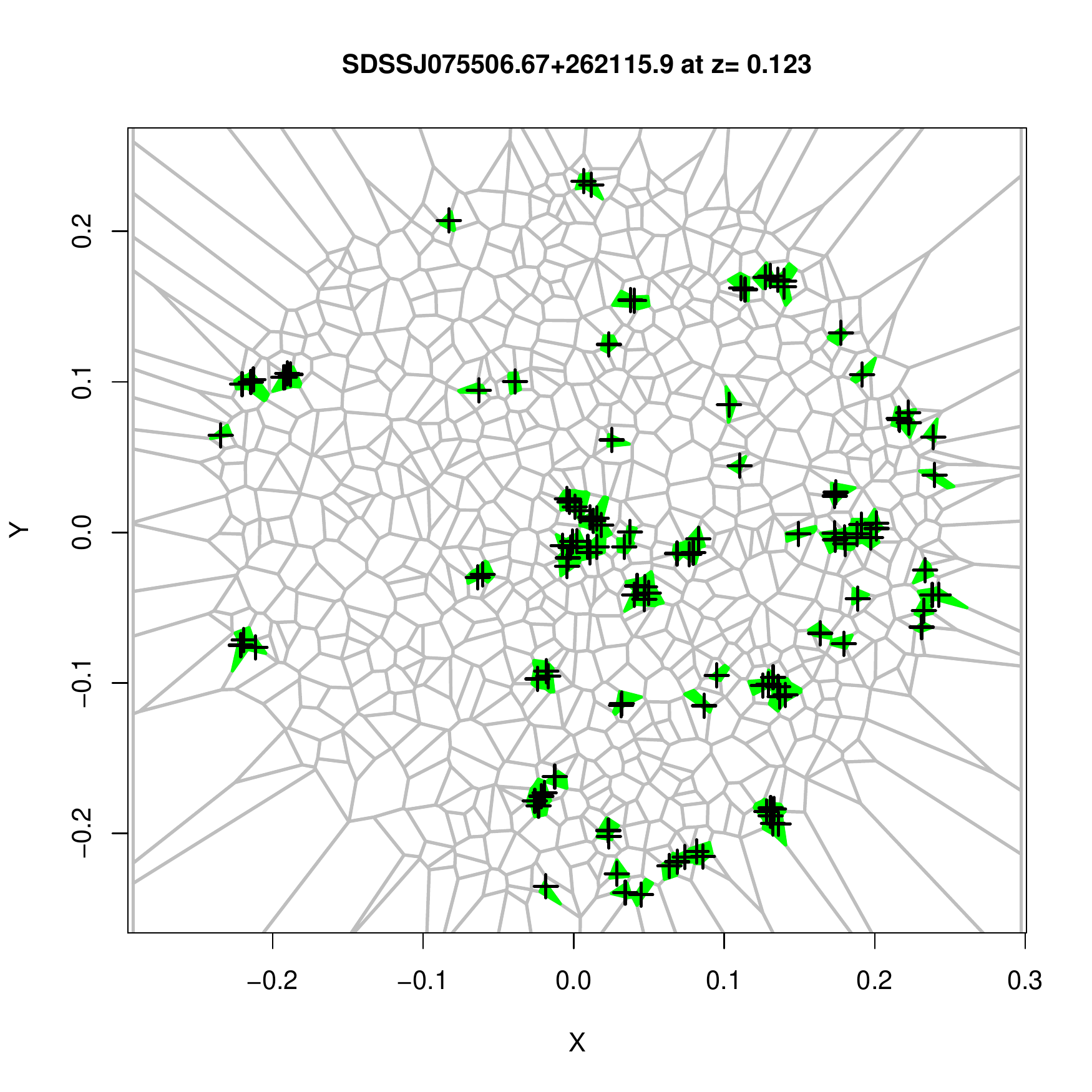}
\caption{The Voronoi cells built for SDSS J075506.67+262115.0 at $z_\mathrm{src}$=0.123 are shown in grey, while regions with area smaller than the average value of top 5\% computed on 100 replicas of the field built assuming the same number of galaxies, with a uniform distribution, are marked in black. Green crosses point the location of sources in the high-density cells. }
\label{fig:figure20}
\end{center}
\end{figure*}

In Figure~\ref{fig:figure20} we show an example of the Voronoi cells computed for the radio galaxy SDSS J075506.67+262115.0 at z=0.123. To mark areas with high source density for each object analyzed, we first counted the number of galaxies within a 2\,Mpc radius and then we simulated 100 replicas of that region, assuming a uniform galaxy distribution. We set a threshold for the area of the Voronoi cells equal to the top 5\% of the simulated fields as galaxy-rich region.

\begin{figure*}
\begin{center}
\includegraphics[height=6.4cm,width=7.8cm,angle=0]{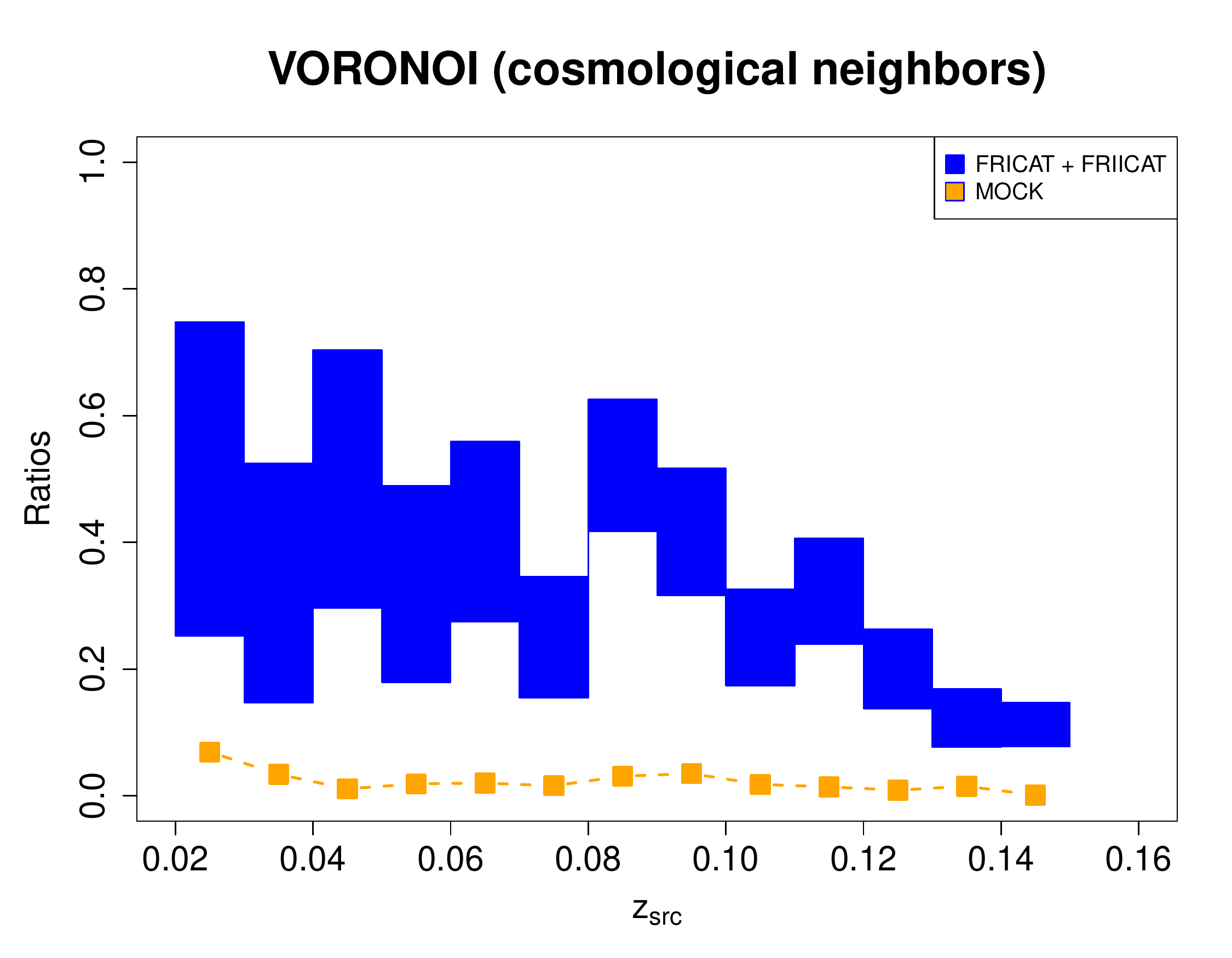}
\includegraphics[height=6.4cm,width=7.8cm,angle=0]{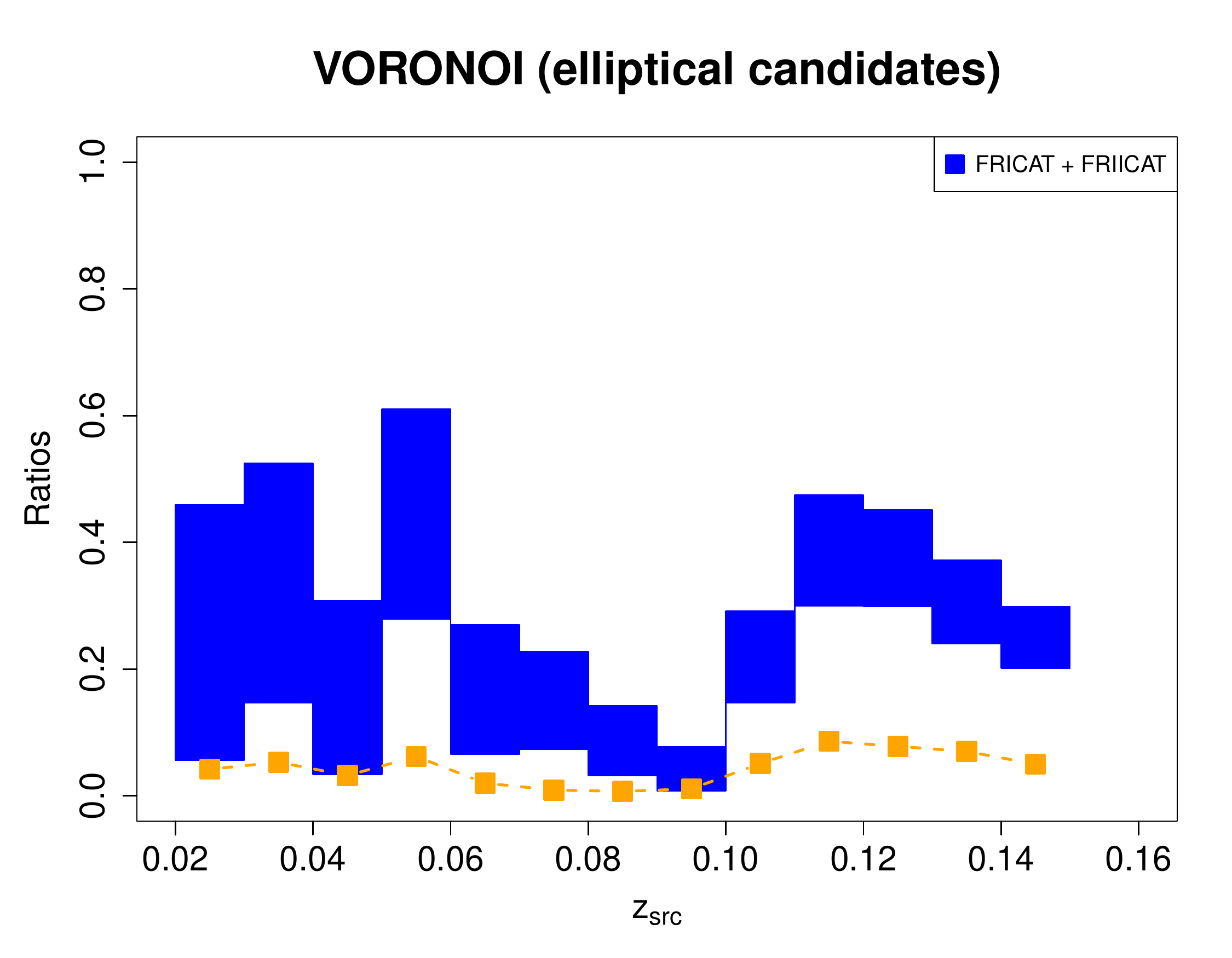}
\includegraphics[height=6.4cm,width=7.8cm,angle=0]{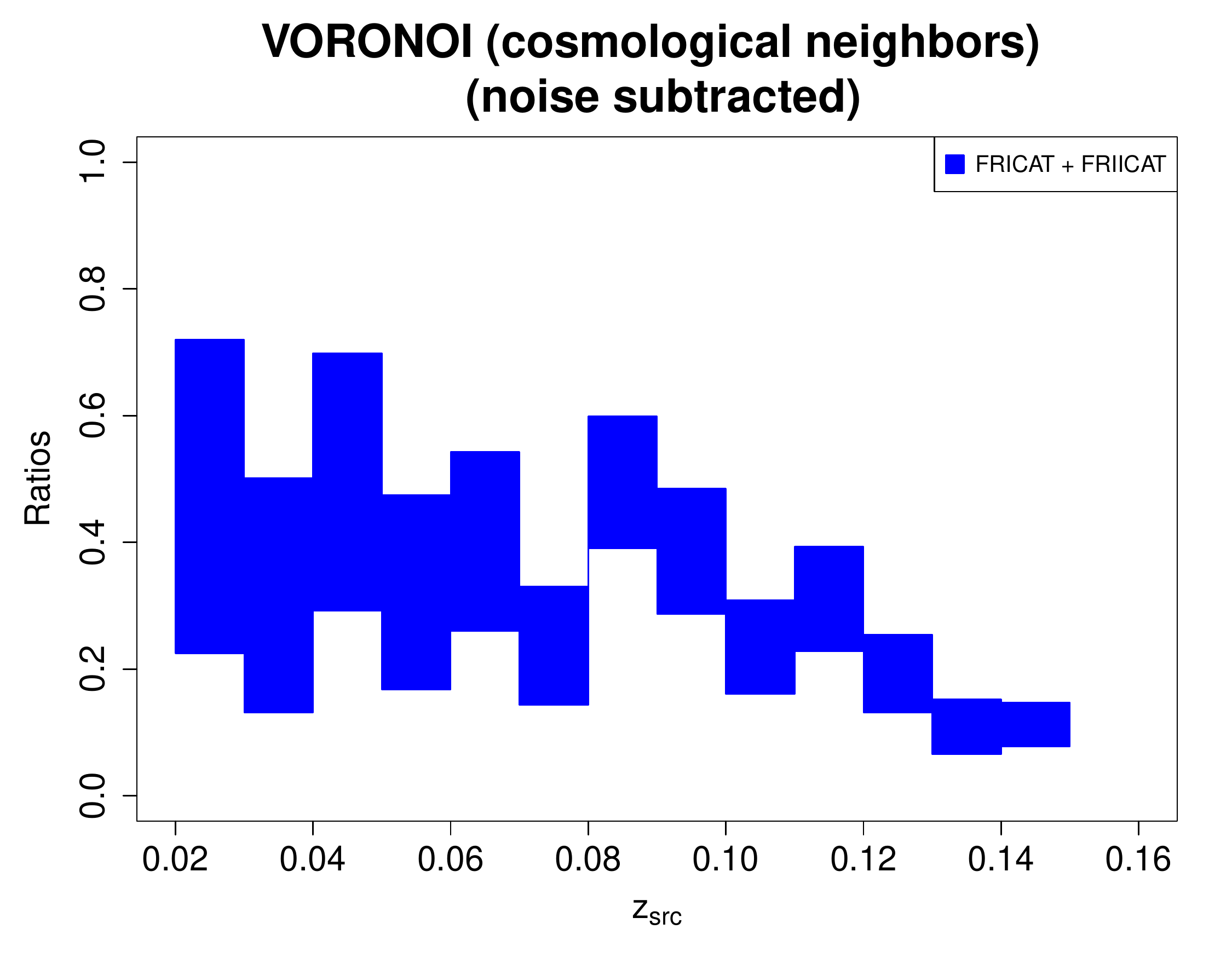}
\includegraphics[height=6.4cm,width=7.8cm,angle=0]{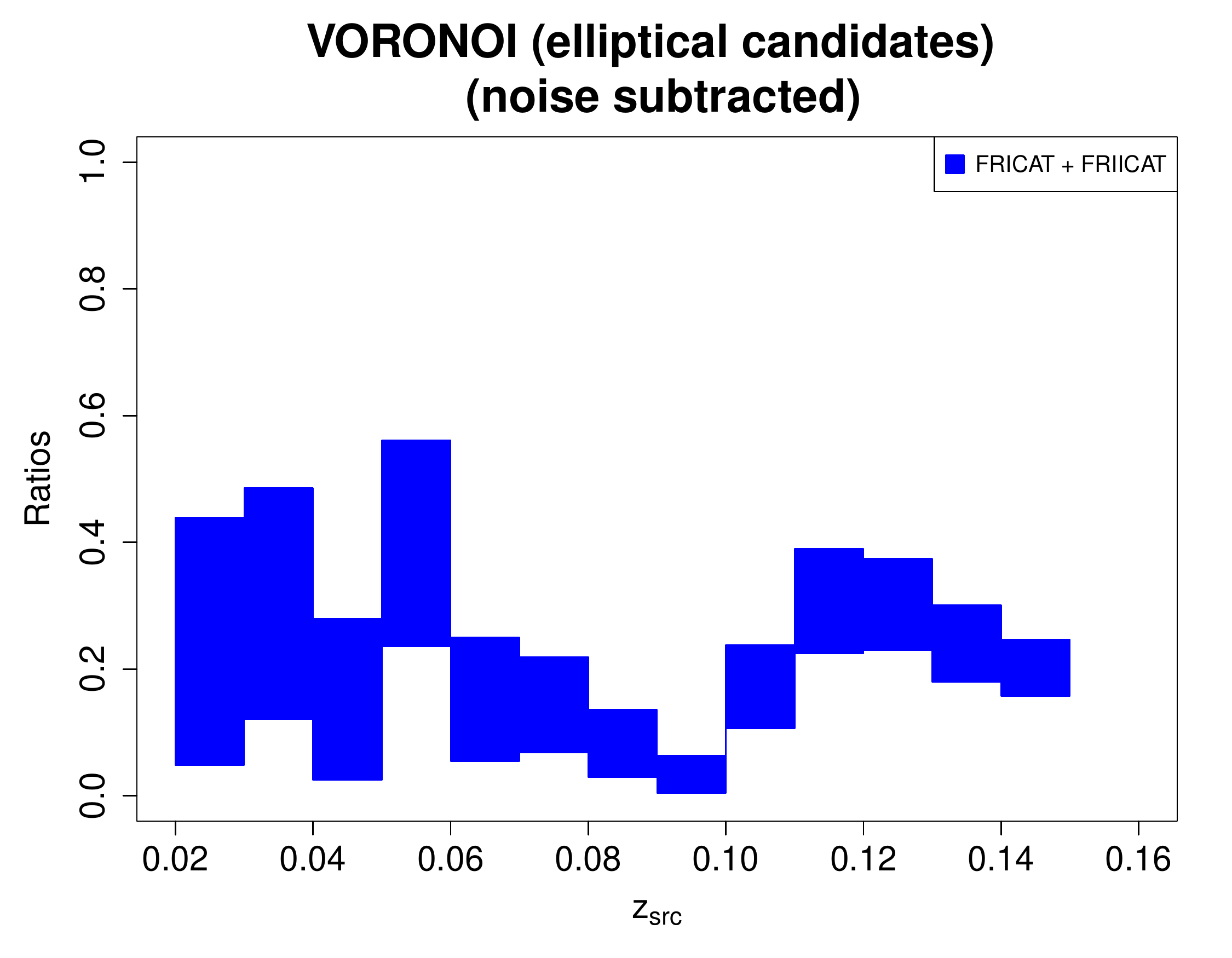}
\caption{Top left and top right panels are the same of Figure~\ref{fig:figure19}, where source fractions are computed with the Voronoi tessellation algorithm only. The gap between real and fake sources is again clear when the definition of galaxy-rich large-scale environments depend on the {\it cosmological neighbors} present in each Voronoi cell (central panel), while it is only evident at $z_\mathrm{src}$$>$0.1 when counting {\it candidate elliptical galaxies} (left panel). Lower panels show the same ratios but taking into account of the noise subtraction as described in \S~\ref{sec:analysis}.}
\label{fig:figure21}
\end{center}
\end{figure*}

Then, as adopted for the previous methods, we considered as galaxy-rich, large-scale environment those Voronoi cells having the number of {\it cosmological neighbors} or {\it candidate elliptical galaxies}, respectively, larger than the top 5\% of those detected in the MOCK sample. Results of this algorithm are shown in Figure~\ref{fig:figure21}, where using {\it cosmological neighbors} the gap between real and fake sources is quite evident while using {\it candidate elliptical galaxies} it becomes less significant, in particular at $z_\mathrm{src}$$>$0.1.

This clustering algorithm is among the most used ones to search galaxy clusters/groups in photometric and spectroscopic surveys \citep[see e.g.,][]{ramella01}. It generally uses both galaxy positions and magnitudes to find clusters as significant density fluctuations above the background. As applied here it is a non-parametric procedure and does not apply any smoothing of the data set.

\subsubsection{The Minimum Spanning Tree}
The minimum spanning tree (MST) is a clustering algorithm used to search for candidate sources in gamma-ray images \citep{campana08,campana13}, as recently occurred for DBSCAN \citep{tramacere13}, but also used to search for groups/clusters of galaxies in photometric surveys \citep{barrow85}. Photon arrival directions or galaxy positions are treated as the nodes of a 2-dimensional graph, over which the tree with the minimal length is constructed. Edges with a length (in this case: angular distance) larger than the MST average value are removed, leaving several disconnected source clusters, which are further selected by their characteristics. Parameters used for the selection are the number of nodes $N$ of the cluster, its clustering degree $g$ (i.e. the ratio between the mean edge length in the MST and the local cluster average edge length) and the magnitude $M$ (defined as $M$$=$$gN$).
\begin{figure*}
\begin{center}
\includegraphics[height=5.2cm,width=5.8cm,angle=0]{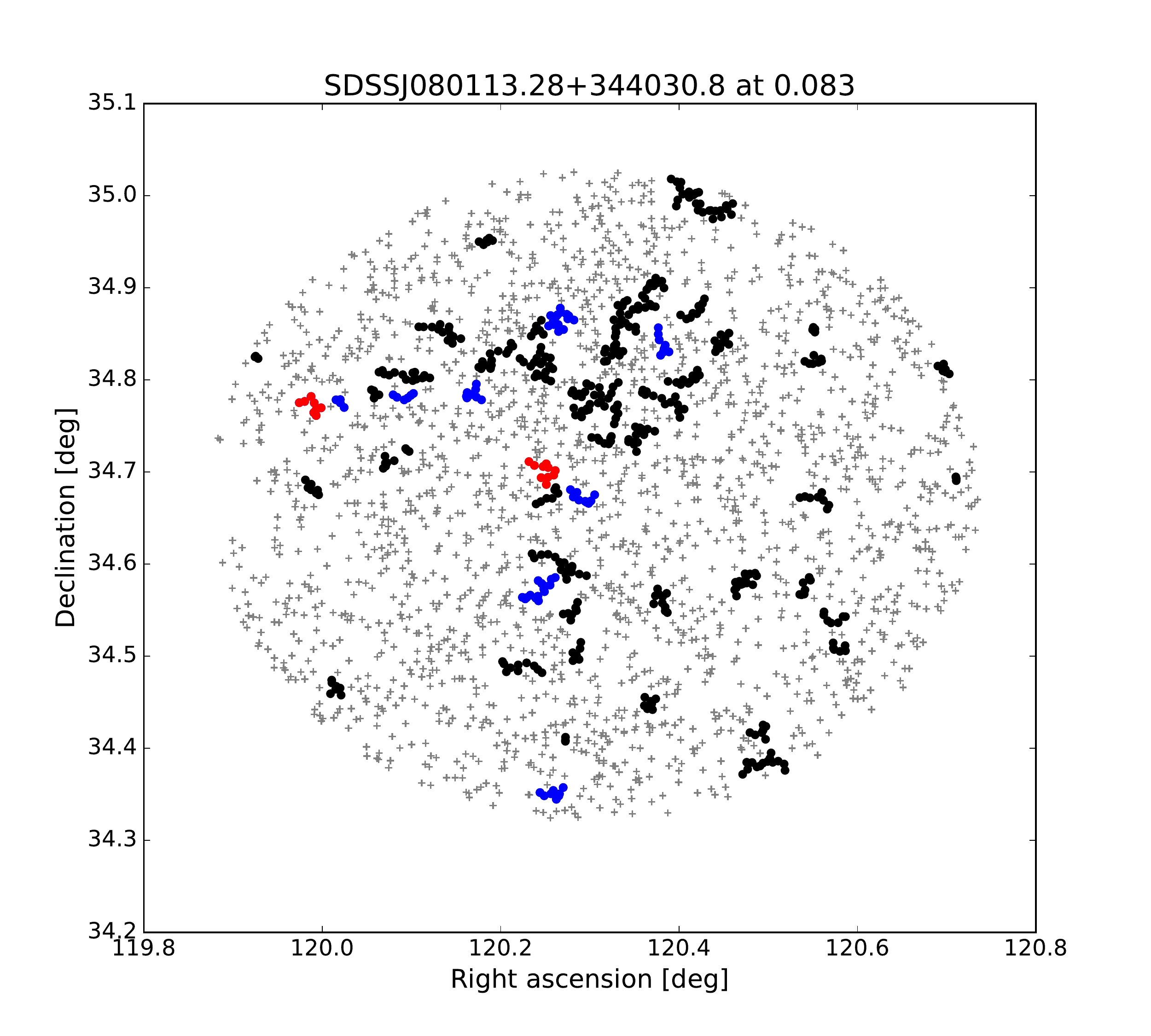}
\caption{The clusters detected using the MST algorithm for the 2\,Mpc region of SDSS J080113.28+344030.8 at $z=$0.083. Those including at least one {\it cosmological neighbors} are shown in red, whole those with at least one {\it candidate elliptical galaxies} appear in blue, the remaining ones are marked in black and overlaid to the background and foreground galaxies in the field.}
\label{fig:figure22}
\end{center}
\end{figure*}

\begin{figure*}
\begin{center}
\includegraphics[height=6.4cm,width=7.8cm,angle=0]{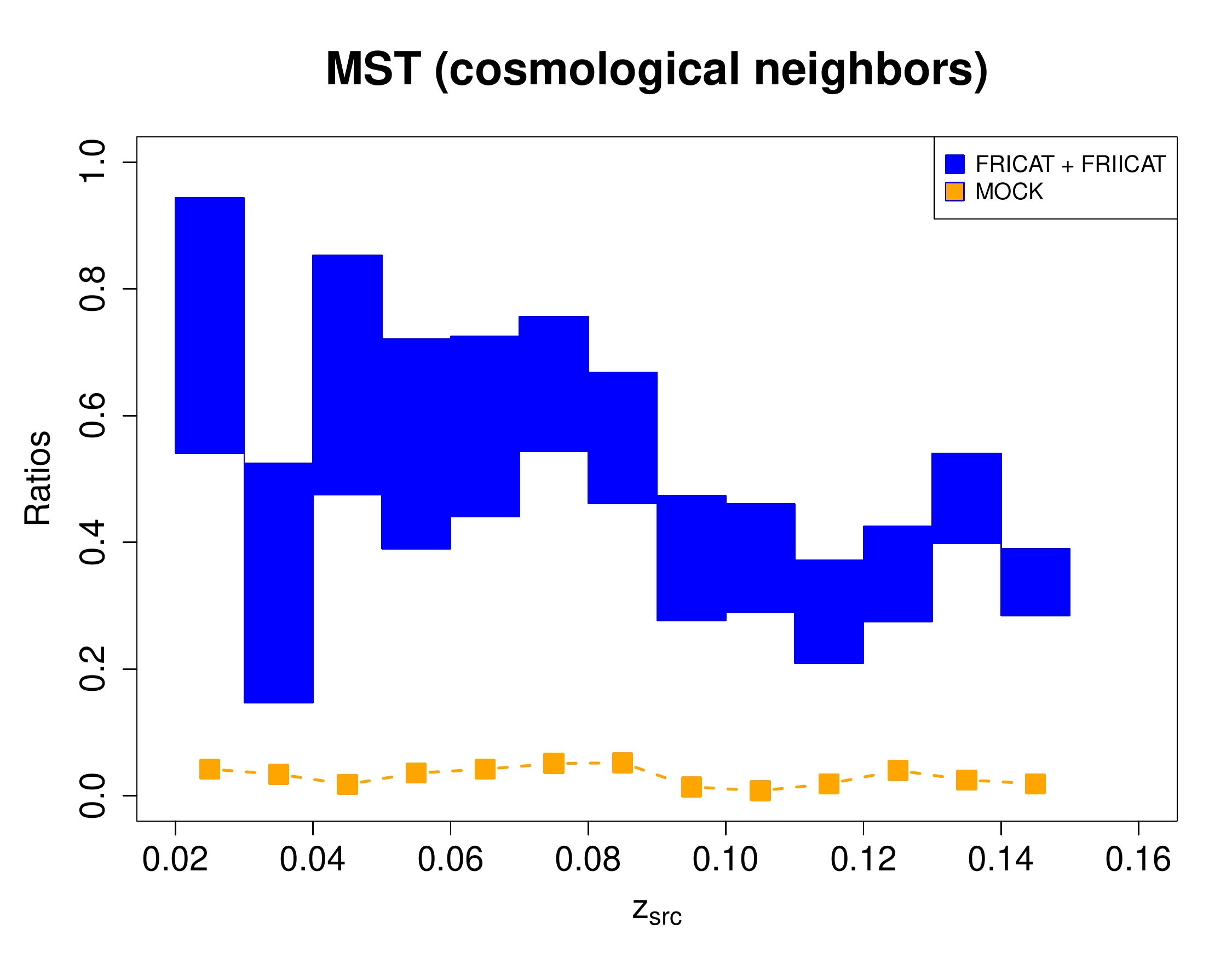}
\includegraphics[height=6.4cm,width=7.8cm,angle=0]{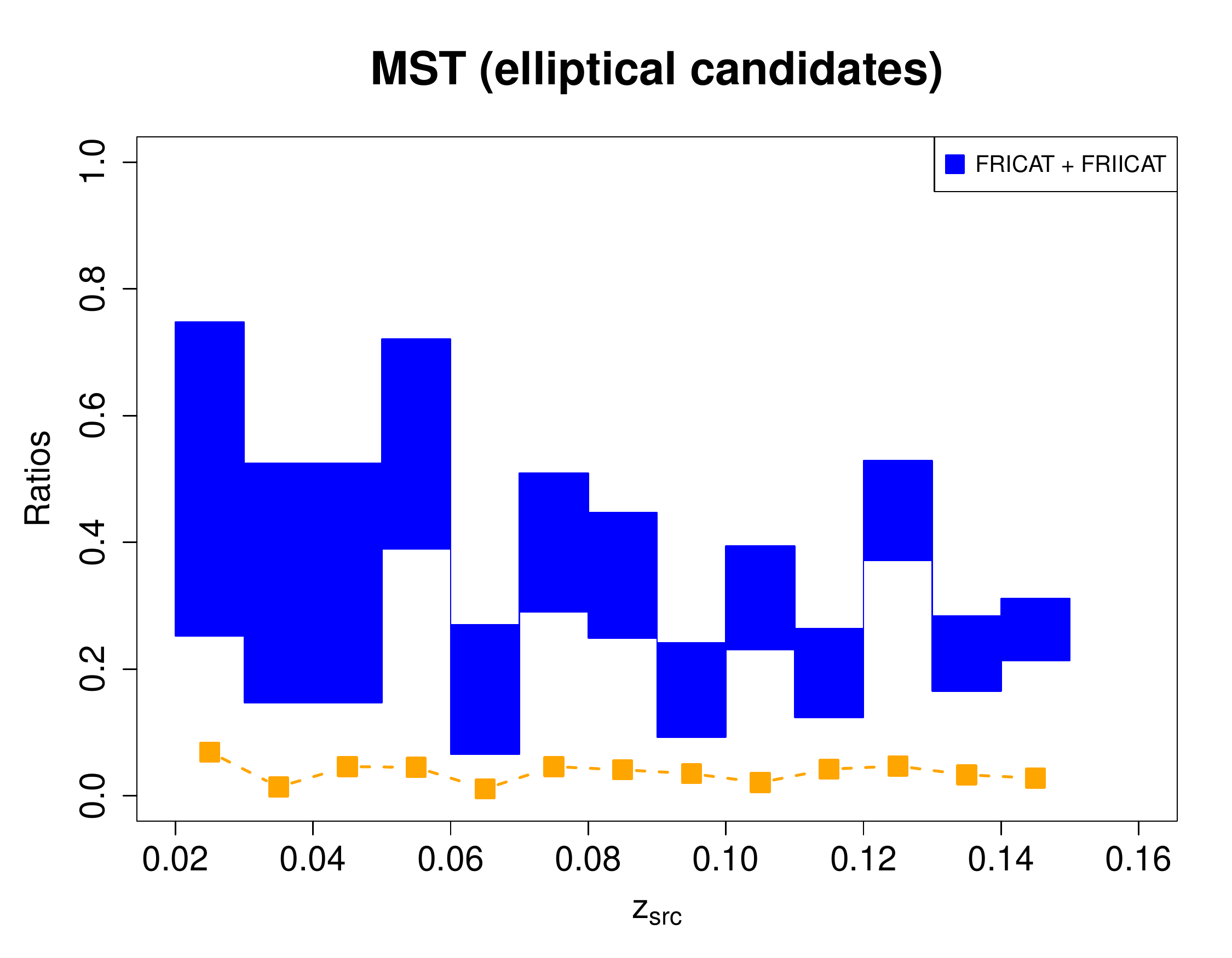}
\includegraphics[height=6.4cm,width=7.8cm,angle=0]{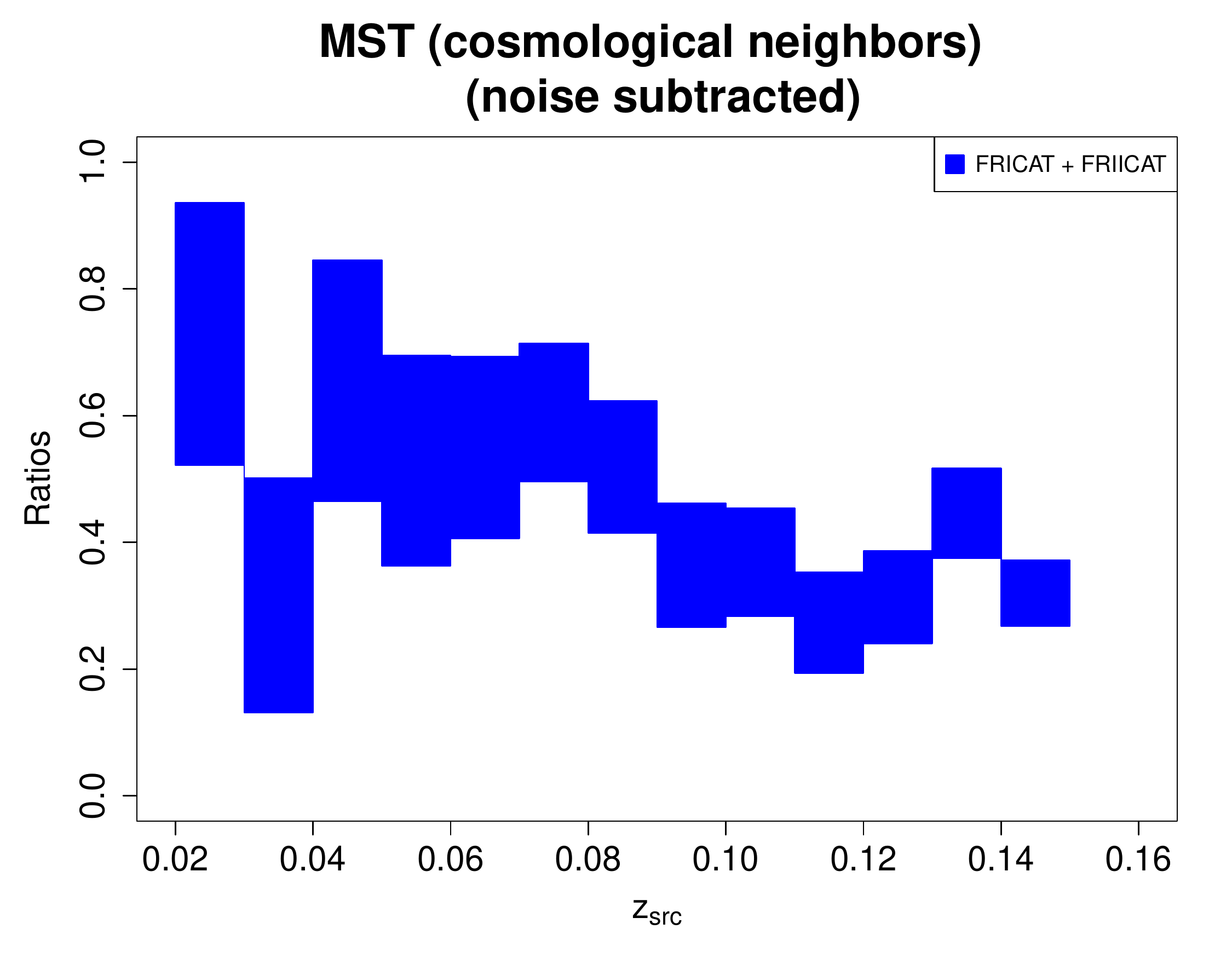}
\includegraphics[height=6.4cm,width=7.8cm,angle=0]{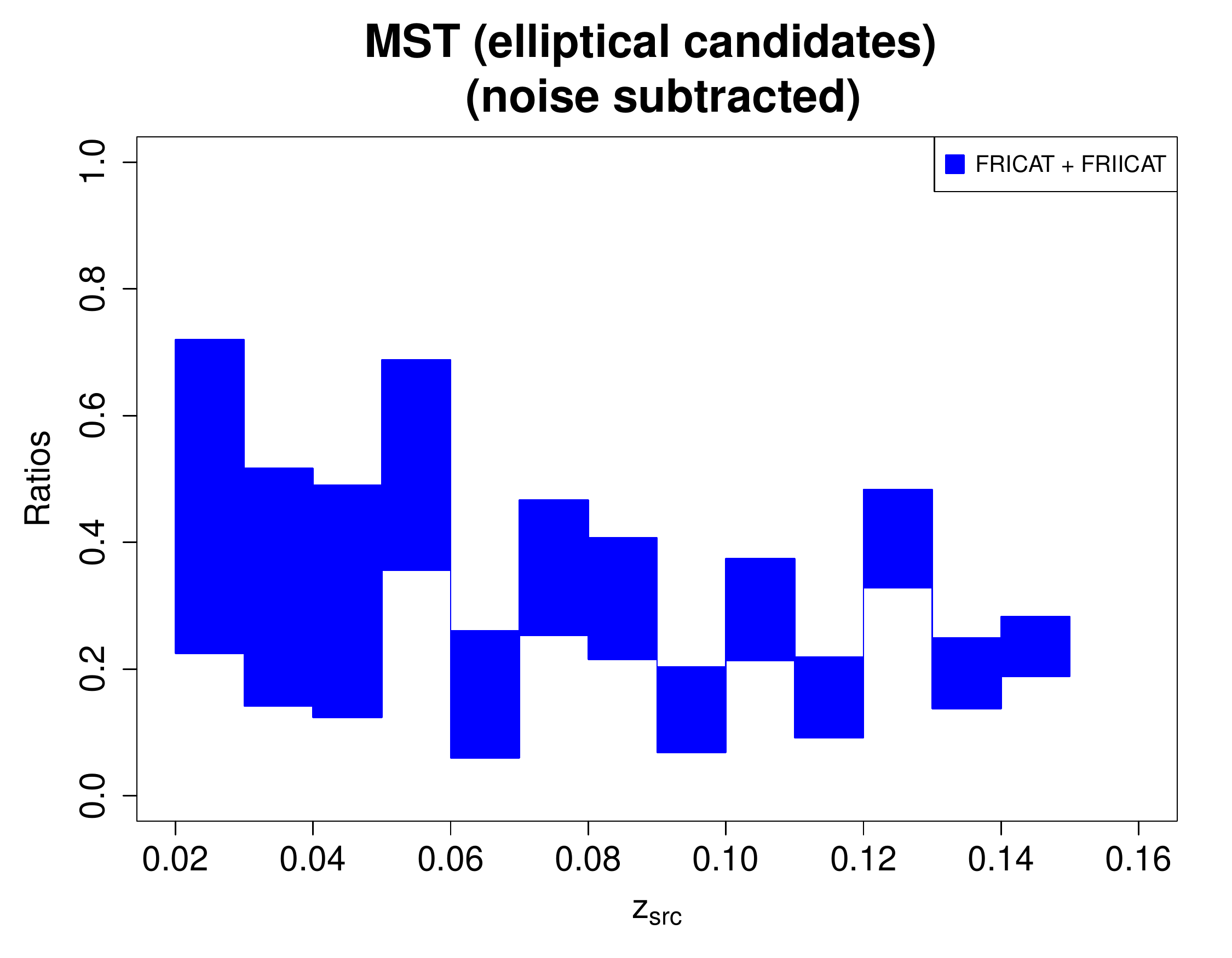}
\caption{Top left and top right panels are the same as Figure~\ref{fig:figure19} and Figure~\ref{fig:figure21}, where source fractions are computed with the MST algorithm only. The gap between real and fake sources appears well marked when the definition of galaxy-rich large-scale environments depend on the {\it cosmological neighbors} counts present in each MST cluster (central panel), while it is only rises at $z_\mathrm{src}$$>$0.1 when adopting the threshold for the number of {\it candidate elliptical galaxies} present in the 5\% of the MOCK sample (left panel). Lower panels show the same ratios but taking into account of the noise subtraction as described in \S~\ref{sec:analysis}.}
\label{fig:figure23}
\end{center}
\end{figure*}

As in the previous cases we chose the case of the MST procedure applied to one radio galaxy (see Figure~\ref{fig:figure22}). Then, thresholds to consider a source surrounded by a galaxy-rich, large-scale environment to the number of {\it cosmological neighbors} or {\it candidate elliptical galaxies}, respectively, belonging to a source cluster, detected applying the MST algorithm, to the MOCK sample in the top 5\% of the cases Results of the MST procedure are indeed shown in Figure~\ref{fig:figure23}.

\subsubsection{Galaxy density estimated with Kernel Density Estimation}
\begin{figure*}
\begin{center}
\includegraphics[height=5.2cm,width=5.8cm,angle=0]{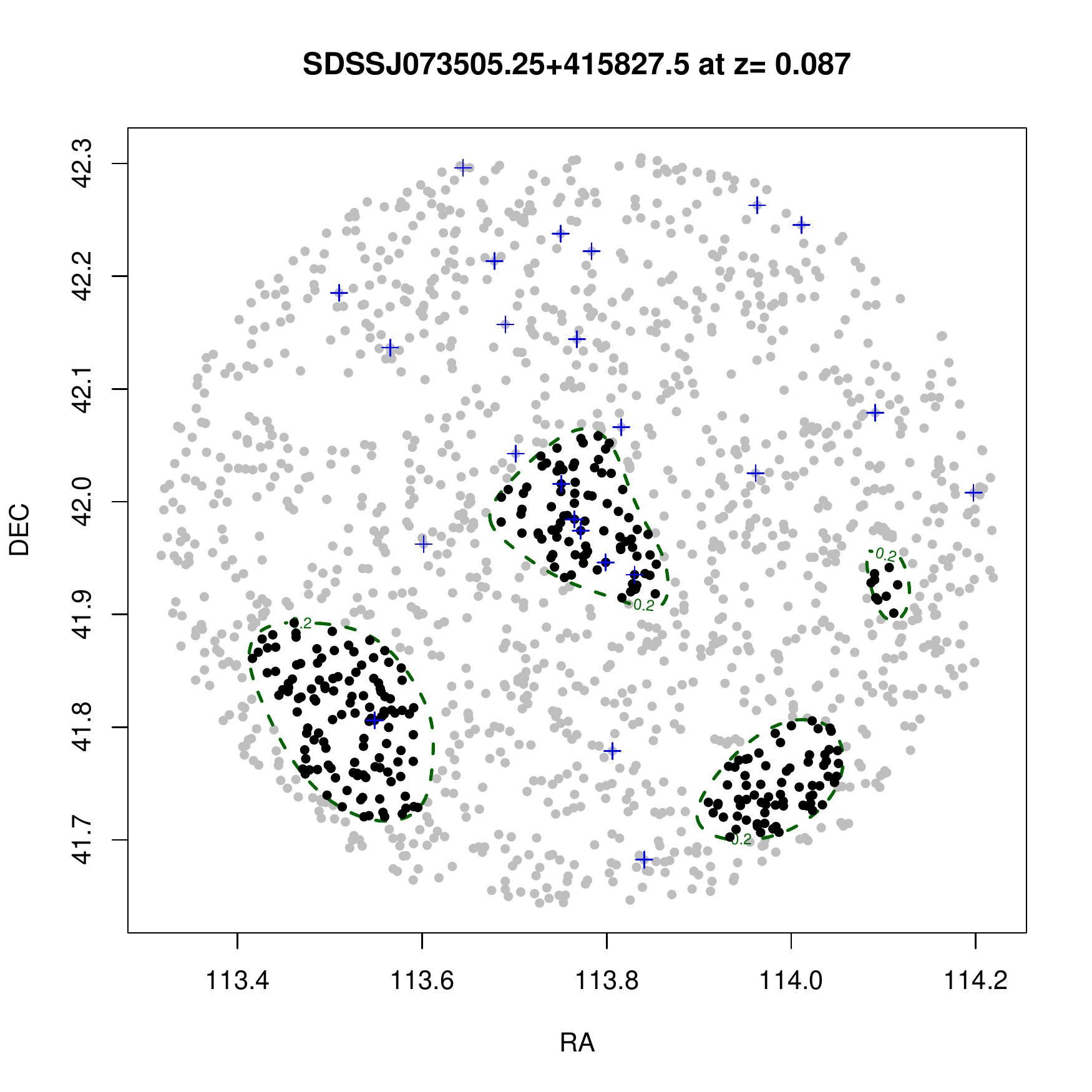}
\includegraphics[height=5.2cm,width=5.8cm,angle=0]{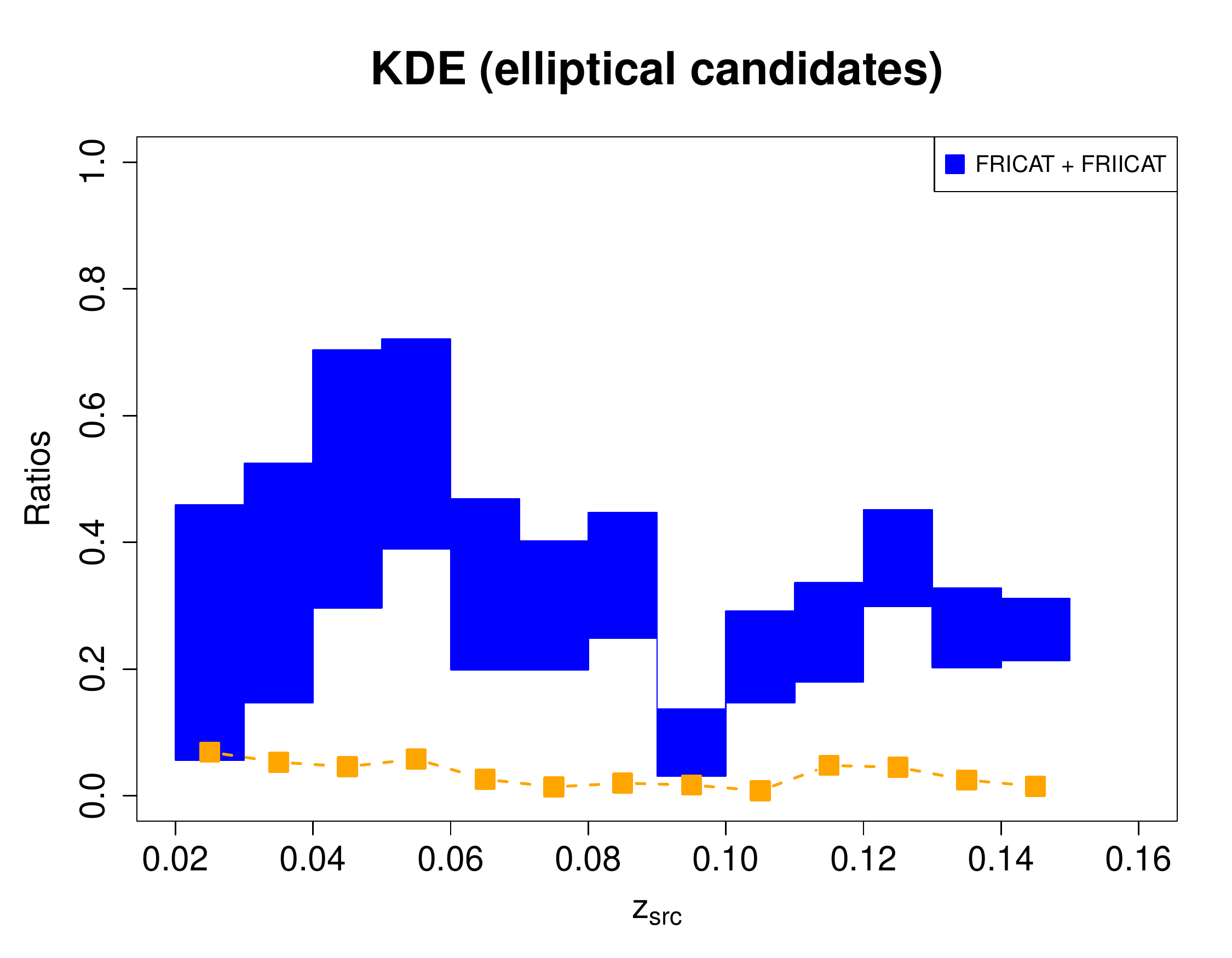}
\includegraphics[height=5.2cm,width=5.8cm,angle=0]{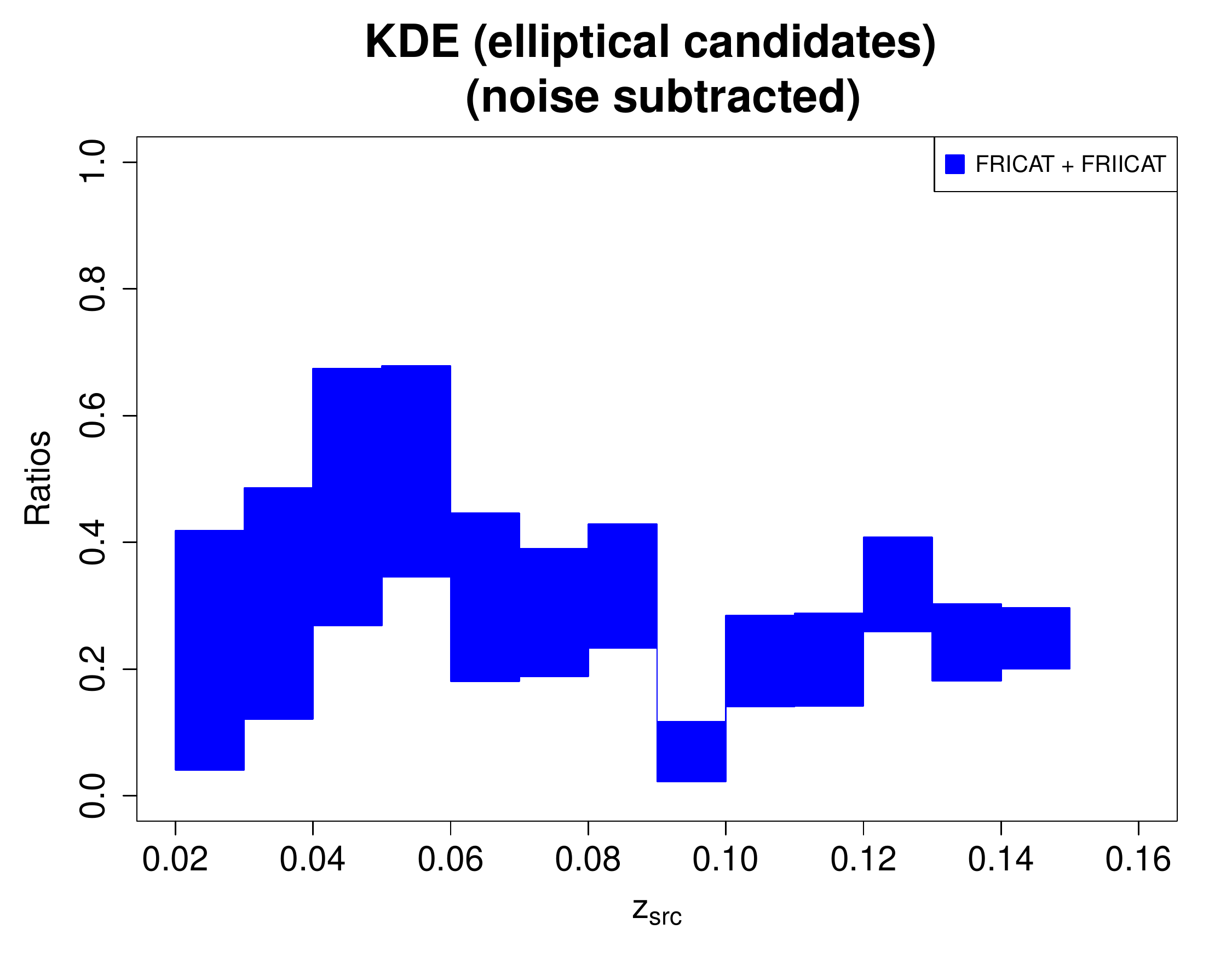}
\caption{All the SDSS galaxies in the 2\,Mpc circular region centered on the position of SDSS J073505.25+415827.5 at $z=$0.087 are shown in grey, while regions with galaxy-density larger than 20\% are marked with green contours computed applying the KDE technique and source included are highlighted in black. Blue crosses mark the location of {\it candidate elliptical galaxies}. Central panel is the same as Figure~\ref{fig:figure19}, Figure~\ref{fig:figure21} and Figure~\ref{fig:figure23} where source ratios are calculated with the KDE algorithm only. The gap between real and fake sources is again significant almost at all redshifts. Fluctuations in the radio galaxies samples are mainly due to their limited source number. Right panel, as in previous figures, show the same ratios of the central panel but noise subtracted.}
\label{fig:figure24}
\end{center}
\end{figure*}

The Kernel Density Estimation (KDE) is a method that provides an effective procedure to estimate the probability function of a multivariate variable without any assumption on the shape of the {\it parent} distribution \citep{richards04}. KDE divides the data set into a square grid and convolve the discrete data with a kernel function, estimating the density for each one of them. Isodensity contours drawn from its application and associated with 20\% of highest density level are plotted in Figure~\ref{fig:figure24} for the radio galaxy SDSS J073505.25+415827.5 at $z_\mathrm{src}$$=$0.087. All sources lying inside the top 20\% contour level are marked in black while contours are labeled in green. Background and foreground galaxies in the field are also shown in grey. This is also an example where the central source belongs to one of the high galaxy-density regions.

The threshold chosen to consider a source surrounded by a galaxy-rich large-scale environment was arbitrarily set to the top 20\% of high source-density regions found applying the KDE algorithm and including at least the number of {\it candidate elliptical galaxy} present in the top 5\% of the distribution of the MOCK sources, for the same z bin. The fractions of sources lying in galaxy-rich large-scale environments are plotted in Figure~\ref{fig:figure24} as function of their redshift where the gap between real and fake objects is again clear, particularly above $z_\mathrm{src}$$=$0.1 as occurs for the previous clustering algorithms.

\subsubsection{Comparison between cosmological over-densities and clustering algorithms}
\begin{figure}[]
\begin{center}
\includegraphics[height=6.2cm,width=8.2cm,angle=0]{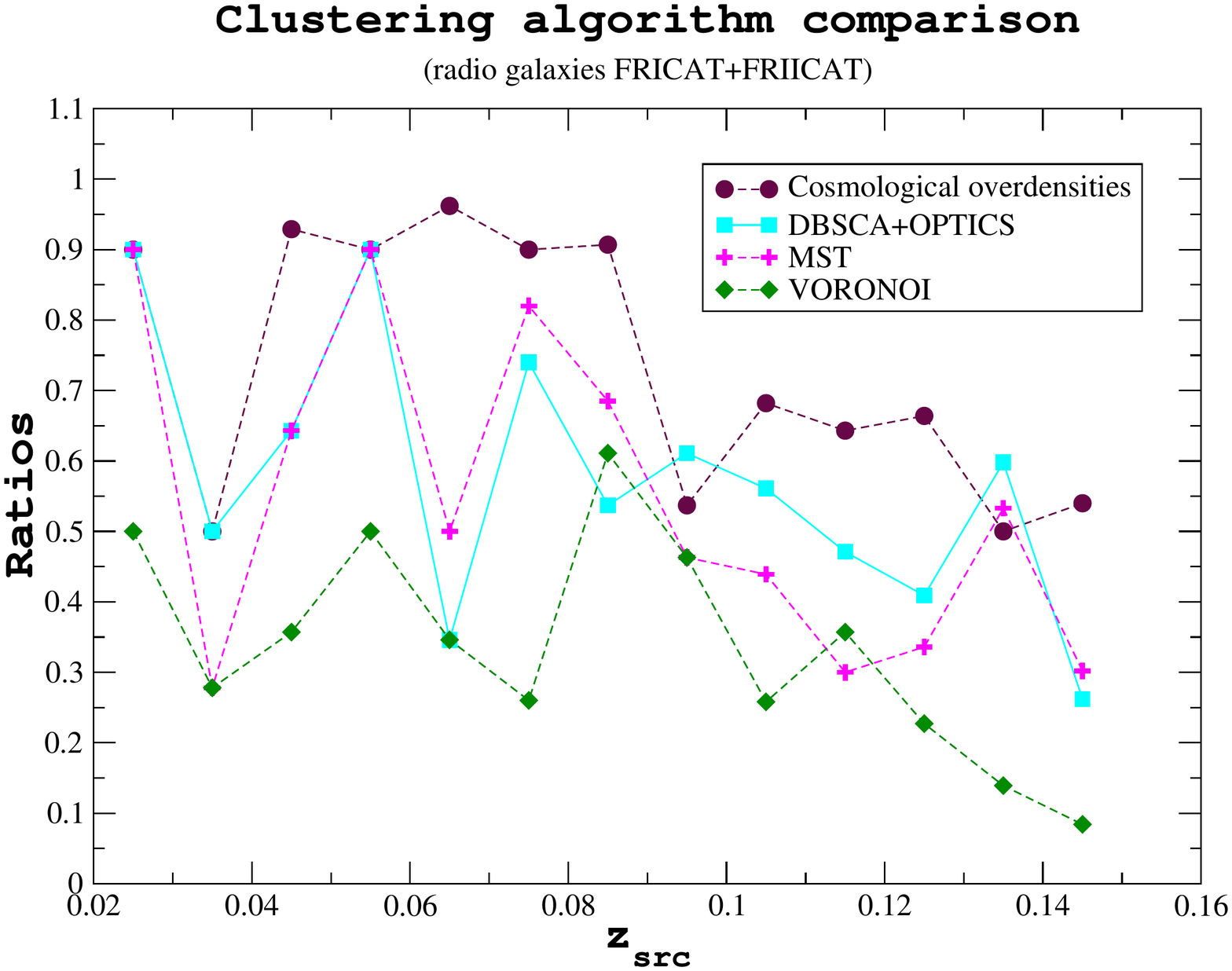}
\caption{The comparison between the results obtained on the whole radio galaxy sample (i.e., FRICAT + FRIICAT) with different clustering algorithms (see Appendix for a full description of these methods). Results achieved with the cosmological over-densities appear to be the best procedure, while Voronoi Tessellation seems to be the less efficient. Density-Based Spatial Clustering of Applications with Noise (DBSCAN) with the OPTICS implementation as well as the Minimum Spanning Tree (MST) seems to have the same results. Independently by the algorithm adopted radio galaxies live in galaxy-rich large-scale environments. Uncertainties are not reported in this plot since the main goal is to compare the efficiency of the procedures.}
\label{fig:figure25}
\end{center}
\end{figure}

The comparison between all three algorithms and their results obtained with the cosmological over-densities are shown in Figure~\ref{fig:figure25}. Here we report the total fraction of radio galaxies (FRICAT and FRIICAT together) living in galaxy-rich large-scle environment as function of their redshift. All these methods are not affected by a strong $z_\mathrm{src}$ dependence that could bias the results at low redshifts as occur for the cluster cross-matches (see Figure~\ref{fig:figure10}). Considering that their threshold is set at 5\% they are all in agreement with the statement that radio galaxies tend to live in dense environment than occurs randomly. MST and DBSCAN appear to have the same efficiency with the latter working better at higher redshifts. Voronoi Tessellation seems to be systematically weaker than all the others finding source clusters, under the assumption that all sources lie in galaxy groups/clusters.

Finally, we note that one of the major drawback of all clustering algorithms adopted, including the FoF method, used to create the T12 cluster catalog, is that they can find regions with higher galaxy density with respect to that of background/foreground sources in the field but in most of these cases such structures include objects at similar redshifts, that could be physically connected. The use of the cosmological over-density procedure, as well as the use of the clustering algorithms combined with the number of {\it cosmological neighbors} lying in detected clusters mitigates this bias.

\acknowledgments
We thank the anonymous referee for useful comments that led to improvements in the paper. 
F. M. also wishes to thank Prof. M Paolillo for his suggestions on the calculation of the binomial confidence intervals and Dr. C. C. Cheung for their valuable discussions on this project initially planned during the IAU 313 on the Galapagos islands.
This work is supported by the "Departments of Excellence 2018 - 2022" Grant awarded by the Italian Ministry of Education, University and Research (MIUR) (L. 232/2016). This research has made use of resources provided by the Compagnia di San Paolo for the grant awarded on the BLENV project (S1618\_L1\_MASF\_01) and by the Ministry of Education, Universities and Research for the grant MASF\_FFABR\_17\_01. This investigation is supported by the National Aeronautics and Space Administration (NASA) grants GO4-15096X, AR6-17012X and GO6-17081X. F.M. acknowledges financial contribution from the agreement ASI-INAF n.2017-14-H.0.
Funding for SDSS and SDSS-II has been provided by the Alfred P. Sloan Foundation, the Participating Institutions, the National Science Foundation, the U.S. Department of Energy, the National Aeronautics and Space Administration, the Japanese Monbukagakusho, the Max Planck Society, and the Higher Education Funding Council for England. The SDSS Web Site is http://www.sdss.org/. The SDSS is managed by the Astrophysical Research Consortium for the Participating Institutions. The Participating Institutions are the American Museum of Natural History, Astrophysical Institute Potsdam, University of Basel, University of Cam- bridge, Case Western Reserve University, University of Chicago, Drexel University, Fermilab, the Institute for Advanced Study, the Japan Participation Group, Johns Hopkins University, the Joint Institute for Nuclear Astrophysics, the Kavli Institute for Particle Astrophysics and Cosmology, the Korean Scientist Group, the Chinese Academy of Sciences (LAMOST), Los Alamos National Laboratory, the Max- Planck-Institute for Astronomy (MPIA), the Max-Planck- Institute for Astrophysics (MPA), New Mexico State University, Ohio State University, University of Pittsburgh, University of Portsmouth, Princeton University, the United States Naval Observatory, and the University of Washington. TOPCAT and STILTS astronomical software \citep{taylor05} were used for the preparation and manipulation of the tabular data and the images.

~

\end{document}